\documentclass[%
 reprint,
superscriptaddress,
 amsmath,amssymb,
 aps,
]{revtex4-1}

\usepackage{dcolumn}
\usepackage[mathlines]{lineno}
\usepackage[colorlinks=true,citecolor=blue,linkcolor=blue,anchorcolor=green, anchorcolor=green,urlcolor=blue,breaklinks=true]{hyperref}
\usepackage{appendix}
\usepackage{mathrsfs}
\usepackage{xcolor}
\usepackage{float}
\usepackage{ulem}
\usepackage{lipsum} 
\usepackage{amsmath}
\usepackage{commath}
\usepackage{graphicx,epsfig,amssymb,times} 
\usepackage{amsmath,amsfonts}
\usepackage{bm}
\usepackage{epstopdf}
\usepackage[capitalise]{cleveref}

\usepackage[caption=false]{subfig}
\usepackage{float}
\usepackage{natbib}
\usepackage{ulem}
\usepackage{cancel}
\usepackage{verbatim}
\usepackage{enumitem}
\usepackage[utf8]{inputenc}

\let\originalleft\left
\let\originalright\right
\renewcommand{\left}{\mathopen{}\mathclose\bgroup\originalleft}
\renewcommand{\right}{\aftergroup\egroup\originalright}
\mathcode`\*="8000
{\catcode`\*=\active\gdef*{\mathclose{}\,\mathopen{}}}

\newcommand{\be}{\begin{equation}}
\newcommand{\ee}{\end{equation}}
\newcommand{\bea}{\setlength\arraycolsep{2pt} \begin{eqnarray}}
\newcommand{\eea}{\end{eqnarray}}

\begin{document}
\title{Mass parameter and the bounds on redshifts and blueshifts of photons emitted from geodesic particle orbiting in the vicinity of regular black holes}

\author{Ricardo Becerril}
\email{ricardo.becerril@umich.mx}
\affiliation{Instituto de F\'{\i}sica y Matem\'{a}ticas, Universidad Michoacana de San Nicol\'{a}s de Hidalgo,\\
Edificio C-3, 58040 Morelia, Michoac\'{a}n, M\'{e}xico.}

\author{Susana Valdez -Alvarado}
\email{svaldeza@uaemex.mx}
\affiliation{Facultad de Ciencias, Universidad Aut\'onoma del Estado de M\'exico, Instituto Literario 100,\\
C.P. 50000, Toluca, Edo. M\'ex, M\'exico.}

\author{Ulises Nucamendi}
\email{unucamendi@gmail.com}
\affiliation{Instituto de F\'{\i}sica y Matem\'{a}ticas, Universidad Michoacana de San Nicol\'{a}s de Hidalgo,\\
Edificio C-3, 58040 Morelia, Michoac\'{a}n, M\'{e}xico.}
\affiliation{Mesoamerican Center for Theoretical Physics, Universidad Aut\'{o}noma de Chiapas, Ciudad Universitaria, Carretera Zapata Km. 4, Real del Bosque (Ter\'{a}n), 29040 Tuxtla Guti\'{e}rrez, Chiapas, M\'{e}xico}

\author{Pankaj Sheoran}
\email{hukmipankaj@gmail.com}
\affiliation{Instituto de F\'{\i}sica y Matem\'{a}ticas, Universidad Michoacana de San Nicol\'{a}s de Hidalgo,\\
Edificio C-3, 58040 Morelia, Michoac\'{a}n, M\'{e}xico.}

\author{J. M. D\'avila}
\email{jmdavilad@uaemex.mx}
\affiliation{Facultad de Ciencias, Universidad Aut\'onoma del Estado de M\'exico, Instituto Literario 100,\\
C.P. 50000, Toluca, Edo. M\'ex, M\'exico.}

\begin{abstract}
  We obtain the mass parameter for a class of static and spherically symmetric regular black holes (BHs) (namely Bardeen, Hayward and Ay\'{o}n-Beato-Garc\'{i}a BHs) which are solutions of Einstein's field equations coupled to nonlinear electrodynamics (NED) in terms of redshifts and blueshifts of photons emitted by geodesic particles (for instance, stars) orbiting around these BHs. The motion of photons is not governed by null geodesics for these type of spacetime geometries which reflects the direct effects of the electrodynamic nonlinearities in the photon motion; hence, an effective geometry needs to be constructed to study null trajectories 
  \cite{Novello:1999pg}. 
  To achieve the above, we first study the constants of motion from the analysis of the motion of both geodesic particles moving in stable circular orbits and photons ejected from them and reaching a distant observer (or detector) in the equatorial plane for the above mentioned regular BHs. The relationship between red/blueshifts of photons and the regular BH observables is presented. We also numerically find the bounds on the photon shifts for these regular BH cases.
\end{abstract}

\maketitle

\section{Introduction}
\label{sec:intro}
The existence of black holes (BHs) in the cosmos has been confirmed by recently detected gravitational waves (GWs) 
by LIGO and Virgo collaborations
\cite{Abbott:2016blz,Abbott:2016nmj} and with the first direct image of compact object M87 by the Event Horizon Telescope (EHT) collaboration \cite{Akiyama:2019cqa, Akiyama:2019brx, Akiyama:2019sww, Akiyama:2019bqs, Akiyama:2019fyp, Akiyama:2019eap}. BHs, which are characterized by three externally observed classical parameters namely, mass $M$, spin $a$ and charge $Q$, are one of the most interesting compact objects predicted by Einstein's general relativity theory (GRT), as well as by other modified gravitational theories (i.e., f(R) \cite{delaCruzDombriz:2009et}, Gauss-Bonnet \cite{Cvetic:2001bk}, Lovelock \cite{Garraffo:2008hu}, massive gravity \cite{Babichev:2015xha}, modified gravity \cite{Moffat:2014aja}, string theory \cite{Maldacena:1996ky}, and some more.).

Hence, to know about the properties of BHs and the physics in their vicinity, it is important to determine BH observables (i.e. $M$, $a$ and $Q$) in terms of quantities which are astrophysically relevant. With this motivation at hand, Herrera and Nucamendi (hereafter known as HN) came up with an algorithm to obtain the mass and spin parameters of a Kerr BH as a function of red/blueshifts ($z_{r}$ and $z_{b}$, respectively) of photons emitted by geodesic particles (say stars) orbiting in stable circular orbits around it and the radius of their respective orbits \cite{Herrera-Aguilar:2015kea}. In their work they found an explicit expression for the parameter $a$ in terms of $z_{r}$, $z_{b}$, radius of circular orbit and the parameter $M$ of Kerr BH. However, they argued that parameter $M$ cannot be found explicitly as it satisfies an eighth order polynomial in $M$ and one needs to find it numerically. Later, the HN algorithm was used to obtain the observables of boson stars and Reissner-Nordstr\"{o}m BH in \cite{Becerril:2016qxf} and for rotating Kerr-MOG BH in \cite{Sheoran:2017dwb}. As HN theoretical approach has been used only for a few compact objects till now, it is important to develop more theoretical templates using HN algorithm, such as if we have a set of observational data containing red/blueshifts emitted by geodesic particles orbiting around a BH at different radii $\{z_{r},z_{b}, r\}$, then it is possible to constrain the values of BH observables using that data set.

Additionally, it is widely known in GRT that collapsing matter forms a spacetime singularity if the strong energy condition and the existence of global hyperbolicity hold \cite{Hawking:1969sw, Hawking:1973uf}. A singularity is a location in spacetime where the laws of normal physics breaks down. However, it is also well established that these singularities are only due to the limitations of GRT and must be get rid of in a quantum gravity theory. Though we do not have any trustable theory of quantum gravity till now, many phenomenological attempts have been made to get rid of these spacetime singularities and study its effects. In this context, many researchers proposed BH solutions even in GRT which are popularly known as regular BHs \cite{Bardeen:1968,Borde:1996df,AyonBeato:1998ub, Hayward:2005gi,Bronnikov:2005gm}. These types of BH solutions do have an event horizon $r_{eh}$ but no spacetime singularities. In these regular BHs models the singularity theorem is bypassed in a way such that these BH solutions only satisfy the weak energy condition and not the strong one. It is important to note that these (regular) BHs are not the solution of Einstein vacuum field equations, but can be found either by using nonlinear electrodynamics, or by revamping the gravity. Interestingly, the recent research on gravitational wave echoes by Abedi and Afshordi \cite{Abedi:2018npz} also gave a hint about the existence of quantum BHs.

\begin{figure}[h]
   \centering
    \includegraphics[width=\linewidth]{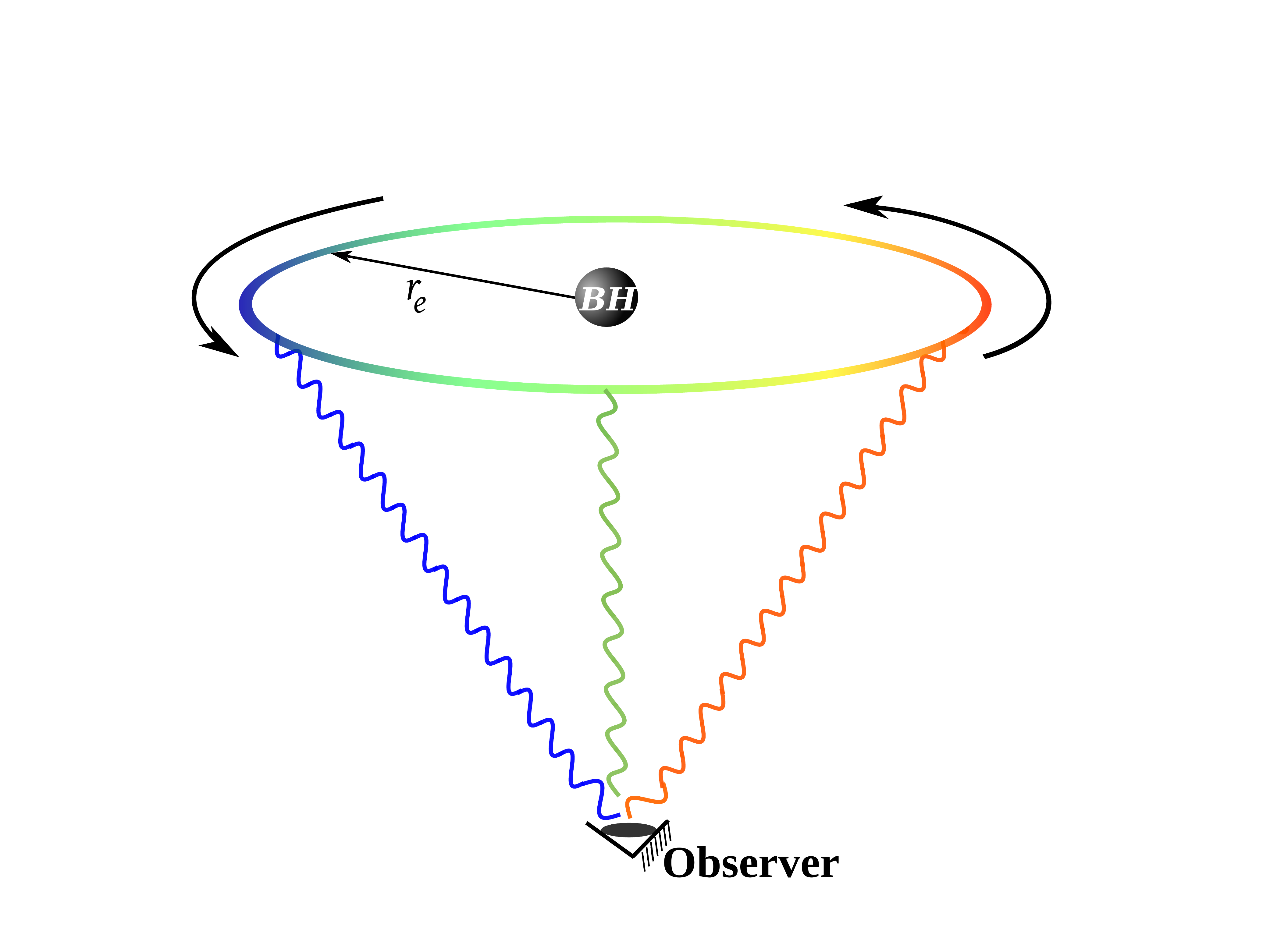}
    \caption{Pictorial representation of the setup in which a massive geodesic test object is moving around a BH in a stable circular orbit. Here, emitter, observer and the trajectory of photons all lie in the equatorial plane $(\theta=\frac{\pi}{2})$.}
    \label{Fig1}
\end{figure}

Motivated by this line of research, in this paper we specifically infer the mass parameter of static and spherically symmetric (SSS)  Bardeen, Hayward and Ay\'{o}n-Beato-Garc\'{i}a (ABG) BHs in terms of red/blueshifts of photons emitted by geodesic particles moving in their stable circular orbits around these regular BHs (see Fig.\ref{Fig1} for pictorial representation of the setup used). In recent times, these three SSS regular BHs gain lot of attentions from the researchers and appeared in many notable works on geodesics of particles \cite{Stuchlik:2014qja,Berry:2020ntz,Amir:2020zab,Abbas:2014oua,Chiba:2017nml,Mondal:2020uwp,Wei:2015qca,Garcia:2013zud,Zhou:2011aa,Hu:2018nhg}, quasinormal modes \cite{Li:2013fka,Flachi:2012nv,Toshmatov:2015wga,Lin:2013ofa,Yekta:2019por}, gravitational lensing \cite{Schee:2017hof,2016arXiv160105749G,Schee:2015nua}, thermodynamics of BHs \cite{Maluf:2018lyu,Ma:2014qma}, BH as a particle accelerator \cite{Ghosh:2014mea,Pradhan:2014oaa}. Also, these regular BHs have been studied in \cite{Myung:2007qt,Kim:2008hm,Halilsoy:2013iza,Macedo:2015qma,Huang:2015oua,Schee:2016mjd,Sharif:2016gyb,Sharif:2018gaj,Frolov:2016pav,Sharif:2014cxa,Perez-Roman:2018hfy,Carballo-Rubio:2018pmi,Ansoldi:2008jw,Bargueno:2020ais,Tsukamoto:2020iez,Ditta:2020npd,2016PhRvD..94l4027F,Bronnikov:2017tnz,Bronnikov:2000vy,Pellicer:1969cf,BRONNIKOV197984,Burinskii:2002pz,Matyjasek:2008kn,Matyjasek:2013dua,Fernando:2015fha}. 
It is worth it to point out here that we consider that, without loss of generality due to spherical symmetry, the whole setup which comprises of the emitter (star), detector (observer) and photons will lie in the equatorial plane ($\theta=\pi/2$) of BH.

The remainder of paper is organized as follows: In Sec. \ref{sec:HN_algorithm}, we give a brief overview of the HN algorithm and note down the general form of some key quantities for a SSS spacetime that are important for later sections. Additionally, in Sec. \ref{Circular Orbits} we also find the expressions like effective potential ($V_{eff}$), conserved energy per unit mass ($E$), conserved orthogonal component of orbital angular momentum per unit mass to the azimuthal rotation ($L$), double derivative of ($V_{eff}$), azimuthal and time components of 4-velocity (i.e., $U^{\phi}$ and $U^{t}$), angular velocity ($\Omega$) of the geodesic particles moving in circular orbits in the equatorial plane around the SSS of compact objects and redshift ($z_{r}$) of the photons emitted from them.  In Sec. \ref{sec:Bardeen BH}, we work with the metric function $f(r)$ for Bardeen regular BHs, discuss the properties of event horizon, find the associated explicit expressions for conserved quantities $E$ and $L$ for geodesics particles moving around these BHs in stable circular orbits with the help of general formulas found in Sec. \ref{sec:HN_algorithm}. Later, we obtain the mass parameter $M$ as a function of frequency shift $z$, radius of stable circular orbits $r_{e}$ (later for simplicity used as $r$) and the parameter $g$. We also verified our results with the Schwarzschild BH in the limit $g\to0$. Finally, in Sec. \ref{sec:Bardeen BH}, we graphically find the bounds on frequency shift $z$ and analyze the mass parameter $M$ for Bardeen regular BH as a function of $r$ and $z$. These analyses must be carried out numerically; hence, in this section, we also provide a numerical algorithm which is employed with the regular Hayward and ABG BHs in Sec. \ref{Hayward BH} and \ref{ABG BH} respectively. 
It is worth mentioning that we also used equations of motion for photons emitted by geodesic particle and traveling along null geodesics in the effective geometry \cite{Novello:1999pg} before getting detected by an observer (residing at a far away location from the emitter). We finally present a summary of our work and conclude with the results in Sec. \ref{Summary and Conclusion}. Further, by completeness, the calculation of effective metrics for Bardeen, Hayward and ABG BHs spacetimes is presented in Appendix \ref{effective_geo}, in which we have explicitly obtained the effective metric components for these regular BHs spacetimes using the prescription of the geometric optic approach in NED theories given originally in \cite{Novello:1999pg} and later in \cite{2019ApJ,Schee_2019,2019EPJC...79...44S}.
Throughout the paper, we work with the sign conventions $(-,+,+,+)$ and use the spherical polar coordinate system. Greek letters denote the spacetime indices, while Latin letters use for space indices only. Unless otherwise stated, geometric units are used for the fundamental constants, $c=G_N=1$.

\section{Brief overview of Herrera-Nucamendi theoretical approach}
\label{sec:HN_algorithm}
The starting point of the HN \cite{Herrera-Aguilar:2015kea} theoretical approach, is the definition of the frequency shift $z$ associated to the emission and detection of photons

\begin{equation}
1+z= \frac{\omega_e}{\omega_d}.
\label{zzz}
\end{equation}

\noindent
Here $\omega_e$ is the frequency emitted by an observer moving with 
a photon emitter particle at the point $e$ and $\omega_d$
the frequency detected by an observer far away from the source
of emission. These frequencies are given by

\begin{equation}
\omega_e = - k_{\mu} U^{\mu} |_e \quad , \quad
\omega_d = - k_{\mu} U^{\mu} |_d,
\label{emission}
\end{equation}

\noindent 
where $U^{\mu}=(U^t,U^r,U^{\theta},U^{\phi})$ is the 4-velocity of a particle that moves in a geodesic trajectory in certain spacetime, particularly we deal with static spherically symmetric spacetimes

\be
    ds^{2}=-f(r)dt^{2}+\frac{1}{f(r)}dr^{2}+r^{2}\left(d{\theta}^2+\text{sin}^{2}\theta d{\phi}^{2}\right).
    \label{metric}
\ee

The geodesic particles are emitting photons which have 4-momentum 
$k^{\mu}=(k^t,k^r,k^{\theta},k^{\phi})$ that move along null geodesics $k_{\mu} k^{\mu}=0$.

In order to compute the frequencies (\ref{emission}) we need to find $U^{\mu}$ as well as $k^{\mu}$. The 4-velocities $U^{\mu}$ can be found by considering the Euler-Lagrange equations

\begin{equation}
\frac{\partial \mathcal{L}}{\partial x^{\mu}} -
\frac{d}{d \tau} \left ( \frac{\partial \mathcal{L}}{\partial \dot{x}^{\mu}}  
\right )=0,
\label{ELE}
\end{equation}

\noindent with the Lagrangian $\mathcal{L}$ given by

\begin{equation}
\mathcal{L}= \frac{1}{2} \left [
g_{tt} \dot{t}^2 + g_{rr} \dot{r}^2 + 
g_{\theta \theta} \dot{\theta}^2+ g_{\phi \phi} \dot{\phi}^2 \right ],
\label{lagrangian}
\end{equation}

\noindent 
being $\dot{x}^{\mu}\equiv \frac{d x^{\mu}}{d\tau}$
and $\tau$ the proper time. Since the metric (\ref{metric}) depends
solely on $r$ and $\theta$ there are two quantities 
that are conserved along the geodesics

\begin{eqnarray}
p_t&=& \frac{\partial \mathcal{L}}{\partial \dot{t}} = 
g_{tt} \dot{t} = g_{tt} U^t = -E, \nonumber \\
p_{\phi} &=& \frac{\partial \mathcal{L}}{\partial \dot{\phi}} = 
g_{\phi \phi} \dot{\phi}= g_{\phi \phi} U^{\phi}= L.
\label{constants}
\end{eqnarray}

From (\ref{constants}) two components of the 4-velocity vector are readily found 

\begin{equation}
U^t = -\frac{E}{g_{t t}}= \frac{E}{f(r)}
\quad , \quad 
U^{\phi} = \frac{L}{g_{\phi \phi}}= \frac{L}{r^2 \text{sin}^2 \theta}.
\label{us}
\end{equation}

\noindent
The normalized 4-velocity condition 

\begin{equation}
-1 = g_{tt} (U^t)^2 + g_{rr} (U^r)^2 + g_{\theta \theta} (U^{\theta})^2+
g_{\phi \phi} (U^{\phi})^2.
\label{condition}
\end{equation}

\noindent renders

\be
(U^{r})^{2} + f(r)\;V_{eff}=0,
\ee

\noindent here $V_{eff}$ is an effective potential given by

\be
V_{eff}=1 + r^2 U^{\theta}-\frac{E^{2}}{f(r)} + \frac{L^2}{r^2 \text{sin}^2 \theta}.
\ee

The 4-momentum $k^{\mu}$ can be obtained in a similar fashion. Using the same Lagrangian (\ref{lagrangian})
one gets two conserved quantities for photons: the energy and the orthogonal component of the angular momentum to the azimuthal rotation respectively   

\begin{equation}
E_{\gamma}= f(r)k^{t} \quad, \quad L_{\gamma}= r^{2}\text{sin}^2 \theta \, k^{\phi}.
\label{ks}
\end{equation}

Thereby $k^{t}$ and $k^{\phi}$ can be written in terms of $E_{\gamma}$ and $L_{\gamma}$ respectively, which are needed in (\ref{emission}) which in turn are required to obtain $1+z$ which reads now as

\begin{equation}
1+z=
\frac{\left ( E_{\gamma}U^t-L_{\gamma}U^{\phi}- U^r k^r/f(r)-
r^2 U^{\theta} k^{\theta} \right)|_e}
{ \left ( E_{\gamma}U^t-L_{\gamma}U^{\phi}- U^r k^r/f(r)-
r^2 U^{\theta} k^{\theta} \right)|_d }.
\label{oneplusz}
\end{equation}

Astronomers report the observational data in terms of a kinematic frequency shift $z_{kin}$ defined as
$z_{kin}=z-z_c$ where $z_c$ known as the central frequency shift, corresponds to a gravitational frequency shift of a photon emitted by a static particle located in on the line going from the center of coordinates to the far away observer (represented by the green zig-zag line in Fig. \ref{Fig1}). Hence

\begin{equation}
1+z_c=\frac{(E_{\gamma} U^t)|_e}{(E_{\gamma} U^t)|_d} = \frac{U^t_e}{U^t_d}. 
\end{equation}

\noindent The kinematic redshift $z_{kin}=(1+z)-(1+z_c)$ can be written as

\begin{equation}
z_{kin} =
\frac{(U^t-bU^{\phi}-\frac{1}{E_{\gamma}f(r)} U^rk^r-\frac{1}{E_{\gamma}}
  r^2 U^{\theta}k^{\theta})|_e}
{(U^t-bU^{\phi}-\frac{1}{E_{\gamma}f(r)}U^rk^r-\frac{1}{E_{\gamma}}
  r^2 U^{\theta}k^{\theta})|_d}-\frac{U^t_e}{U^t_d},
\label{zcine}
\end{equation}

\noindent 
where we have introduced the quantity $b\equiv L_{\gamma}/E_{\gamma}$  known as the apparent impact parameter of photons. As the photon energy $E_{\gamma}$ and the orthogonal component of orbital angular momentum $L_{\gamma}$ to the azimuthal rotation are conserved along null trajectories from the point of emission until detection, the value of the impact parameter is preserved i.e., $b_{e}=b_{d}$. 

The analysis may be carried out either with $z$ given by (\ref{oneplusz}) or with $z_{kin}$, in this paper we work with the latter. The expression (\ref{zcine}) is rather simplified for circular orbits ($U^r=0$) in the equatorial plane ($U^{\theta}=0$)

\begin{equation}
z_{kin}= \frac{U_e^t U_d^{\phi} b_d- U^t_d U^{\phi}_e b_e}
{U^t_d(U^t_d-b_d U^{\phi}_d)}.
\label{zkin}
\end{equation}

What is not yet included in (\ref{zkin}) is the light bending due to gravitational field, that is to say, we still need to find $b=b(r_c)$ where $r_c$ is the radius of the circular orbit of the photons emitter. To construct this mapping, we consider photons emitted at both sides of the compact object as shown in Fig. \ref{Fig1} by the red and blue zigzag lines. At those points $k^r=0$ and $k^{\theta}=0$, whereas $k^t$ and $k^{\phi}$ are already known and are given in (\ref{ks}). From
$k_{\mu} k^{\mu} = 0$ one attains

\begin{equation}
b_{\pm}= \pm \sqrt{- \frac{g_{\phi \phi}}{g_{tt}}}= \pm \frac{r}{\sqrt{f(r)}}.
\label{bmm}
\end{equation}

However, if the observer is very far from the photons emitter (i.e. $r \rightarrow \infty$), all the spatial components of the detector's four-velocity (i.e., $U^{r},U^{\theta}$ and $U^{\phi}$) vanish in this asymptotic limit, except the component $U^{t}=1=E$. In this case, the kinematic frequency shift
(\ref{zkin}) reduces to

\be 
\label{zka}
z_{kin}= - U^{\phi} b_{e}.
\ee

According to (\ref{bmm}), the impact parameter $b(r)$ may have two different signs, so does the kinematic frequency shift $z_{kin}$. The frequency shift corresponding to receding emitter is known as a redshift ($z_{r}>0$), while the frequency shift of an approaching emitter is known as a blueshift ($z_{b}<0$).

\subsection{Circular orbits}
\label{Circular Orbits}

For equatorial orbits the effective potential acquires a simple form (for the case of equatorial orbits, the orthogonal component of the orbital angular momentum to the azimuthal rotation $L$ is equal to the total orbital angular momentum.)

\begin{equation}
V_{eff} = 1 + \frac{E^2}{g_{t t}}+\frac{L^2}{g_{\phi \phi}}=1 - \frac{E^2}{f(r)} + \frac{L^2}{r^2}.
\label{potentialreduce}
\end{equation}
For circular orbits $V_{eff}$ and its derivative $\frac{d V_{eff}}{dr}$ vanish.  From these two conditions one finds two general expressions for the constants of motion $E^2$ and $L^2$ for any static spherically symmetric spacetime

\begin{equation}
E^2=-\frac{g_{tt}^2 g_{\phi \phi}^{\prime}}
{g_{tt} g_{\phi\phi}^{\prime}-g_{tt}^{\prime} g_{\phi \phi}}=\frac{2f^2(r)}{2f(r)-r f^{\prime}(r)},
\label{energy}
\end{equation}

\begin{equation}
L^2=\frac{g_{\phi \phi}^2 g_{tt}^{\prime}}
{g_{tt} g_{\phi\phi}^{\prime} - g_{tt}^{\prime}g_{\phi \phi} }=\frac{r^3f^{\prime}(r)}{2f(r)-rf^{\prime}(r)},
\label{angularM}
\end{equation}

\noindent 
where primes denote derivative with respect to $r$. Stability of circular orbits is guaranteed provided that $V_{eff}^{\prime \prime}>0$ holds.

The general expression for $V_{eff}^{\prime \prime}$ reads

\begin{eqnarray}
V_{eff}^{\prime \prime} &=& -E^2 \left [ 
\frac{g_{tt}^{\prime \prime}g_{tt}-2 (g_{tt}^{\prime})^2}{g_{tt}^3}
\right ]
-L^2 \left [ 
\frac{g_{\phi \phi}^{\prime \prime}g_{\phi \phi}-2 (g_{\phi
    \phi}^{\prime})^2}{g_{\phi \phi}^3} \right ] \nonumber \\
&=& \frac{g_{\phi \phi}^{\prime} g_{tt}^{\prime\prime}
 - g_{tt}^{\prime} 
g_{\phi \phi}^{\prime \prime}}{g_{tt}g_{\phi \phi}^{\prime}-g_{tt}^{\prime} g_{\phi \phi}} 
+\frac{2 g_{tt}^{\prime} g_{\phi \phi}^{\prime}}{g_{\phi \phi}g_{tt}} \nonumber \\
&=& \frac{2\left[r f(r) f^{\prime \prime}(r)+3f(r)f^{\prime}(r)-2r f^{\prime}(r) \right]^2}{r f(r)\left[2f(r)-rf^{\prime}(r) \right]}.
\label{segundita}
\end{eqnarray}

\noindent The explicit expressions for the energy and angular momentum, (\ref{energy}) and (\ref{angularM}), were used in the last step. If we employed them now in (\ref{us}) one obtains expressions for the 4-velocities in terms of $f(r)$ only

\begin{equation}
    U^{\phi}= \sqrt{\frac{f^{\prime}(r)}{r(2f(r)-r f^{\prime}(r))}} \quad , \quad U^t=\sqrt{\frac{2}{2f(r)-r f^{\prime}(r)}}.
    \label{vels}
\end{equation}
    
\noindent 
The angular velocity of particles in these circular paths can be readily found

\begin{equation}
\Omega = \sqrt{-\frac{g_{tt}^{\prime}}{g_{\phi \phi}^{\prime}}}= \sqrt{\frac{f^{\prime}(r)}{2r}}.
\label{Omega}
\end{equation}

Since we have an explicit expression for both, $U^{\phi}_e$ and $b_e$, the frequency shift becomes

\begin{eqnarray}
z = U^{\phi}_e b_{e_{+}}&=& 
\sqrt{\frac{-g_{\phi \phi}g_{tt}^{\prime}}{g_{tt} (g_{tt} g_{\phi\phi}^{\prime} - 
g_{tt}^{\prime} g_{\phi \phi} )}} \nonumber\\
&=& \sqrt{\frac{rf^{\prime}(r)}{f(r)(2f(r)-r f^{\prime}(r))}}.
\label{zfinalFB}
\end{eqnarray}

It has been argued that photons motion in nonlinear electrodynamics regular black holes is not governed by null geodesics of the spacetime geometry. In order to see reflected the direct effects of the electrodynamic nonlinearities in the photons motion, the null geodesics should be studied using an effective geometry. The original approach to derive effective metrics can be found in (\cite{Novello:1999pg}). Bardeen and Hayward regular BH were first presented as toy models, they were not exact solutions to Einstein equations; there were no known physical sources associated with any of them. Later on, they were interpreted as singularity-free solutions of the Einstein field equations coupled to a
suitable nonlinear electrodynamics. Our third example in this paper, Ayon-Beato-Garcia (ABG) regular black hole, was constructed from the start in the framework of general relativity as a solutions of the Einstein field equations coupled to nonlinear electrodynamics (\cite{AyonBeato:2000zs}). Some studies of null geodesics in these toy models have been made without considering effective metrics. 
Since we are going to be dealing with regular BHs where light travel along null geodesics with an effective metric $\widetilde{g}_{\mu \nu}$, the previous expression (\ref{zfinalFB}) needs to be modified to take this fact into account. It was in the derivation of the apparent impact parameter where the null geodesic equation was employed, then (\ref{zfinalFB}) needs to be replaced by

\begin{equation}
z = U^{\phi}_e b_{e_{+}} =
\sqrt{\frac{-\widetilde{g}_{\phi \phi}g_{tt}^{\prime}}{\widetilde{g}_{tt} (g_{tt} g_{\phi\phi}^{\prime} - 
g_{tt}^{\prime} g_{\phi \phi} )}}.
\label{zfinalFBeffective}
\end{equation}

The computation of the effective metrics $\widetilde{g}_{\mu \nu}$ for Bardeen, ABG and Hayward BHs can be found in the Appendix. In the next section, we will analyze the relationship between the mass parameter of three regular BHs in terms of the red/blueshifts of light emitted by geodesic particles orbiting in circular trajectories of radius $r_c$. The analysis will be carried out using both expressions (\ref{zfinalFB}) and (\ref{zfinalFBeffective}), the former does not take into account the effective metric, the later does. In the literature one can find null geodesics studies that inappropriately, do not take into account the fact that an effective metric must be employed. By carrying out the analysis with both expressions (\ref{zfinalFB}) and (\ref{zfinalFBeffective}), we can compare both results and see the differences of neglecting the effective metric in regard to the computation of the mass parameter in terms of circular orbits radii and red/blueshifts. For each of our three working examples, we will find bounds of these frequency shifts as well.

\section{Bardeen Regular Black Hole}
\label{sec:Bardeen BH}
Our first working example is the Bardeen spacetime. This was the first regular BH {\it model} in general relativity, it was later on reinterpreted as the gravitational field of a nonlinear magnetic monopole, namely as a magnetic solution to Einstein equations coupled to a nonlinear electrodynamics \cite{AyonBeato:2000zs}. Hence, here we will work it as a toy model first and then as a solution of Einstein equation with NED. For this spacetime, the function $f(r)$ reads

\begin{equation}
    f(r)=1-\frac{2M}{r} \left [ \frac{r^2}{r^2+g^2} \right ]^{3/2} \equiv 1- \frac{2M}{r}R_B(r,g).
    \label{f_B}
\end{equation}

\begin{figure}[h!]
    \includegraphics[scale=0.8]{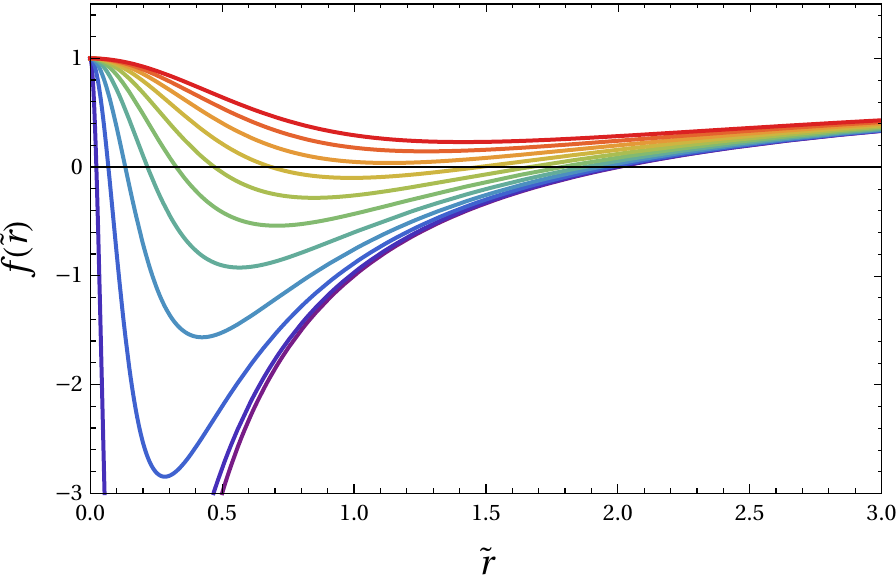}
    \caption{Plot of Bardeen's $f(\tilde{r})$ as a function of $\tilde{r}=r/M$ for different values of parameter $\tilde{g}$. Here, $\tilde{g}= g/M$ changes from zero (Purple) to unity (red) in steps of 0.1.
    As $\tilde{g}$ increases its value, the roots of $f(\tilde{r})$ get closer and then cease to exist. For $\tilde{g} > 0.76$ approximately, $f(\tilde{r})$ is always positive.}
    \label{F_Bardeen}
\end{figure}

\begin{figure}
    \centering
    \includegraphics[scale=0.55]{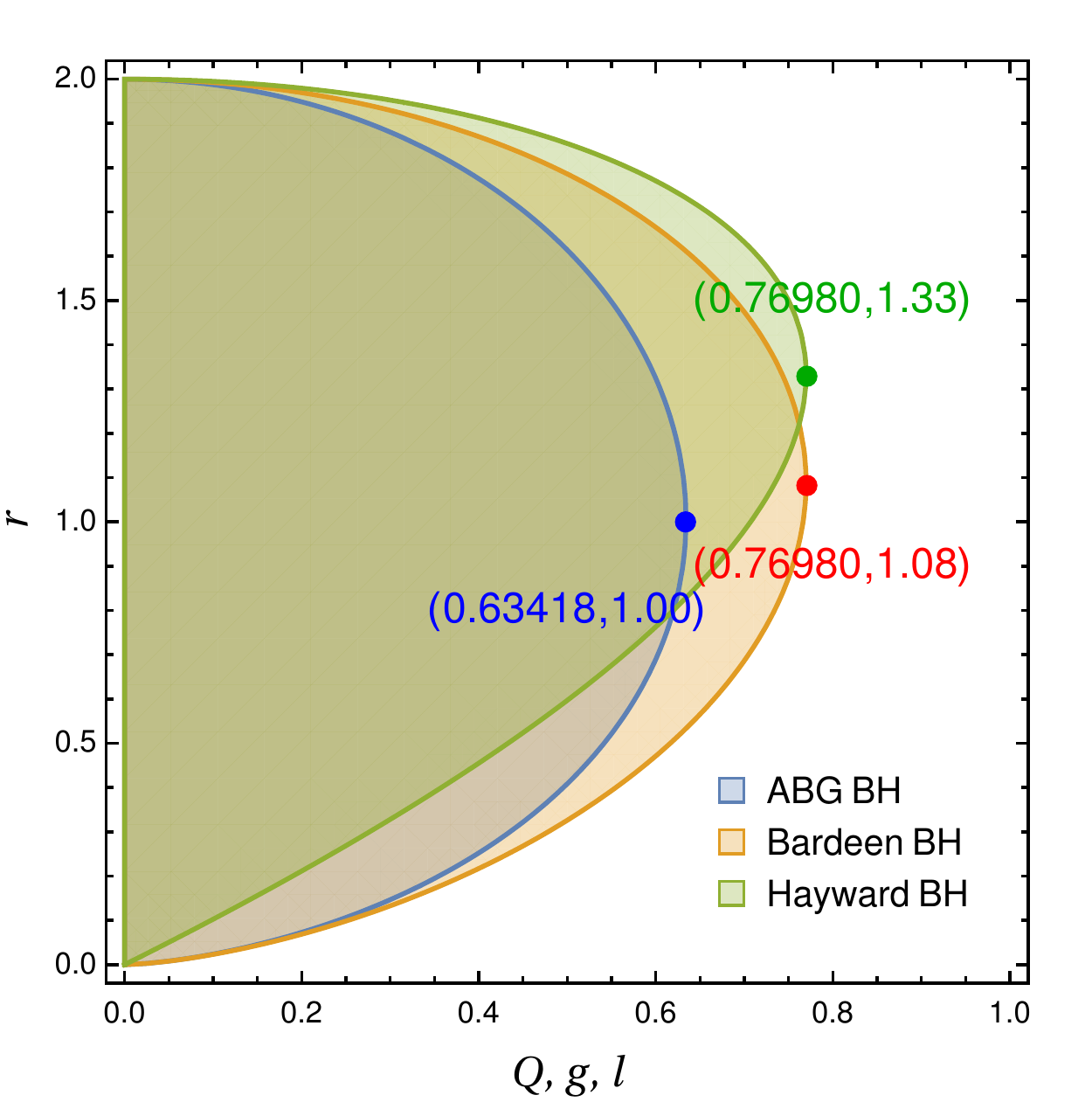}
    \caption{Plot shows the bound on the parameters $Q,g$ and $l$ of ABG, Bardeen and Hayward BHs, respectively. Here, the BHs exist in the shaded region region only. Whereas, dots (blue, red and green) represent the location of extremal (ABG, Bardeen and Hayward) BHs case. It worth to mention that parameters $r,Q,g$ and $l$ are normalized with the mass parameter $M$ of the respective BHs.}
    \label{fig:bounds1}
\end{figure}

\noindent As the parameter $g \to 0$, the function 
$R_B(r,g) \to 1$ and (\ref{f_B}) becomes the Schwarzschild metric.  A plot of $f(r)$ for different values of $g$ is shown in Fig. \ref{F_Bardeen}, it has two roots which disappear as $g$ increases. Consequently Bardeen spacetime possesses an exterior (and interior) event horizon $r_H^{ext}$ ($r_H^{int}$) for certain values of $g$ and $M$. In order to locate these event horizons one has to find the roots of $f(r)=0$. It is convenient to introduce the variables $\tilde{r}=r/M$ and $\tilde{g}=g/M$; hence, finding the roots of $f(r)=0$ is akin to finding the roots of \\
\begin{equation}
    \tilde{r}^6 + (3 \tilde{g}^2-4) \tilde{r}^4 + 3 \tilde{g}^4 \tilde{r}^2 + \tilde{g}^6 = 0.
    \label{rHB}
\end{equation}
There are two roots real and positive if and only if $0 < \tilde{g} < 0.7698$, otherwise there are no real and positive roots at all. At $\tilde{g}_c= 0.7698$ the two roots collide as shown in Fig. \ref{fig:bounds1}, it is also shown the bounds on the parameters $l$ and $Q$ of the Hayward and ABG BHs as well. Shaded regions corresponds to the existence of BHs, the dots point out the location of the critical cases. The upper (orange) line represents the external event horizon. We will work in the region outside the exterior horizon $r > r_H^{ext}$, whose value depends on $g$ and $M$, that is 
$r_H^{ext}= r_H^{ext}(g,M)$ and is found by solving (\ref{rHB}). Hence $r>r_H^{ext}$ is a condition that we ought to keep in mind. We must also mention that only when the mass $M$ exceeds the critical mass $M_c=3\sqrt{3}g/4$ we have the pair of event horizons.\\
What we are seeking is to find an analytical formula for the mass parameter $M=M(z,r,g)$, where $r$ is the radius of circular orbits followed by geodesics particles emitting photons whose frequency shift is $z$ as detected by a far away observer, $g$ is a parameter of the compact object which is interpreted as the magnetic monopole charge. To reach this goal, we insert function (\ref{f_B}) into (\ref{zfinalFB}) and it is found that
\begin{equation}
    z^2 = \frac{MR_B (r^2-2 g^2)}{(r-2MR_B)\left [ g^2+ r(r-3MR_B)\right ]}, 
    \label{z2Bardeen} \\
\end{equation}

\noindent which is a relationship between the redshift (blueshift) $z$, the mass parameter $M$ and the radius $r_c$ of a particle's circular orbit that emits light. (\ref{z2Bardeen}) makes sense provided that $z^2>0$. It turns out that existence of circular orbits demands $r^2-2g^2 >0$ and $g^2+r(r-3MR_B)>0$; thus (\ref{z2Bardeen}) is consistent as long as $r-2MR_B>0$. In the limit $g \to 0$, this latter condition is akin to working outside the event horizon $r>2M$ and the former condition is akin to $r>3M$ which guarantee circular orbits for the Schwarzschild BH. (\ref{z2Bardeen}) leads us to quadratic equation for the mass parameter $M$, its analytical expression is then given by

\begin{widetext}
\begin{equation}
    M_{\pm}= \frac{r}{12 z^2 R_B(r,g)} \left[ (1+5z^2)+\frac{2g^2}{r^2} (z^2-1) \pm \frac{1}{r^2} \sqrt{\mathcal{H}(r,z,g)} \right ],
    \label{MBar}
\end{equation}

where the $\mathcal{H}$ and $R_B$ are given by

\begin{equation}
\mathcal{H}(r,z,g) = (1+10z^2+z^4)(r^2-2g^2)\left [ r^2 - \frac{2g^2(z^2-1)^2}{1+10z^2+z^4} \right ] \quad , \quad R_B(r,g)=\frac{r^3}{(r^2+g^2)^{3/2}}. \nonumber
\end{equation}
\end{widetext}    

Observing that $0 < (z^2-1)^2/(1+10z^2+z^4) \le 1$ and due to the condition $r^2 - 2g^2 >0$, it turns out that $\mathcal{H} >0$ always; hence, the mass parameter is never a complex quantity. 

Nonetheless, (\ref{MBar}) still poses a problem, since for a particle following a circular orbit and emitting light with a shift $z$, it is not physically acceptable to have two values of the mass parameter. Lets figure out how to overcome this difficulty. 

Using the conditions for existence of circular orbits, namely $V_{eff}=0$ and $V^{\prime}_{eff}=0$, explicit expressions for $E^2$ and $L^2$ were found in the previous section. For the Bardeen case, these are

\begin{equation}
    E^2= \frac{(r-2M R_B)^2(r^2+g^2)}{r^2 \left [ g^2+r (r-3MR_B) \right ]},
    \label{E2Bardeen}
\end{equation}

\begin{equation}
    L^2= \frac{Mr(r^2-2g^2)R_B}{\left [ g^2+r (r-3MR_B) \right ] },
    \label{L2Bardeen}
\end{equation}

\noindent (\ref{E2Bardeen}) requires that $g^2+r(r-3MR)>0$ whereas (\ref{L2Bardeen}) requires additionally that $r^2-2g^2 >0$. On the other hand, circular orbits are stable provided that $V^{\prime \prime}>0$, from (\ref{segundita}) $V^{\prime \prime}$ reads

\begin{equation}
    V^{\prime \prime}_{eff} = \frac{\left [ r^3(r-6MR_B)+8g^2(r^2-g^2) \right ] 2Mr^4}{(r^2+g^2)^4R[r-2MR_B][g^2+r(r-3MR_B)]},
    \label{VppBardeen}
\end{equation}

\noindent which is positive as long as 

\begin{equation}
    r^3(r-6MR_B) + 8g^2(r^2-g^2) > 0.
    \label{c4Bar}
\end{equation}

\noindent Therefore, in addition to the condition (\ref{c4Bar}), the conditions

\begin{eqnarray}
    && r-2MR_B >0 \quad , \quad  r^2 -2 g^2 >0 , \nonumber \\ 
    && g^2+r(r-3MR_B) > 0, 
    \label{c1c2c3Bar}
\end{eqnarray}

\noindent must be simultaneously satisfied.
    
It can be verified that as $g \to 0$ the mass parameter (\ref{MBar}) reduces to the one for Schwarzschild BH \cite{Becerril:2016qxf}, namely

\begin{equation}
    M_{\pm}(r,z)=r \frac{1+5z^2 \pm \sqrt{1+10z^2+z^4}}{12 z^2}.
    \label{MSCH}
\end{equation}

\begin{figure}
    \includegraphics[scale=0.36]{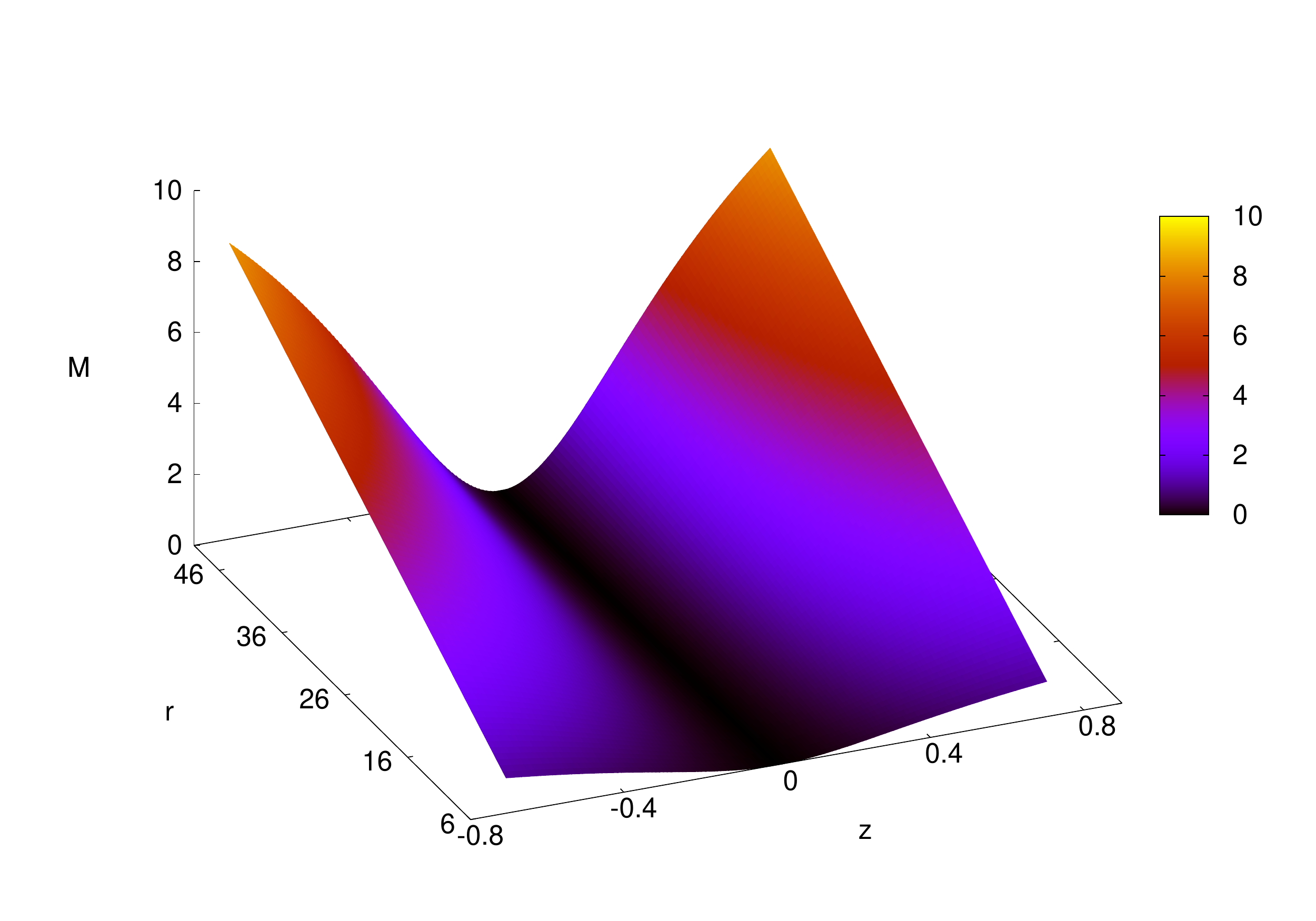} 
    \caption{The mass parameter $M$ for the Schwarzschild black hole is shown as a function of he frequency shift $z$ ( redshift $z > 0$ and blueshift $z < 0$) and the radius $r$ of an eventual circular orbit of a photon emitter. $M$ and $r$ are in geometrized units and scaled by $pM_{\odot}$ where $p$ is an arbitrary factor of proportionality.}
    \label{MasaSchwarzschild}
\end{figure}

\noindent Stability for circular orbits (\ref{VppBardeen}), in this limit, implies $r>6M$. In \cite{Becerril:2016qxf} it was proven that $r>6M$ led us to the conclusion that only the minus sign in (\ref{MSCH}) is allowed and that there is a bound for the frequency shift, explicitly $|z|<1/\sqrt{2}$. The function $M=M(r,z)$ given by (\ref{MSCH}) is in geometrized units ($G_N=c=1$). We scale $M$ and $r$ by an arbitrary multiple of the solar mass $p M_{\odot}$, for Sgr A, 
$p=2.72 \times 10^6$. Its graph is shown in figure \ref{MasaSchwarzschild}.

For Bardeen BH, solely the minus sign in (\ref{MBar}) is allowed as well. To see that this is indeed the case, we substitute the expression (\ref{MBar}), for the mass parameter $M_{\pm}$, into the stability condition (\ref{c4Bar}). After some algebra (\ref{c4Bar}) becomes

\begin{eqnarray}
    \mathcal{L}_{HS} & \equiv & -(1+3z^2)(r^2-2g^2)(r^2 -2 g^2 \mathcal{G}(z) ) > \pm r^2 \sqrt{H},  \nonumber \\
    &&\text{with} \quad \mathcal{G}(z) = \frac{4z^2}{1+3 z^2}.  
    \label{CondicionSignoBardeen}
\end{eqnarray}

\noindent Recalling the condition $r^2 - 2 g^2>0$, for the case $|z| \leq 1$, it is apparent that $0 \leq  \mathcal{G}(z) \leq 1$, then one has $r^2 > 2 g^2 > 2 g^2 \mathcal{G}(z)$; therefore, the left-hand side in Eq. (\ref{CondicionSignoBardeen}) $\mathcal{L}_{HS}$ is negative and the positive sign appearing in  (\ref{CondicionSignoBardeen}) is out of the question; therefore $M_{+}(r,z)$ must be discarded. Nonetheless, for $|z|>1$, the function $\mathcal{G}(z)$ is bounded as $1 <  \mathcal{G}(z) < 4/3$ and verifying analytically that (\ref{CondicionSignoBardeen}) holds for just the minus sign is not straightforward. We checked numerically that (\ref{CondicionSignoBardeen}) holds only with the minus sign for a variety of intervals  $0\leq g \leq g_{max}$, $r_{min} \leq r \leq r_{max}$ and $1 < z \leq z_{max}$ satisfying $r^2 -2 g^2 >0$ ($z_{max}$ going beyond unity). This establishes the uniqueness of the mass parameter given by (\ref{MBar}), only $M_{-}(r,z)$ is compatible with the stability condition of circular orbits.  The stability condition (\ref{CondicionSignoBardeen}), together with (\ref{c1c2c3Bar}) should allow us to find the bounds for $z$ and plot $M=M_{-}(g,r,z)$. For $g \to 0$, (\ref{MBar}) becomes \ref{MSCH} for Schwarzschild BH.
For the case of a magnetic monopole, using the relationship (\ref{zfinalFBeffective}), that takes into account the fact that light travels along null geodesics with the effective metric, one obtains an expression for the redshift (blueshift) $z$, and a quadratic equation for the mass parameter that we call $\widetilde{M}$ to distinguish it from (\ref{MBar}), they read

\begin{widetext}
\begin{equation}
    z^2= \frac{2 \widetilde{M} r^2 (r^2-2g^2) (r^2+g^2)^{5/2}}{(3r^2-4g^2) \left [(r^2+g^2)^{3/2}-2\widetilde{M} r^2 \right ] \left [ (r^2+g^2)^{5/2}-3\widetilde{M}r^4\right ]}.
    \label{z2BardeenEffective}
\end{equation}

For the quadratic equation $A_{m} \widetilde{M}^2 + B_{m} \widetilde{M} + C_{m}=0$, the coefficients are given by

\begin{eqnarray}
    A_{m} &=& 6 r^6 (3 r^2-4g^2) z^2, \nonumber \\
    B_{m} &=& -r^2 (r^2 + g^2)^{3/2} \left [ 2(r^2-2g^2)(r^2+g^2)+(3r^2-4g^2)(5r^2+2g^2) z^2 \right ], \nonumber \\
    C_{m} &=& (r^2+g^2)^4 (3r^2-4g^2)z^2.
\end{eqnarray}
The explicit expression for the $M$ reads

\begin{equation}
    \widetilde{M}_{\pm}=\frac{r^2 (r^2 + g^2)^{3/2} \left [ 2(r^2-2g^2)(r^2+g^2)+(3r^2-4g^2)(5r^2+2g^2) z^2 \right ] \pm \sqrt{\triangle}}{12 r^6 (3 r^2-4 g^2) z^2 }. 
    \label{MBarEff}
\end{equation}

The discriminant $\triangle=B_m^2-4A_mC_m$ 

\begin{equation}
    \triangle = r^4 (r^2-2g^2)(r^2+g^2)^3\left [ 4(r^2-2g^2)(r^2+g^2)^2 + 4(3 r^2-4g^2)(r^2+g^2)(5r^2+2g^2)z^2 + (r^2-2g^2)(3r^2-4g^2)^2 z^4 \right ],
\end{equation}

is always positive since $r^2-2g^2>0$.

\end{widetext}
One faces again the undesirable possibility of having two values for the mass parameter. We will verify whether the conditions to have circular stable orbits will determine the uniqueness of the solution, as mentioned before, this proof ought to be done numerically. 

We shall carry out the analysis considering Bardeen spacetime as a toy model first and then considering it as an exact solution of Einstein field equations with NED, that is to say, working first with the expression (\ref{MBar}) and second with (\ref{MBarEff}) and then comparing both outcomes. We shall work in the region outside the exterior event horizon $r > r_H^{ext}$ which exists only for $g<0.7698 M$, otherwise one has a globally regular spacetime. These class of solutions have been proven to exist, yet they are given numerically \cite{Breitenlohner:1993es,Bartnik:1988am}. The numerical algorithm to perform this analysis is provided next.

\begin{enumerate}
    \item Construct a domain $\mathcal{D}$ which is a set of points $\{ (g_i,r_j,z_k) \} = P_{ijk}$.
    \item For each point $P_{ijk}$ compute $M_{+}$ and $M_{-}$ with (\ref{MBar}).   
   \item Test whether (\ref{c4Bar}) and (\ref{c1c2c3Bar}) are simultaneously fulfilled. 
   \begin{enumerate}
       \item If the conditions are fulfilled. Then store $M(P_{ijk})$ and $z=z(g_i,r_j)$, the former to plot $M=M(g,r,z)$, the later to be constructing the bounds of $z$. 
       \item If the conditions are not satisfied. For the current point $P_{ijk}$ there is not a physically acceptable value of the parameter $M$ and this point $P_{ijk}$ can be disregarded. 
   \end{enumerate}
   \item We can distinguished between a Bardeen BH or a globally regular spacetime by testing whether $g < (0.317)(2M)$ is satisfied.
   
    \begin{enumerate}
        \item  It is satisfied (BH). Then one has to find $r^{ext}_H=r^{ext}_H(g_i,r_j,z_k)$ and verified that $r_j > r_H$, if it is the case, continue, otherwise remove this set $\{P_{ijk},M(P_{ijk})\}$ from the plot $M=M(g,r,z)$, select another point and go back to 2.
        \item It is not satisfied (Globally regular spacetime). Then any $r>0$ is acceptable at this stage, store  $M(P_{ijk})$ and $z=z(g_i,r_j)$ and continue.
    \end{enumerate}
    
   \item Go back to 2. until every point in the domain $\mathcal{D}$ has been tested and data stored.
\end{enumerate}

\begin{figure}
    \includegraphics[scale=0.68]{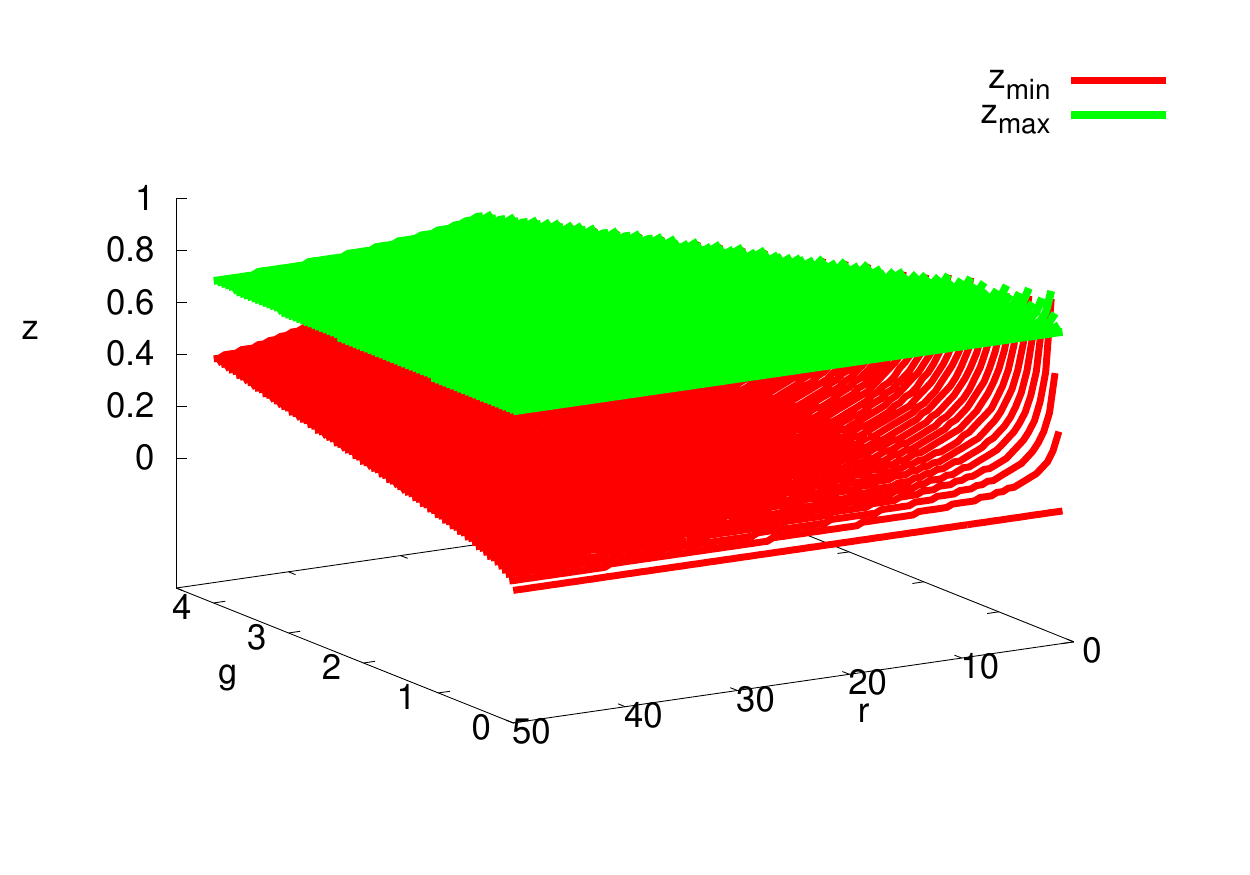}
    \includegraphics[scale=0.68]{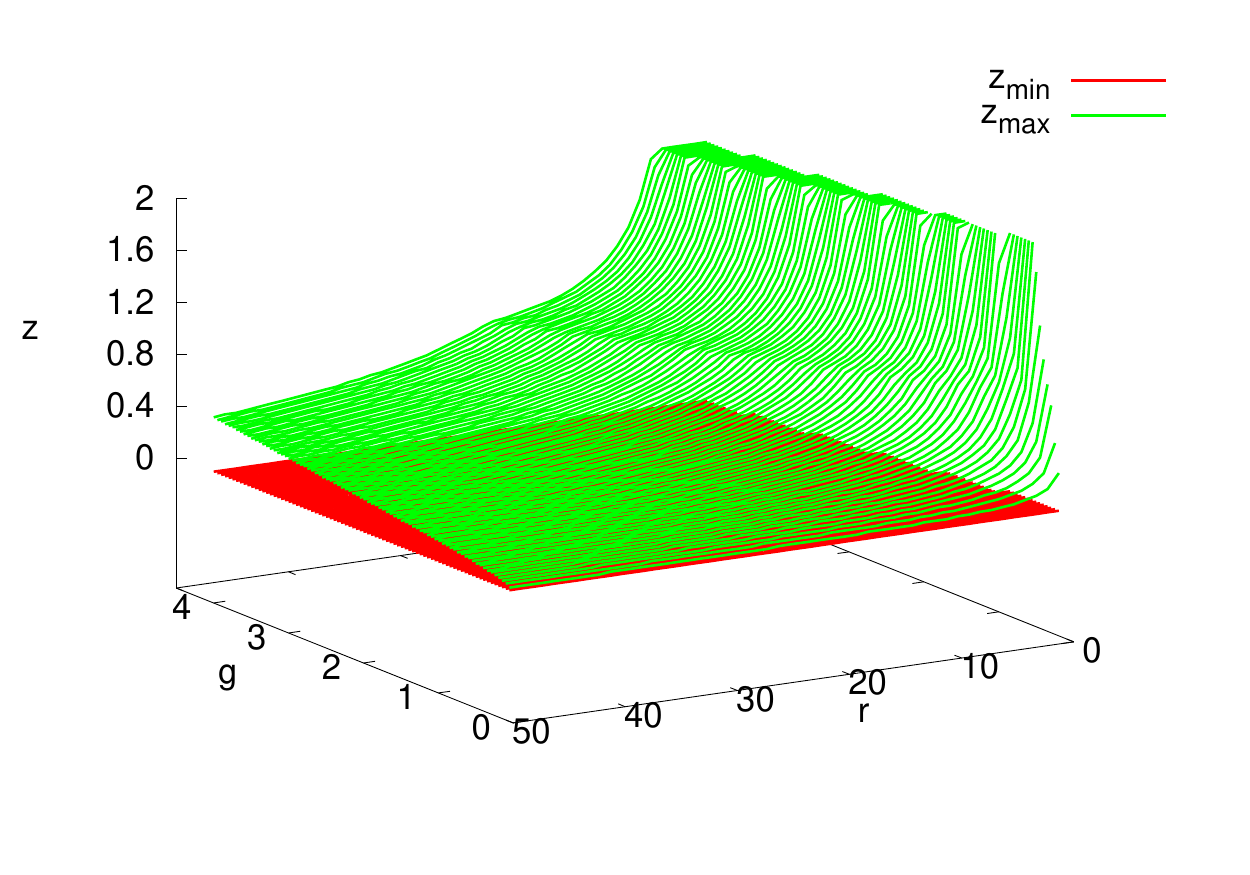}
     \includegraphics[scale=0.68]{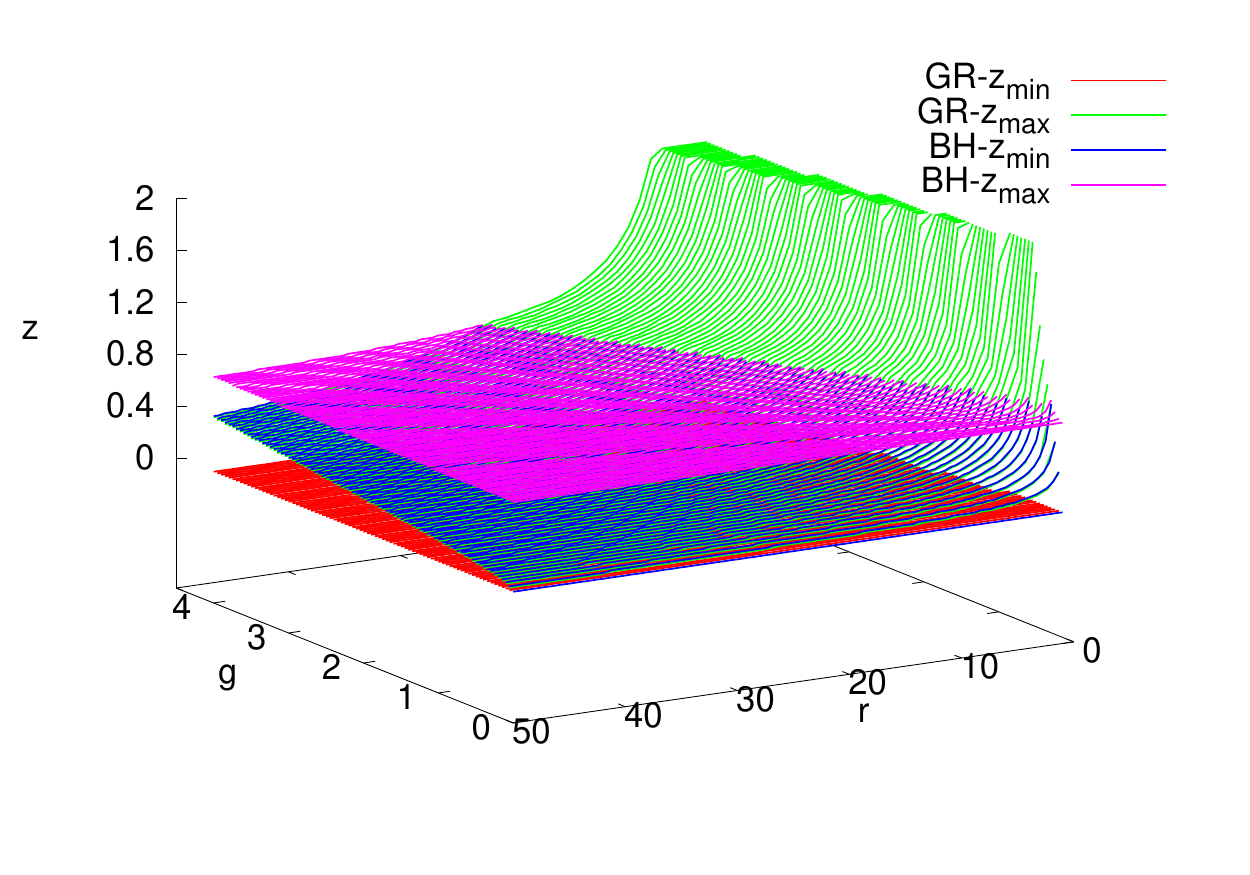}
    \caption{In the upper plot, we presents bounds of $z$ for Bardeen's BH constructed with the {\it original metric} $g_{\mu \nu}$. The only allowed frequency shifts that could be detected by a far away observer is located in the gap  $z_{min}(g,r) < z < z_{max}(g,r)$. The value of $z$ does not go beyond $0.85$. In the middle plot, we present bounds of $z$ for globally regular Bardeen's spacetime, only $z \in [z_{min},z_{max}]$ are allowed. The two surfaces GR-$z_{max}$ and BH-$z_{min}$ corresponding to Bardeen's globally regular and BH spacetime respectively, coincide when $z<0.85$. Whereas, when $z>0.85$ only globally regular spacetime is allowed.
    This is shown in the lower plot where we have superimposed both bounds for Bardeen BH and globally regular spacetimes. }
    \label{Cota_Bardeen}
\end{figure}

\begin{figure}
   \includegraphics[scale=0.56]{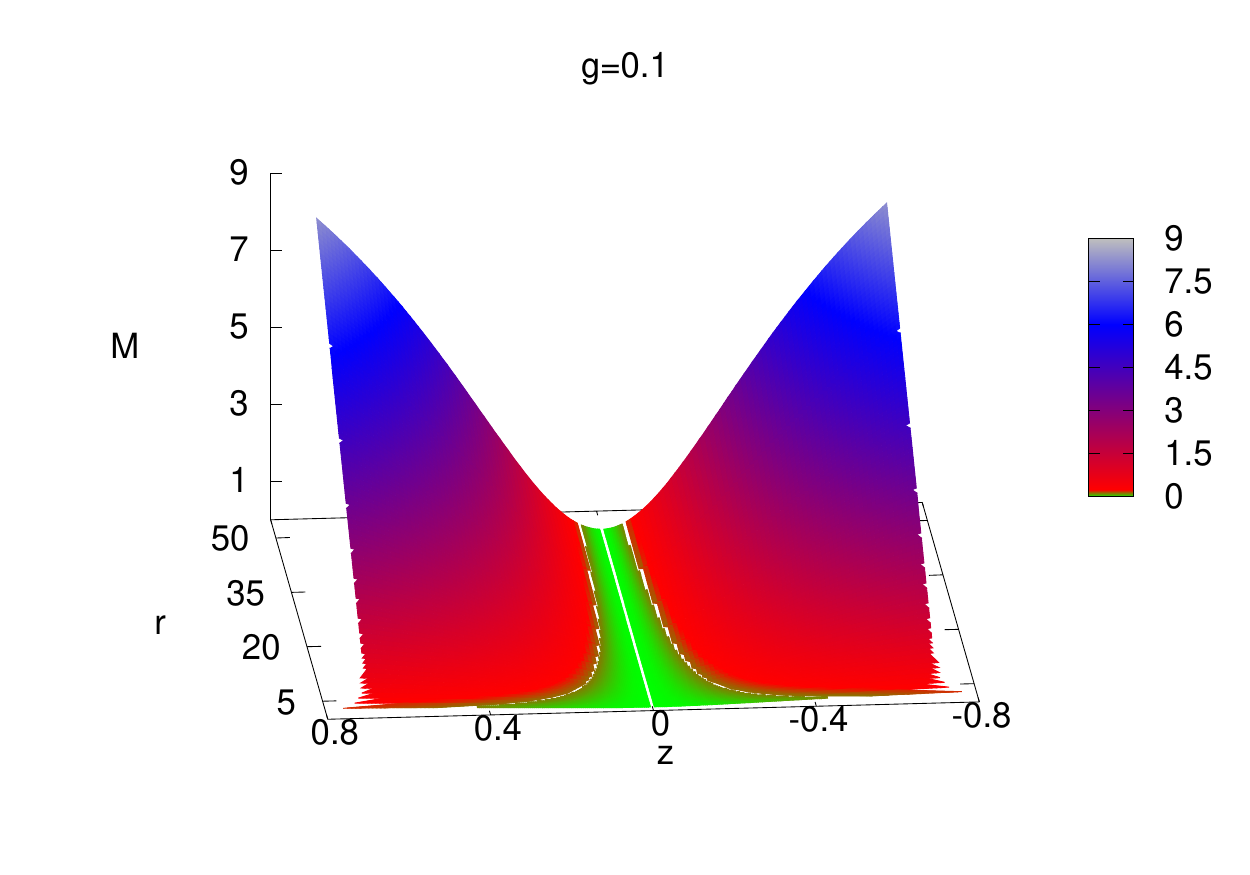} \\ \vspace{-0.49cm}
    \includegraphics[scale=0.56]{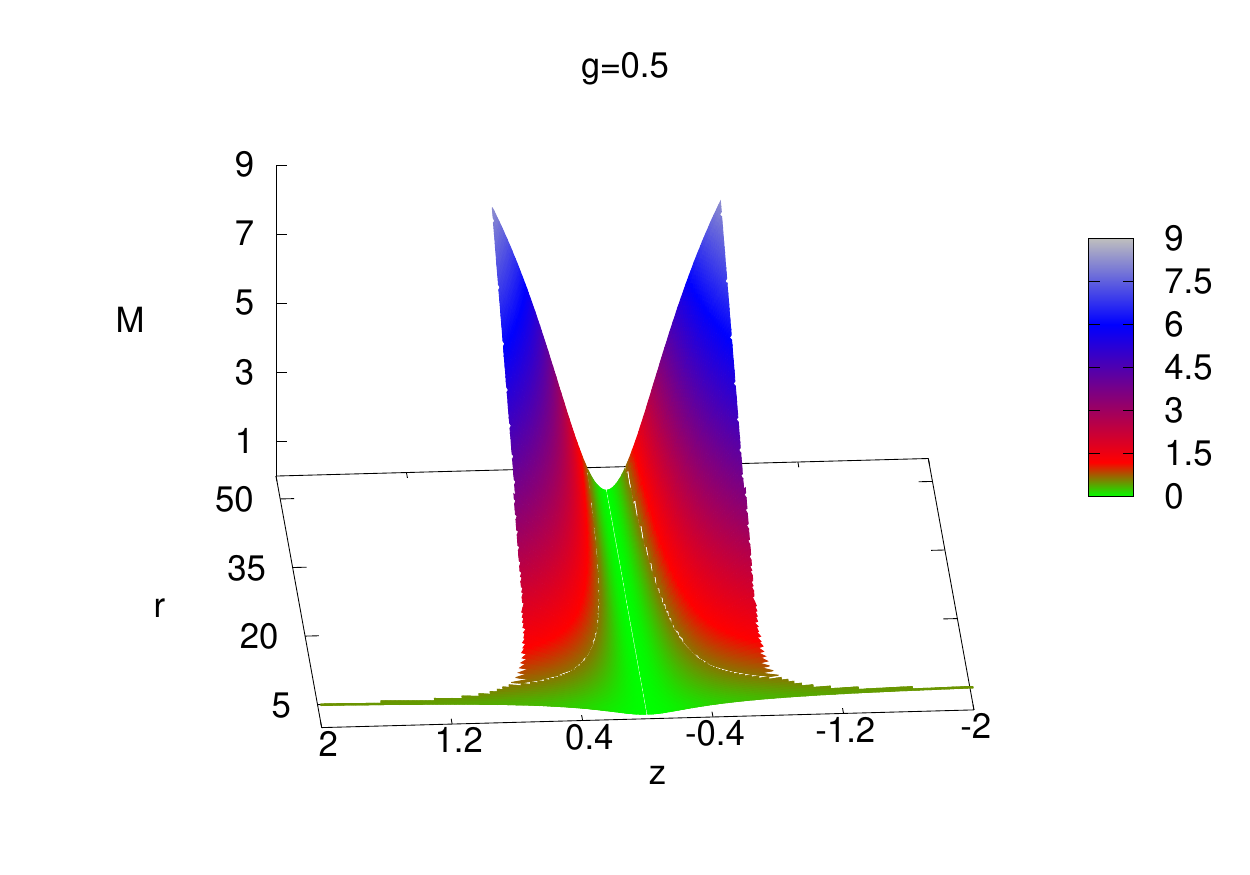} \\
    \vspace{-0.49cm}
    \includegraphics[scale=0.56]{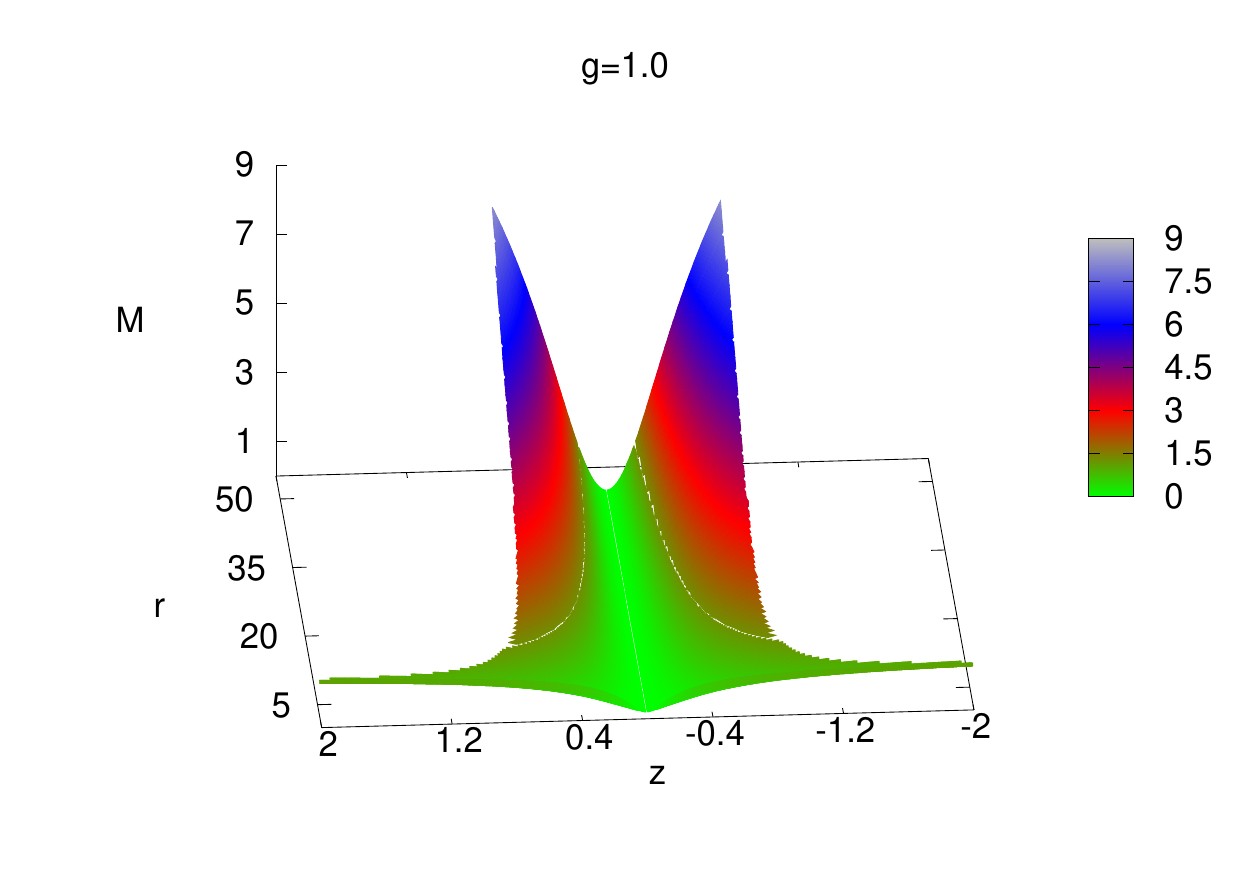}  \\
    \vspace{-0.49cm}
    \includegraphics[scale=0.56]{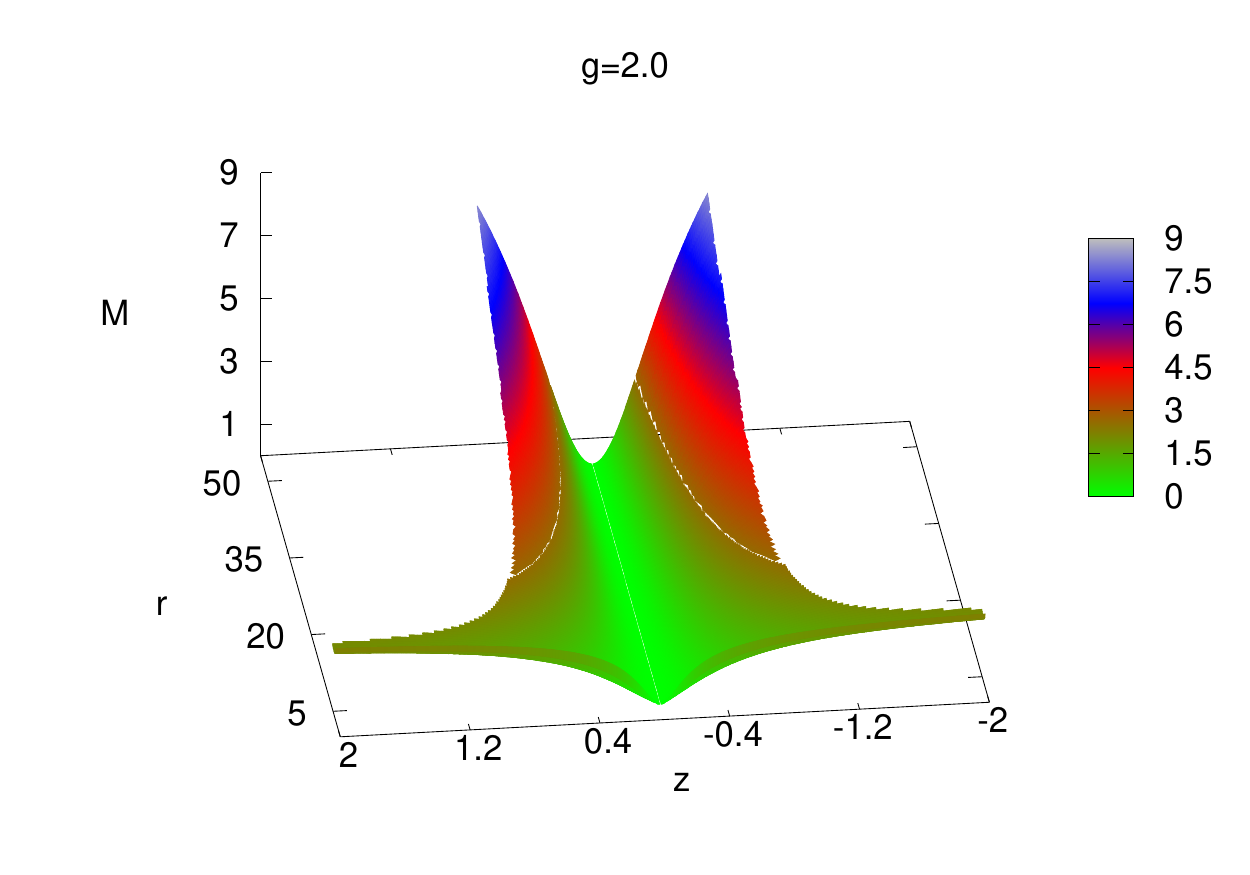}
    \caption{The mass parameter $M$ for Bardeen's BH $g/M < 0.7698$ (red$\dashrightarrow$blue colors) is shown as a function of the frequency shift $z$ ( redshift $z > 0$ and blueshift $z < 0$) and the radius $r$ of an eventual circular orbit of a photon emitter for four values of $g$. The green surface corresponds to the mass parameter for the globally regular spacetime sector $g/M> 0.7698$. $M$ and $r$ are in geometrized units and scaled by $pM_{\odot}$ where $p$ is an arbitrary factor of proportionality. Here, the numerical algorithm employed yields the known critical mass $M_c=(3\sqrt{3}g)/4$ that separates the BH from the GR spacetime (see text for details).}
    \label{Masas_Bardeen}
\end{figure}

{\bf A. Bardeen spacetime as a toy model}\\

It turns out that with $M_{+}$ there is not a single point in the domain $\mathcal{D}$, no matter how large it is constructed, for which all conditions are simultaneously fulfilled. On the other hand, working with $M_{-}$, there is a subset  $\mathcal{D}_M^{BH} \subset \mathcal{D}$ where these conditions are simultaneously satisfied and those are considered physically acceptable. Thus, a measurement of the redshift $z$ of light emitted by a
particle that follows a circular orbit of radius $r$ in the equatorial plane around a Bardeen BH ($\tilde{g}=g/M < 0.7698$) will 
have a mass parameter determined by $M=M_{-}(r,z,g)$ given in (\ref{MBar}) whose domain is the subset $\mathcal{D}_M^{BH}$. Part of this subset is shown in Fig. \ref{Cota_Bardeen}, where the bounds of $z$ in terms of $g$ and $r$ are plotted, only for $z$ between the lower (red) surface $z_{min}(g,r)$ and the upper (green) surface $z_{max}(g,r)$, the conditions for circular stable orbits of photon emitters together with $r>r_H^{ext}$ are simultaneously satisfied. For $g=0$ the Schwarzschild bound $|z|<1/\sqrt{2}$ is certainly recovered. The gap between the bounding surfaces 
$|z_{sup}-z_{inf}|$ narrows as $g$ increases its value. The two surfaces $z_{max}(g,r)$ and 
$z_{min}(g,r)$, collide along a curve. On that curve (it looks a line in Fig. \ref{Cota_Bardeen}), $z$ increases its value as $g$ increases, yet $z$ never goes beyond $0.85$.  
For $\tilde{g}> 0.7698$ one has globally regular (GR) spacetimes, there is a subset $\mathcal{D}_M^{GR}$ where the conditions for circular stable orbits of photon emitters are simultaneously fulfilled. Part of this subset is shown in Fig. \ref{Cota_Bardeen} where we also superimposed both bounds for Bardeen's BH and globally regular spacetimes.

\begin{figure}
    \includegraphics[scale=0.6]{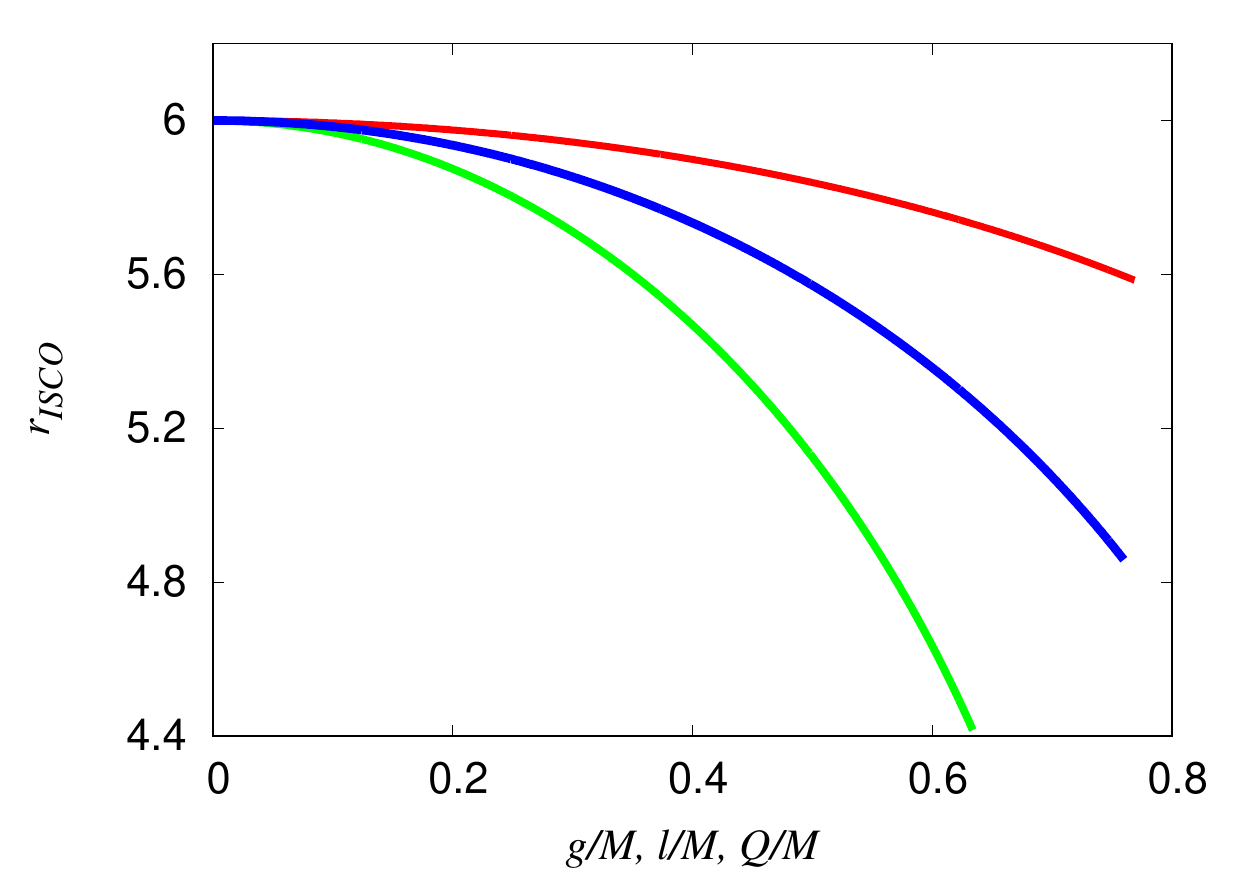} 
    \caption{Stable circular orbits for the Schwarzschild spacetime are found for 
    $ \tilde{r} = r/M > r_{ISCO} = 6 $. For Bardeen BH, this $r_{ISCO}=r_{ISCO}(\tilde{g})$ is shown in the upper curve (red color), where $\tilde{g}=g/M < 0.7698$. We also present the behavior of $r_{ISCO} = r_{ISCO}(\tilde{l})$ for the Hayward (middle red curve) and  $r_{ISCO}=r_{ISCO}(\tilde{Q})$ for the ABG (lower green curve) BH as function of their corresponding parameters $l/M$ and $Q/M$ respectively.}
    \label{LowerLimit}
\end{figure}

We scale $M$ and $r$ by an arbitrary multiple of the solar mass, i.e. by $pM_{\odot}$, being $p$ an arbitrary proportionality constant.  Fig. \ref{Masas_Bardeen} shows this numerically generated scaled relationship $M=M(r,z,g)$ which is symmetric with respect to the frequency shift $z$ ($z_{red} > 0, z_{blue}<0$) for $g=0.1, 0.5, 1.0, 2.0$. For $g=0.1$ the graph $M=M(r,z)$ is nearly the one shown above in Fig. \ref{MasaSchwarzschild} for Schwarzschild BH, yet there is a gap in the $z$ axis that splits the regions for $z>0$ and $z<0$. $z$ is no longer bounded as $|z|<1/\sqrt{z}$ as in the Schwarzschild case, it allows a little bit larger values, up to $0.85$. The mass parameter allowed for the globally regular spacetime sector (green color) is considerably smaller than for BH sector (red-blue color) yet, since there is no event horizon, in principle, stable circular orbits of photon emitters may allow smaller radii than those for the BH sector.  As $g$ increases, the surface $M(r,z)$ for the BH region, splits more notoriously into two parts for the red and blueshifts; this gap width varies with $r$ according to Fig. \ref{Cota_Bardeen}. The green sector for GR spacetimes and the red-blue sector for BHs is separated by the critical mass $M_c=3\sqrt{3} g/4$.
For $g=0$ (Schwarzschild) one has stable circular orbits for $\tilde{r}=r/M > 6 = \tilde{r}_s$. As mentioned above, for $\tilde{g} = g/M < \tilde{g}_c = 0.7698$ Bardeen spacetime is a BH. As $\tilde{g}$ grows,
the value of this lower limit $\tilde{r}_s$ for stability decreases as shown in the upper red curve in Fig. \ref{LowerLimit}. In the same figure one may see the behaviour of $\tilde{r}_s=\tilde{r}_s(\tilde{l})$ for the Hayward (middle red curve) and  $\tilde{r}_s=\tilde{r}_s(\tilde{Q})$ for the ABG (lower green curve) BH as function of their corresponding parameters $l$ and $Q$ respectively. It is for the Bardeen BH that this stability limit is lowered the most. \\

{\bf Bardeen spacetime as an exact solution of Einstein field equations with NED}\\

We performed an analogous analysis to the one just done above. Yet, we take into account now that, since light moves in null geodesics with an effective metric, then the mass parameter expression to be employed is (\ref{MBarEff}). It turns out again that, solely with $\widetilde{M}_{-}$, there is a subset of the domain $\mathcal{D}$ where the conditions of having circular and stable orbits are all simultaneously fulfilled; therefore, the solution is unique as it should be. 
Bardeen's effective metric $\widetilde{g}_{\mu \nu}$ takes into account the effects of the electromagnetic nonlinearities on the motion of light, not on the motion of particles with mass, it is related with the original Bardeen's geometry by 

\begin{equation}
    \widetilde{g}_{tt}= g_{tt}, \quad \widetilde{g}_{rr}= g_{rr}, \quad
    \widetilde{g}_{\theta \theta}= g_{\theta \theta}/\Phi, \quad  \widetilde{g}_{\phi \phi}= g_{\phi \phi}/\Phi, \nonumber
\end{equation}
where $\Phi(r)$ reads
\begin{equation}
    \Phi=1+\frac{1}{2} \left [ \frac{r^2-6g^2}{r^2+g^2}\right ],
\end{equation}
whose derivation is described in the appendix. Since $\Phi \to 3/2$ as $g \to 0$ (eliminating the electromagnetic fields) in this limit, the effective metric does not becomes exactly the Schwarzschild. By the same token, the expression for $\widetilde{M}(r,z,g)$ as $g \to 0$ does not become exactly the one corresponding to Schwarzschild.

\begin{figure}
    \includegraphics[scale=0.68]{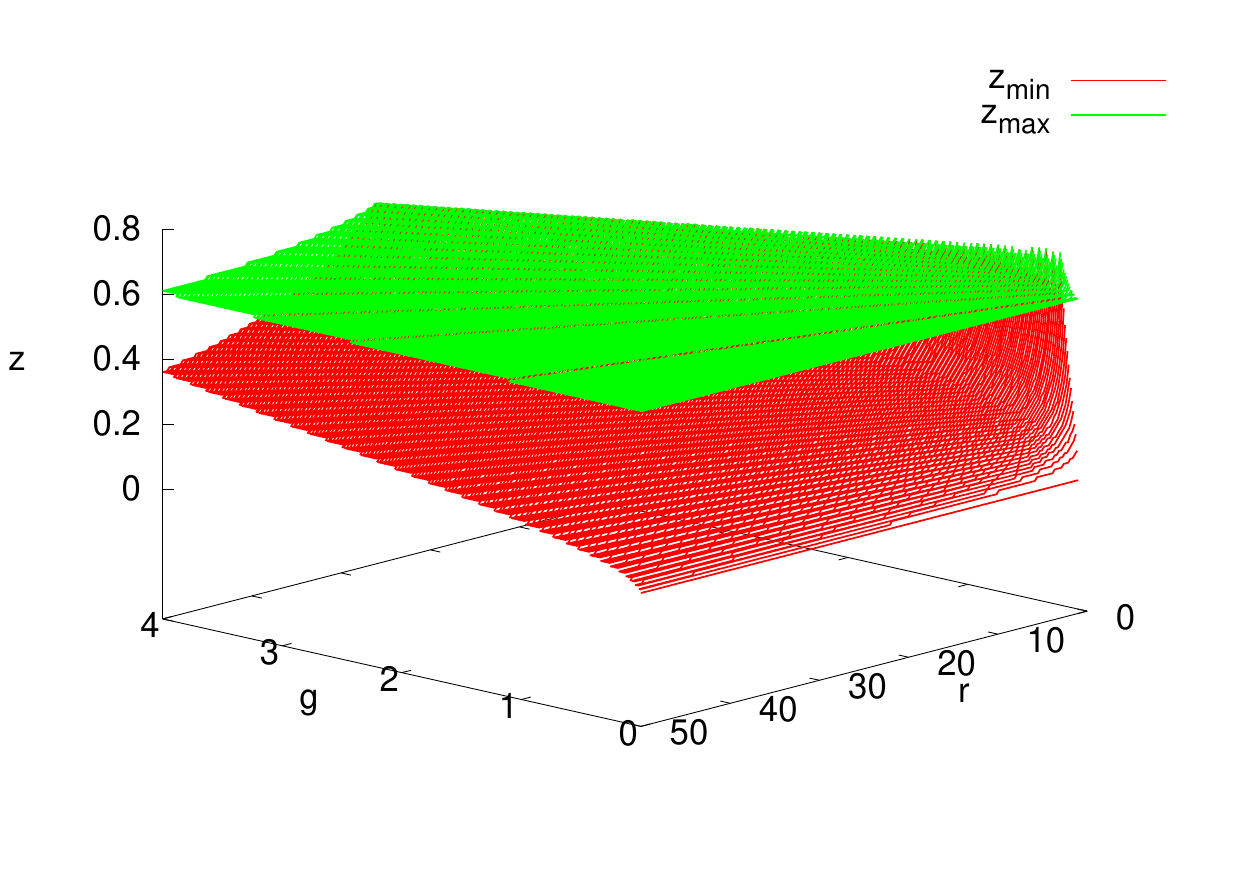}
    \includegraphics[scale=0.68]{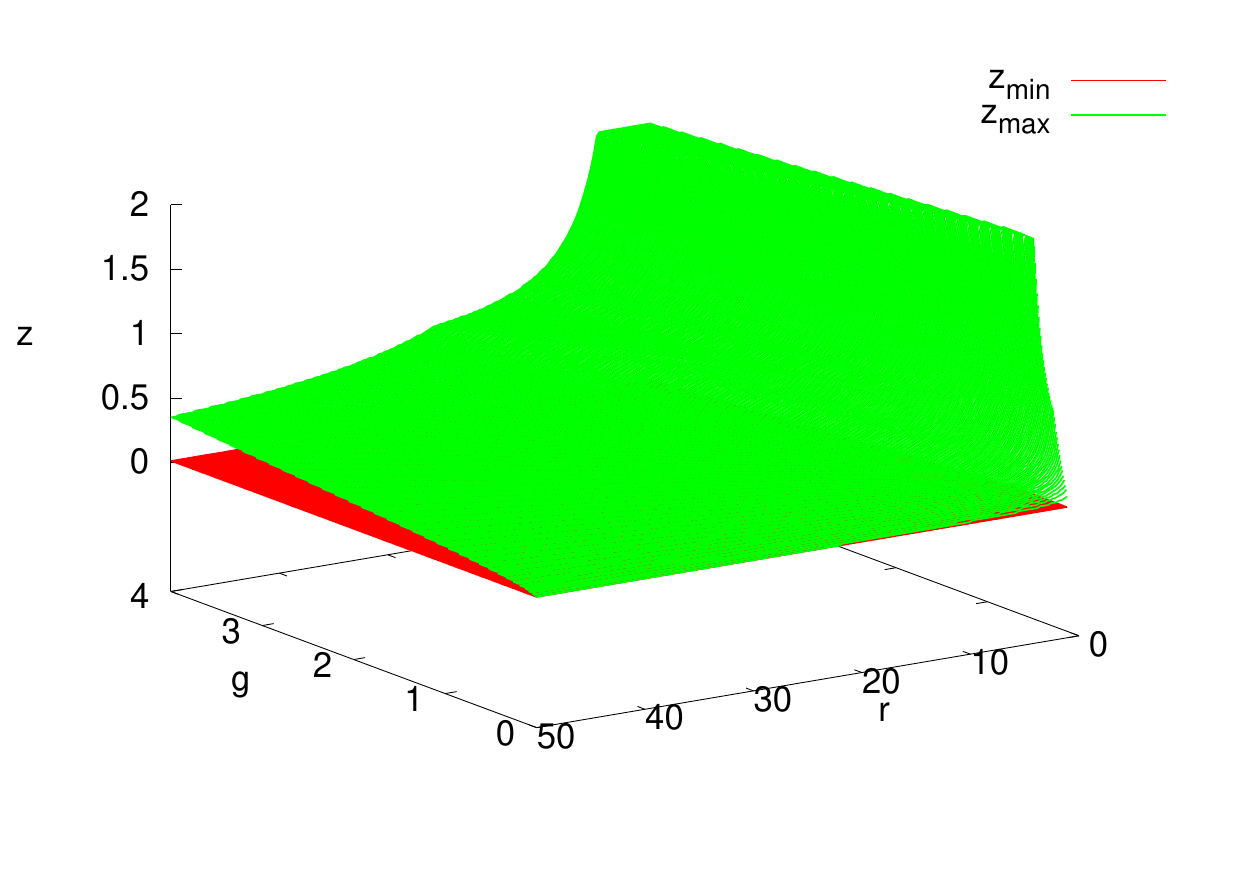}
     \includegraphics[scale=0.68]{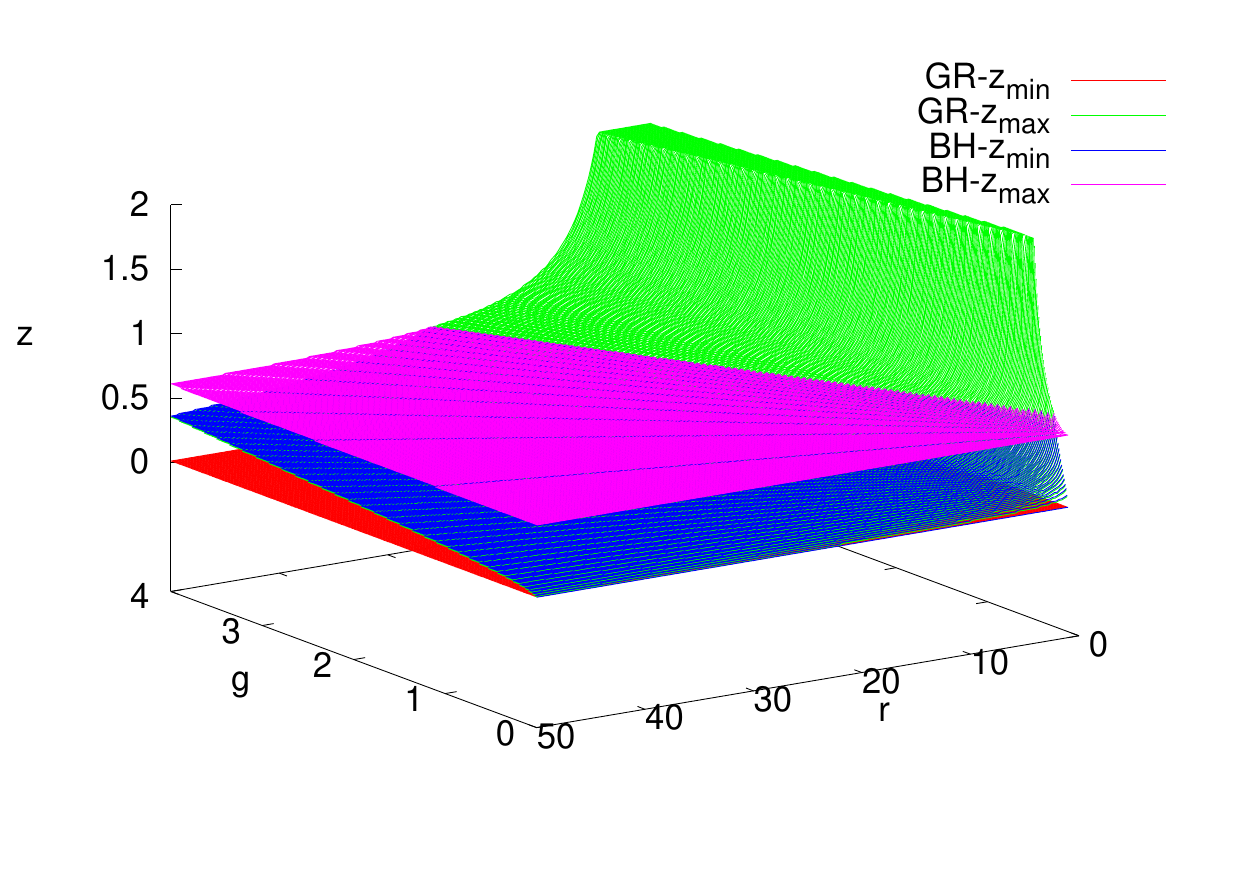}
    \caption{In the upper plot, we presents bounds of $z$ for Bardeen's BH considering that photons move in the corresponding {\it effective metric}. The only allowed frequency shifts that could be detected by a far away observer is located in between the lower and upper surfaces. The value of $z$ never goes beyond unity. In the middle plot, we present bounds of $z$ for globally regular Bardeen's spacetime, only $z \in [z_{min},z_{max}]$ are allowed. The two surfaces $z_{max}$ and $z_{min}$ corresponding to Bardeen's globally regular and BH spacetime respectively, coincide when $z<0.71$. Whereas, when $z>0.71$ only globally regular spacetime is allowed.
    This is shown in the lower plot where we have superimposed both bounds for Bardeen BH and globally regular spacetimes. }
    \label{CotasBardeenEff}
\end{figure}

\begin{figure}
   \includegraphics[scale=0.56]{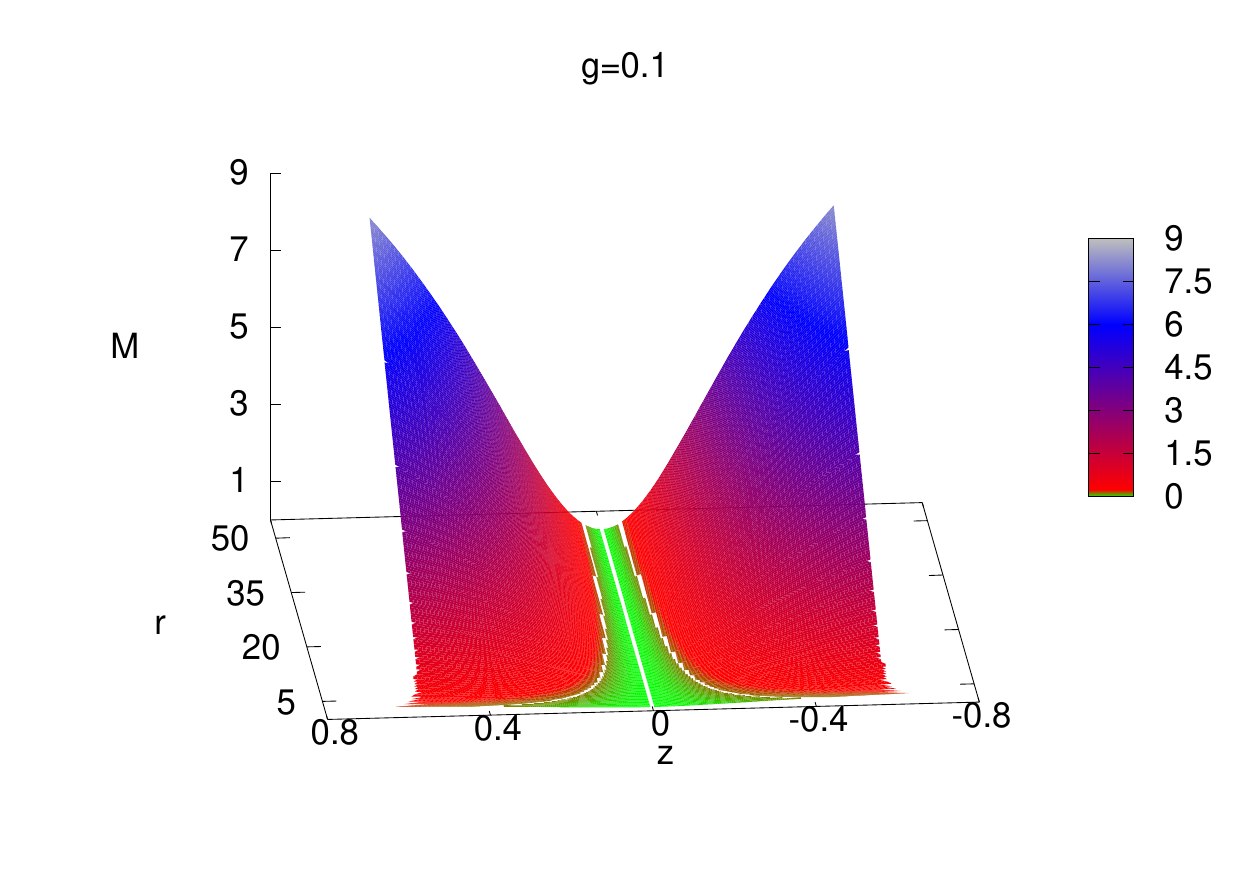} \\ 
   \vspace{-0.49cm}
    \includegraphics[scale=0.56]{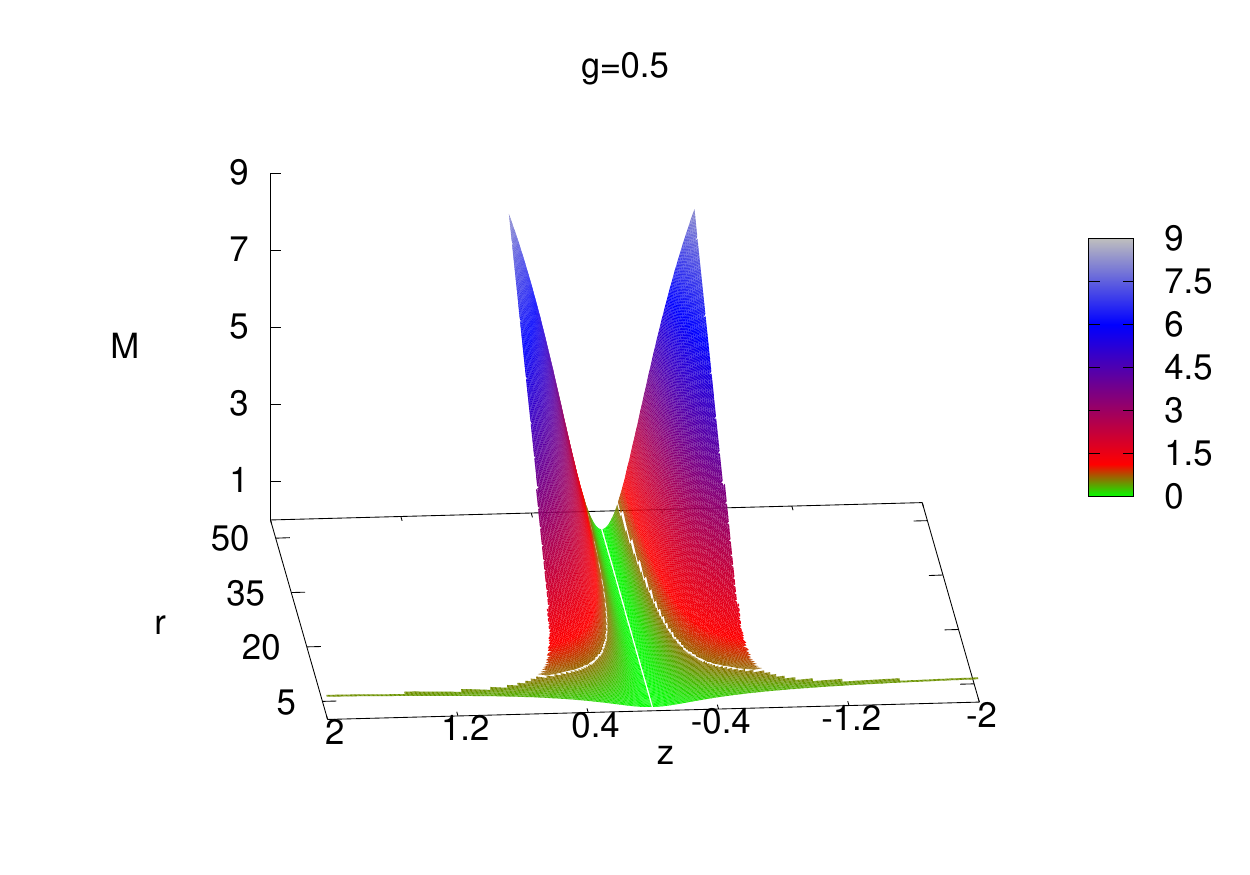} \\
    \vspace{-0.49cm}
    \includegraphics[scale=0.56]{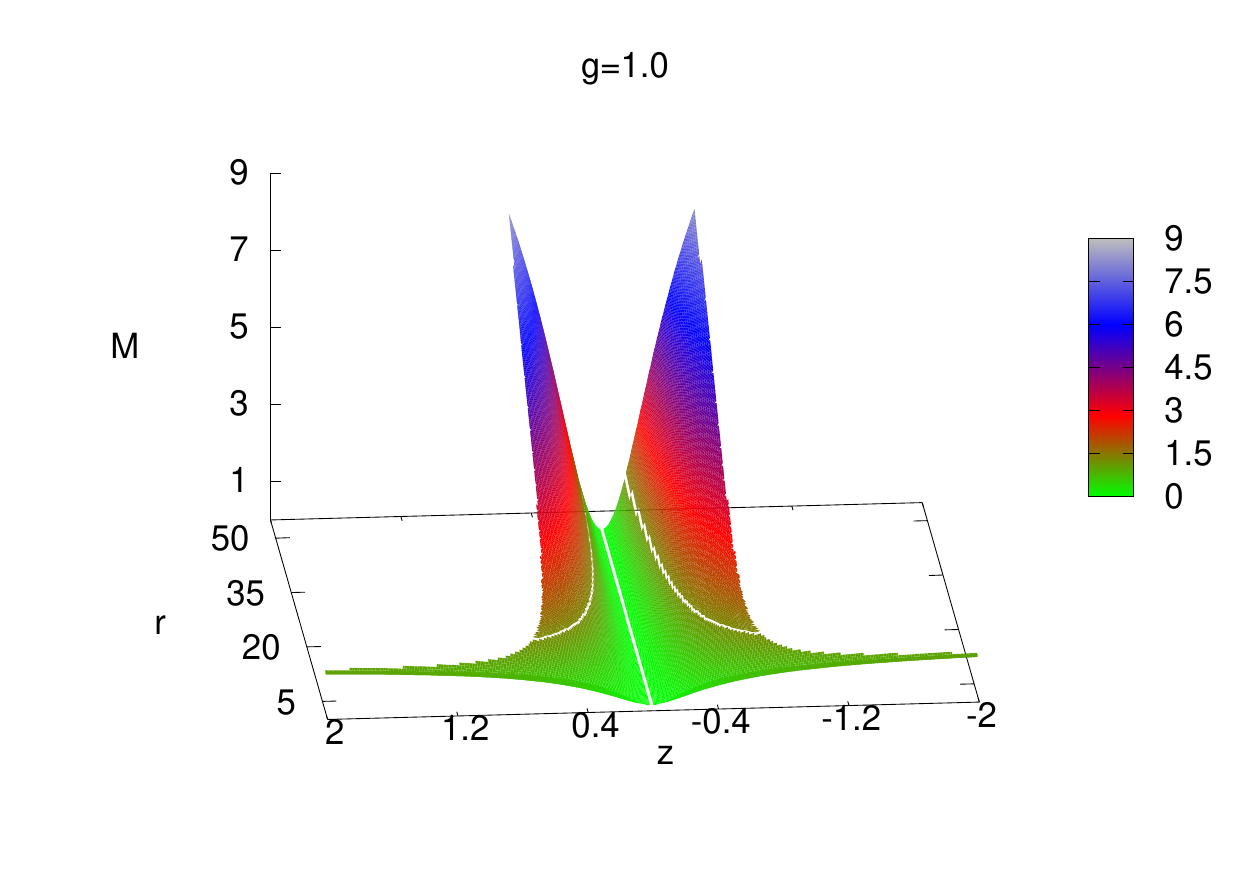}  \\
    \vspace{-0.49cm}
    \includegraphics[scale=0.56]{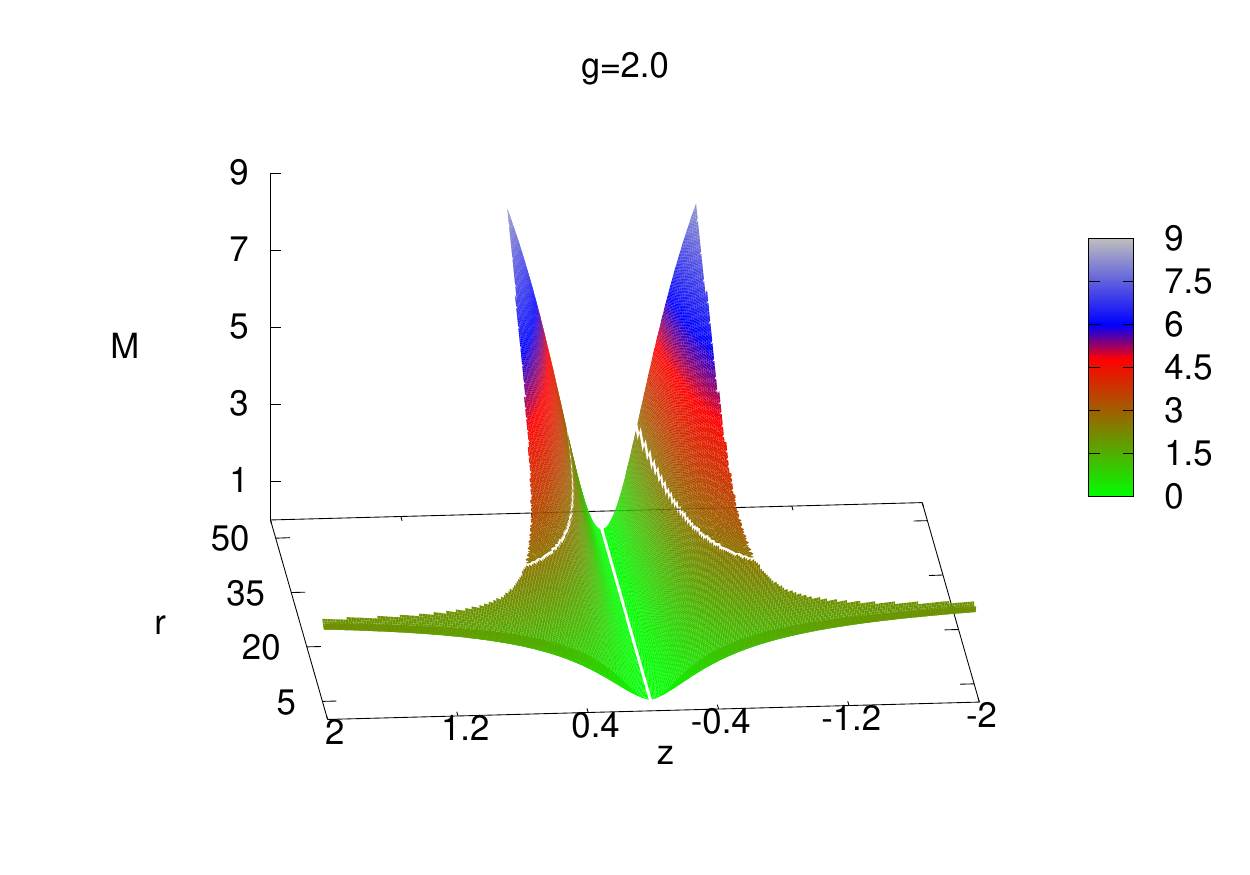}
    \caption{The mass parameter $M$ for Bardeen's BH $g/M < 0.7698$ (red$\dashrightarrow$blue colors) is shown as a function of the frequency shift $z$ ( redshift $z > 0$ and blueshift $z < 0$) and the radius $r$ of an eventual circular orbit of a photon emitter for four values of $g$. The fact that photons travel in the corresponding {\it effective metric} was taken into account to generate these plots. The green surface corresponds to the mass parameter for the globally regular spacetime sector $g/M> 0.7698$. $M$ and $r$ are in geometrized units and scaled by $pM_{\odot}$ where $p$ is an arbitrary factor of proportionality. Here, the numerical algorithm employed yields the known critical mass $M_c=(3\sqrt{3}g)/4$ that separates the BH from the GR spacetime (see text for details).}
    \label{Masas_BardeenEff}
\end{figure}

Figure \ref{CotasBardeenEff} shows a portion of this subset of points $\{ (r,g,z) \}$, in which, plots of the z-bounds in terms of $g$ and $r$ are displayed. Only with values of $z$ in the gap between the lower (red) surface $z_{min}(g,r)$ and the upper (green) surface $z_{max}(g,r)$ one finds $M(r,z,g)$ to be compatible with all the conditions for circular and stable orbits for photon emitters. This plot can be compared with figure \ref{Cota_Bardeen} constructed with the original metric $g_{\mu \nu}$ not with the effective $\widetilde{g}_{\mu \nu}$. The upper bounds shown in figure \ref{CotasBardeenEff}, that is to say the upper surface $z_{max}(r,g)$, reaches its highest values (for a given fixed $g$) along the line where the two surfaces $z_{max}(r,g)$ and $z_{max}(r,g)$ merge, on this line $z$ does not exceed $0.71$ which is slightly higher than Schwarzschild upper bound $1/\sqrt{2}$. 
Figure \ref{Masas_BardeenEff} presents the scaled relationship $M=M(r,z,l)$ for three $g=0.1,0.5,1.0,2.0$. As mentioned above, using Bardeen's original metric $g_{\mu \nu}$, $z$ is bounded approximately by $0.85$, whereas using the effective metric, it is bounded by approximately $0.71$. We also observed that as $g$ climbs up, the scaled mass parameter value yields a larger value. In figure \ref{Masas_Bardeen} (working with the original metric), looking at the data file, one may pick randomly a point in the domain, for instance $M_o(r=50,g=0.1,z_{max}(50,0.1)=8.266$ value is $8.266$ and see that increases its value as $g$ grows, $M_o(r=50,g=2,z_{max}(50,2))=8.402$. Whereas working with the effective metric (figure \ref{Masas_BardeenEff}), the largest value of the mass parameter is 
$M_{eff}(r=50,g=0.1,z_{max}(50,0.1)) =8.247$ for $g=0.1$ and $M:{eff}(r=50,g=2,z_{max}(50,0.1))=8.3928$. In regard to the mass parameter, generally speaking, there is a slight reduction when using the effective metric in relation to the original one.

We end this section calculating the angular velocity $\Omega$ of photon emitters orbiting in stable circular orbits. The relationships for $U^t$ and $U^{\phi}$ given in (\ref{vels}) are employed to compute $\Omega=U^{\phi}/U^t=\sqrt{f^{\prime}/r^4}$. For Bardeen metric it reads
\begin{equation}
    \Omega(g,r,z)=\sqrt{\frac{M(r^2-2g^2)}{(g^2+r^2)^{5/2}} }.
    \label{omegaB}
\end{equation}
\noindent The angular velocity becomes a function of $g,r,z$ after substituting $M_{-}$ given in (\ref{MBar}).  $\Omega$ is indeed a real quantity since $r^2 - 2 g^2>0$ and it is valid regardless Bardeen spacetime is considered a toy model or an exact solution. For $g=0$, the angular velocity for Schwwarzschild BH is recovered

\begin{equation}
    \Omega(r,z)=\sqrt{\frac{M}{r^3} }=\sqrt{\frac{\mathcal{F}_{-}(z)}{r^2} },
\end{equation}

\noindent where $\mathcal{F}_{-}=(1+5z^2-\sqrt{1+10z^2+z^4})/(12z^2)$.\\

\section{Hayward spacetime}
\label{Hayward BH}

Our second working example is the Hayward regular BH. Inasmuch as this regular spacetime had not been associated with any NED field, later on it was, here we work it as a toy model first and then as a solution of Einstein equation with NED, just as we did it in the previous section. Its $g_{tt}=-f(r)$ metric component reads
\begin{equation}
f(r)= 1 - \frac{2Mr^2}{r^3 + 2l^2 M} \equiv 1- \frac{2M}{r} R_H. 
\end{equation}

\noindent As it happens with the Bardeen spacetime, the function 
\begin{equation}
R_H(r,l)=\frac{r^3}{r^3+ 2 l^2 M},
\end{equation}

\noindent $R_H(r,l) \to 1$ as the parameter $l \to 0$ and the Schwarzschild BH is recovered. The function $f(r)$ behaves in a similar fashion as Bardeen's, i.e. it has two roots for $\tilde{l}=l/M<0.7698$ which implies the existence of an exterior $r_H^{ext}$ and interior $r_H^{int}$ event horizons, these horizons are found by solving $f(\tilde{r})=0$ (where $\tilde{r}= r/M$), which is equivalent to finding the roots of 
\begin{equation}
    \tilde{r}^3 - 2 \tilde{r}^2 + 2 \tilde{l}^2 =0.
    \label{r_H_Hayward}
\end{equation}
\begin{figure}
    \includegraphics[scale=0.8]{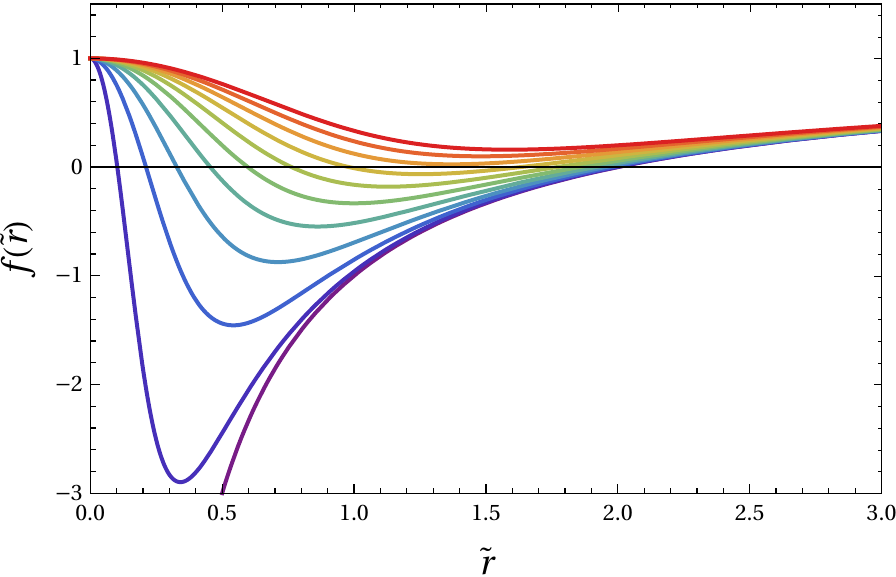}
    \caption{Plot of $f(\tilde{r})$ for Hayward spacetime as a function of $\tilde{r}=r/M$ for different values of parameter $\tilde{l}$. Here, $\tilde{l}= l/M$ changes from zero (purple) to unity (red) in steps of 0.1.
    As $\tilde{l}$ increases its value, the roots of $f(\tilde{r})$ get closer and then cease to exist. For $\tilde{l} > 0.7698$, $f(\tilde{r})$ is always positive.}
    \label{F_Hayward}
\end{figure}
For $\tilde{l} \neq 0$ the cubic has two real and positive roots and are given by Cardan's formula, provided that $\tilde{l}< \tilde{l}_c= 0.7698$. At the critical value $\tilde{l}_c$ the two horizons $r_H^{ext}$ and $r_H^{int}$ collide as it is shown in Fig. \ref{fig:bounds1}. For $\tilde{l}=0$ one is left with the Schwarzschild horizon $\tilde{r}=2$. We will work in the region outside the exterior horizon $r>r_H^{ext}$. The critical mass in this case is $M_c=3\sqrt{3} l/4$. \\

{\bf A. Hayward spacetime as a toy model} \\

Once again, in order to find an explicit expression for the mass parameter $M=M(z,r,l)$ we substitute the Hayward function $f(r)$ into (\ref{zfinalFB}) to attain a relationship between the redshift $z$, the Hayward parameter $l$, the radius of the photon emitter circular orbit $r$ and the mass parameter $M$  

\begin{equation}
    z^2= \frac{M r^5 R^{-1}_H(r,l) (r^3-4 l^2 M)}{\left [(r-2M)r^2+ 2l^2M \right ] \left [ r^5(r-3M)+4l^4M^2+4l^2M r^3 \right ]}.
    \label{z2Hay}
\end{equation}
\noindent Expressions for $E^2$, $L^2$ and $V^{\prime \prime}$ are computed as well

\begin{equation}
    E^2= \frac{(r^2 (r-2M)+2l^2 M)^2}{\left [ r^5(r-3M)+4l^4M^2+4l^2M r^3 \right ]},
    \label{E2H}
\end{equation}
    
\begin{equation}
    L^2= \frac{Mr^4(r^3-4 l^2 M)}{\left [ r^5(r-3M)+4l^4M^2+4l^2M r^3 \right ]},
    \label{L2H}
\end{equation}

\begin{equation}
    V^{\prime \prime}= \frac{2M \left [ r^5 (r-6M)+ 22l^2r^3M-32l^4M^2 \right ]}{\left [ r^5(r-3M)+4l^4M^2+4l^2M r^3 \right ] \left [(r-2M)r^2+ 2l^2M \right ]}.
    \label{VppH}
\end{equation}

Existence of circular orbits is guaranteed as long as $E^2,L^2 > 0$, that is to say, the following two conditions 

\begin{eqnarray}
    & & r^5(r-3M)+4l^4M^2 + 4l^2r^3M >0, \nonumber \\
    & & r^3 -4l^2 M > 0,
    \label{COHay}
\end{eqnarray}

\noindent simultaneously hold. Equation (\ref{z2Hay}) additionally requires that 

\begin{equation}
    r^2(r-2M)+ 2l^2 M > 0,
    \label{COz2Hay}
\end{equation}

\noindent which is satisfied since we are working outside the exterior horizon. Lastly, circular orbits stability $V^{\prime\prime} > 0$ is akin to demanding that

\begin{equation}
    r^5 (r-6M) + 22l^2 r^3 M-32l^4M^2 > 0.
    \label{COddV}
\end{equation}

It is apparent that as $l \to 0$, the condition for stability of circular orbits $V^{\prime \prime}>0$ implies $r>6M$ as in the Schwarzschild BH case. 

From (\ref{z2Hay}) one arrives at a cubic equation to get $M=M(z,r,l)$ ($l\neq0$), explicitly

\begin{equation}
    C_3 M^3 + C_2 M^2 + C_1 M + C_0 = 0, 
    \label{MHay}
\end{equation}
where 
\begin{eqnarray}
    C_3 &=& 8 l^4 (r^2+z^2(l^2-r^2)), \nonumber \\
    C_2 &=& z^2( 12 l^4 r^3 -14 l^2 r^5+6 r^7) + 2l^2r^5,\nonumber \\
    C_1 &=& z^2 (6 l^2 r^6- 5r^8) -r^8, \nonumber \\
    C_0 &=& z^2 r^9. \nonumber
\label{Coeficientes_M}
\end{eqnarray}

It can be verified that for $l=0$ (\ref{MHay}) becomes the mass parameter formula (\ref{MSCH}) for Schwarzschild. 
Given a cubic equation, there could be either three real roots or one real root and two complex. 
We have verified numerically that only one real value of the mass parameter allows the conditions for stability of circular orbits to be fulfilled together with the condition 
$r>r_H^{ext}$; furthermore, these conditions are satisfied solely for $|z|<0.75$. This analysis was performed using a combination of bisection and Newton-Raphson methods to solve (\ref{MHay}). We have also attained the bounds of $z=z(l,r)$. We just basically followed the five steps of the numerical algorithm described in the previous section. 

\begin{figure}
   \includegraphics[scale=0.68]{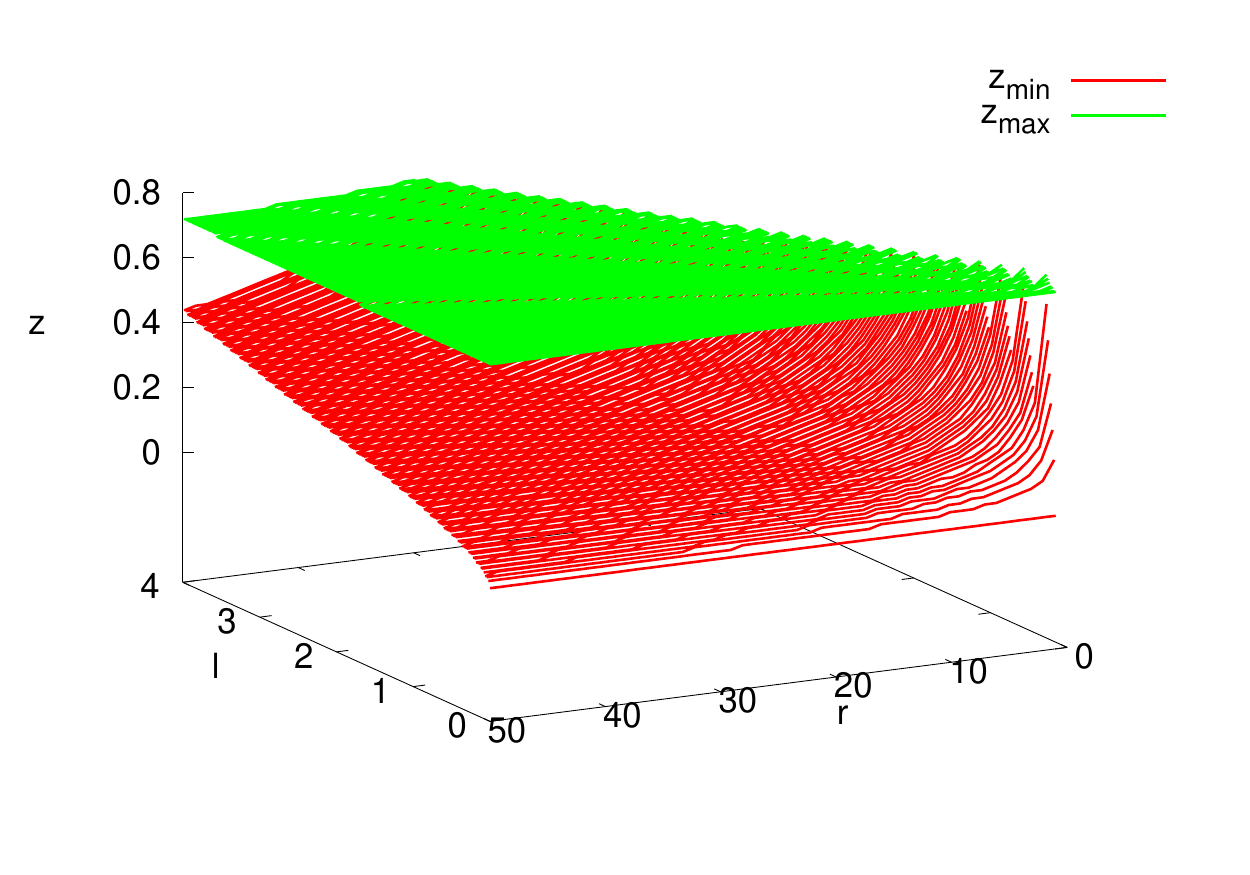}
    \includegraphics[scale=0.68]{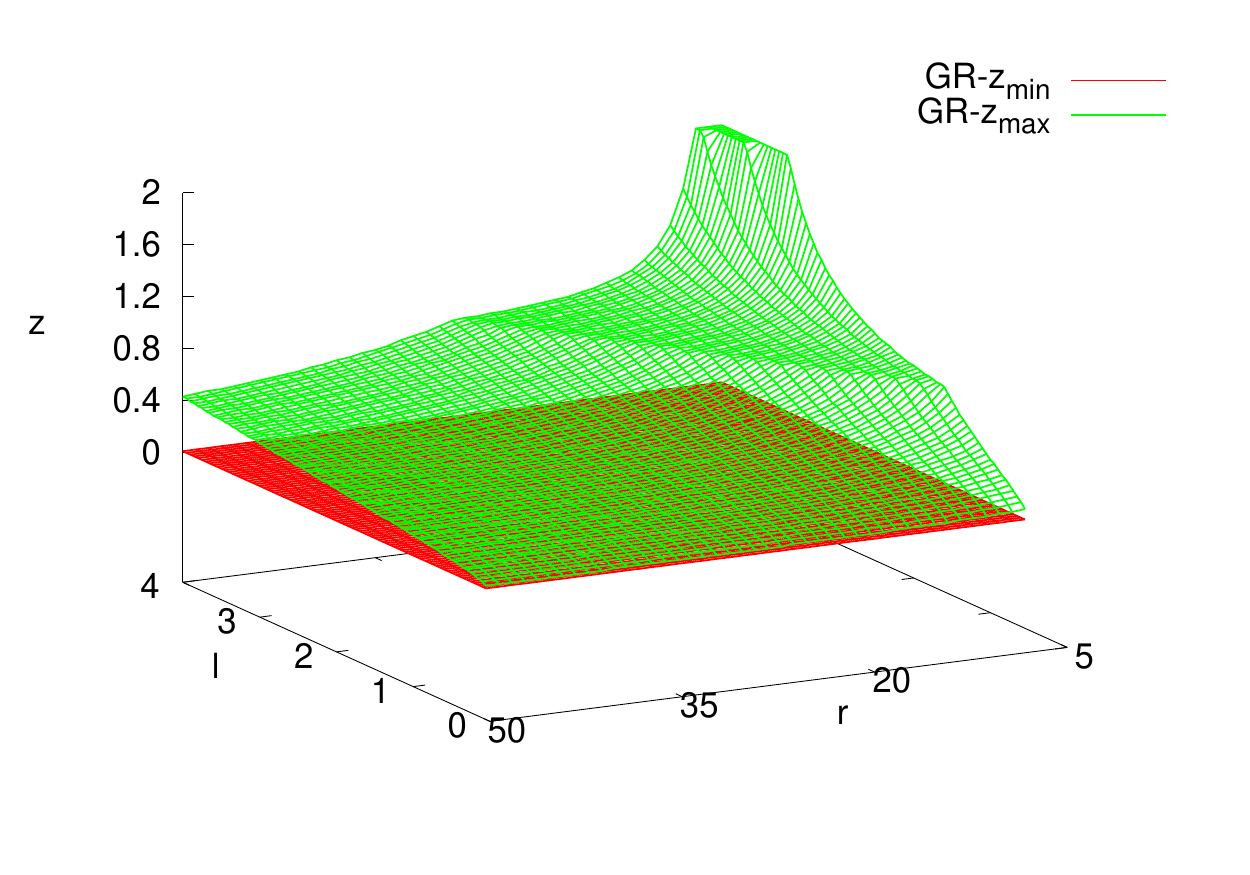}
     \includegraphics[scale=0.68]{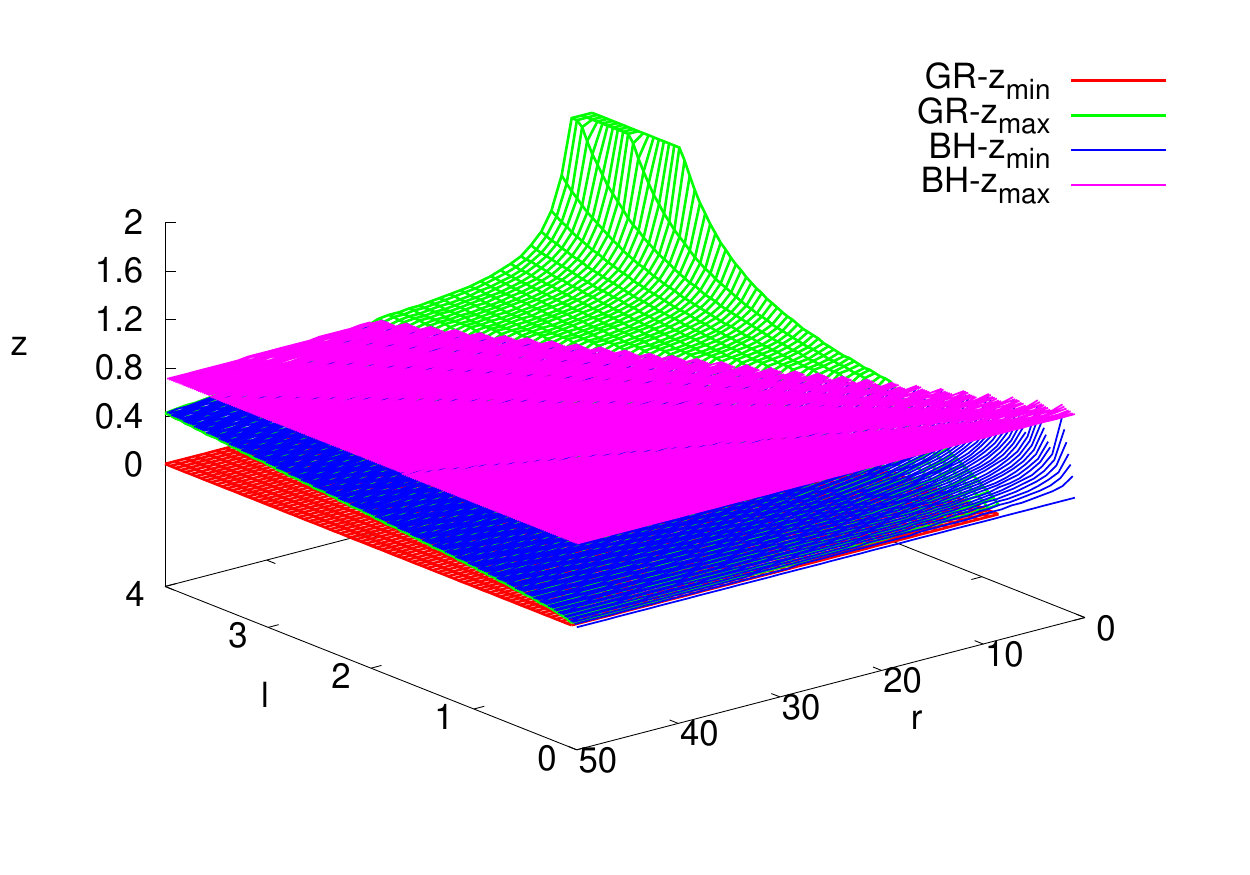}
    \caption{In the upper plot, we presents bounds of $z$ for Hayward's BH when using the {\it original metric}. The only allowed frequency shifts that could be detected by a far away observer is located in the gap  $z_{min}(r,l) < z < z_{max}(r,l)$. It turns out that $z<0.75$. In the middle plot, we present bounds of $z$ for globally regular Hayward's spacetime. Only $z \in [z_{min},z_{max}]$ are allowed. The two surfaces $z_{max}$ and $z_{min}$ corresponding to Hayward's globally regular and BH spacetimes respectively, coincide when $z<0.75$. Whereas, when $z>0.75$ only globally regular spacetime is allowed.
    This is shown in the lower plot where we have superimposed both bounds for Hayward's BH and globally regular spacetimes.}
    \label{Cota_Hayward}
\end{figure}
\begin{figure}[!h]
    \centering
    \includegraphics[scale=0.58]{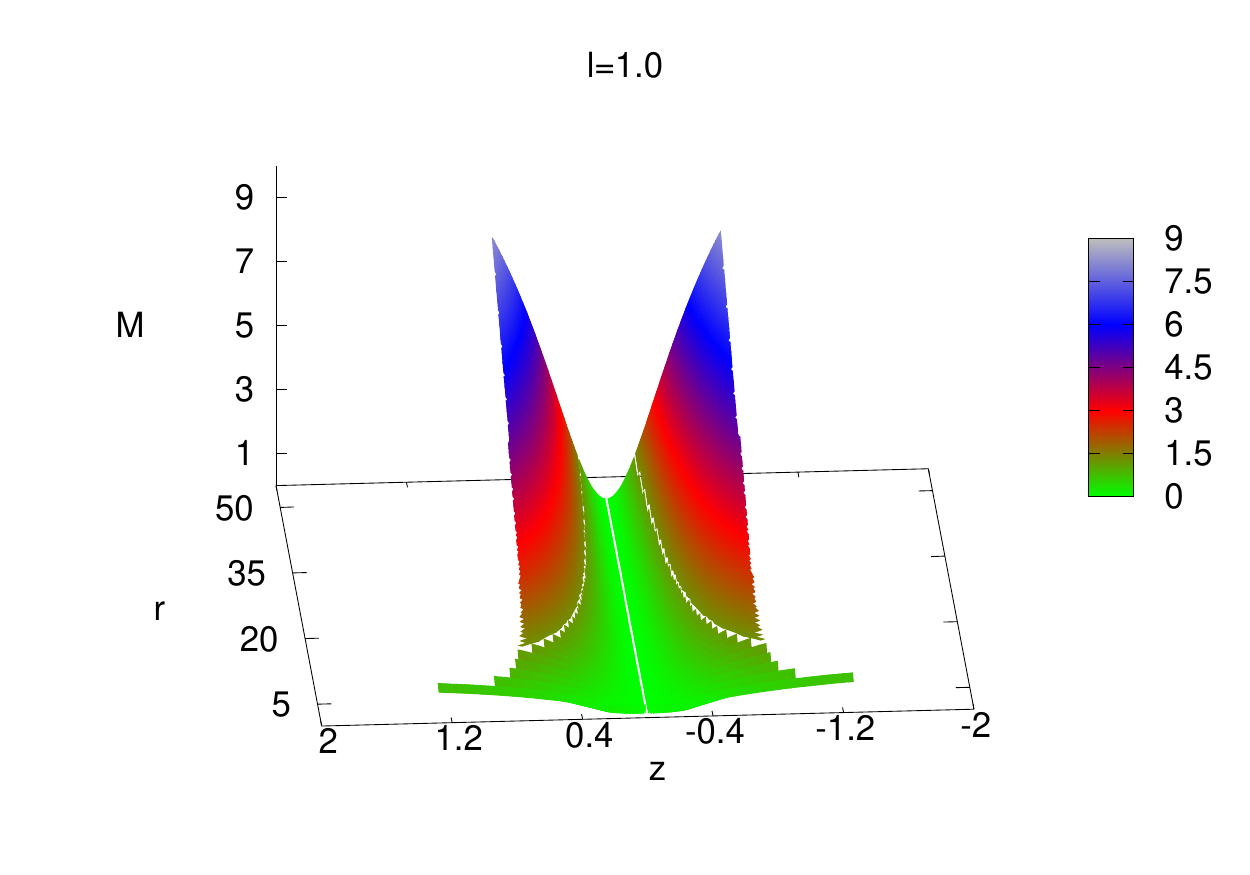} \\
    \vspace{-0.8cm}
    \includegraphics[scale=0.58]{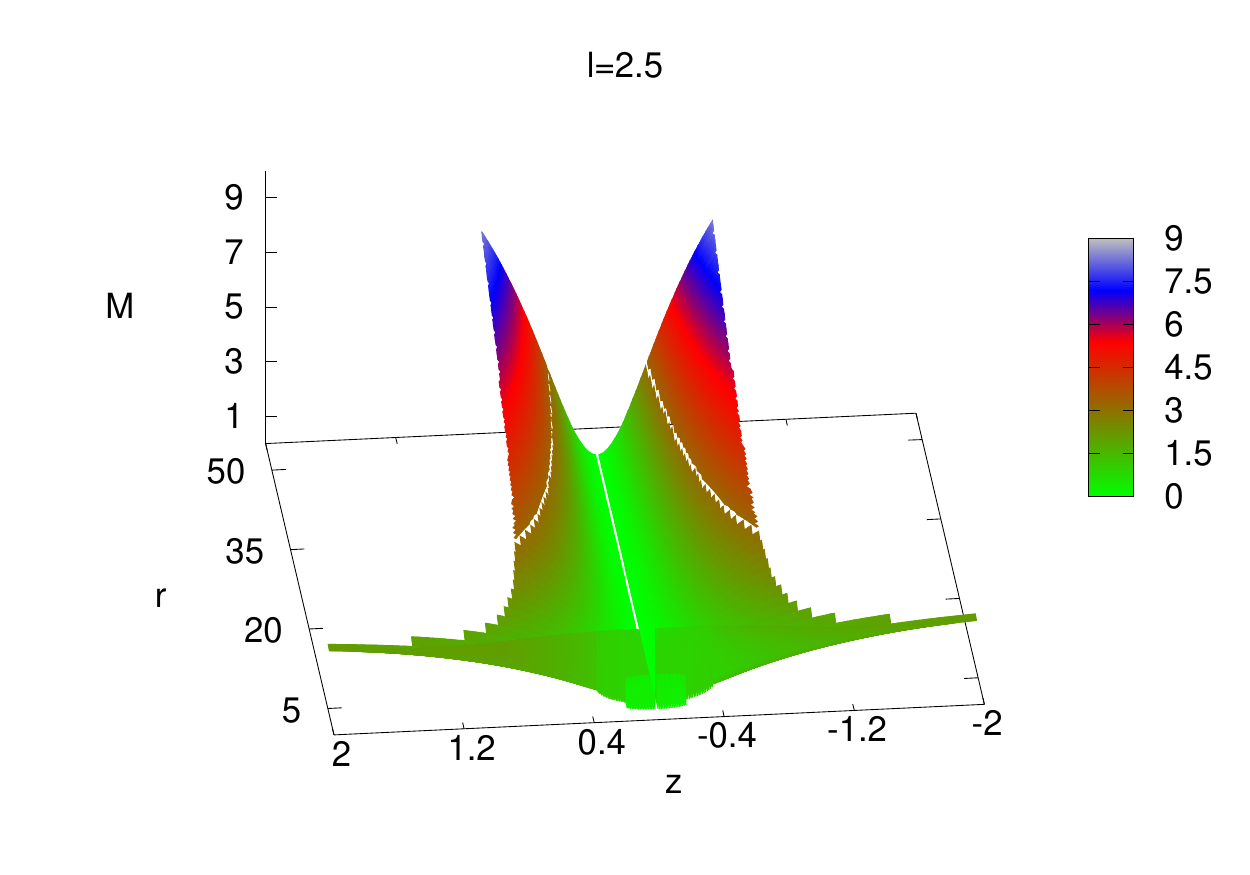} \\
    \vspace{-0.8cm}
    \includegraphics[scale=0.58]{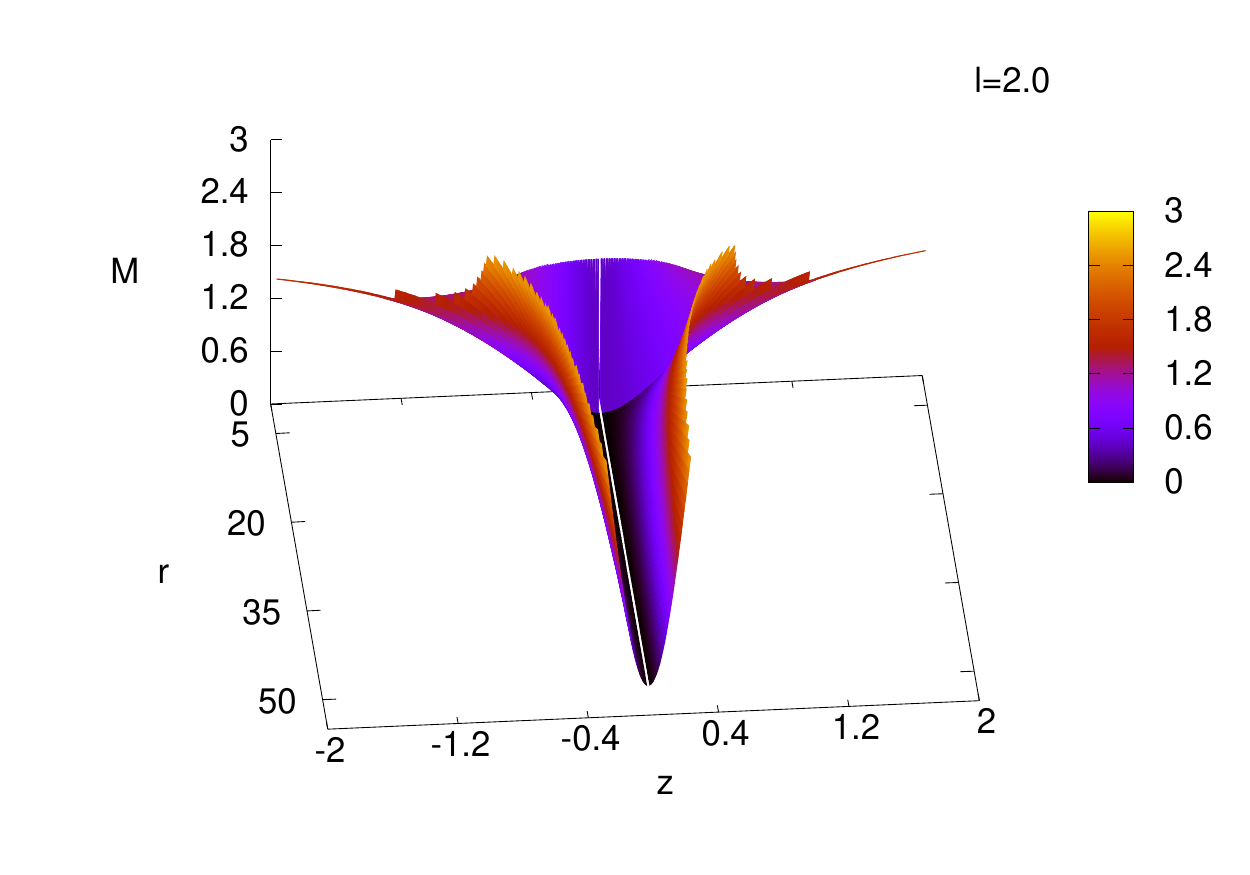} \\
    \vspace{-0.8cm}
    \includegraphics[scale=0.58]{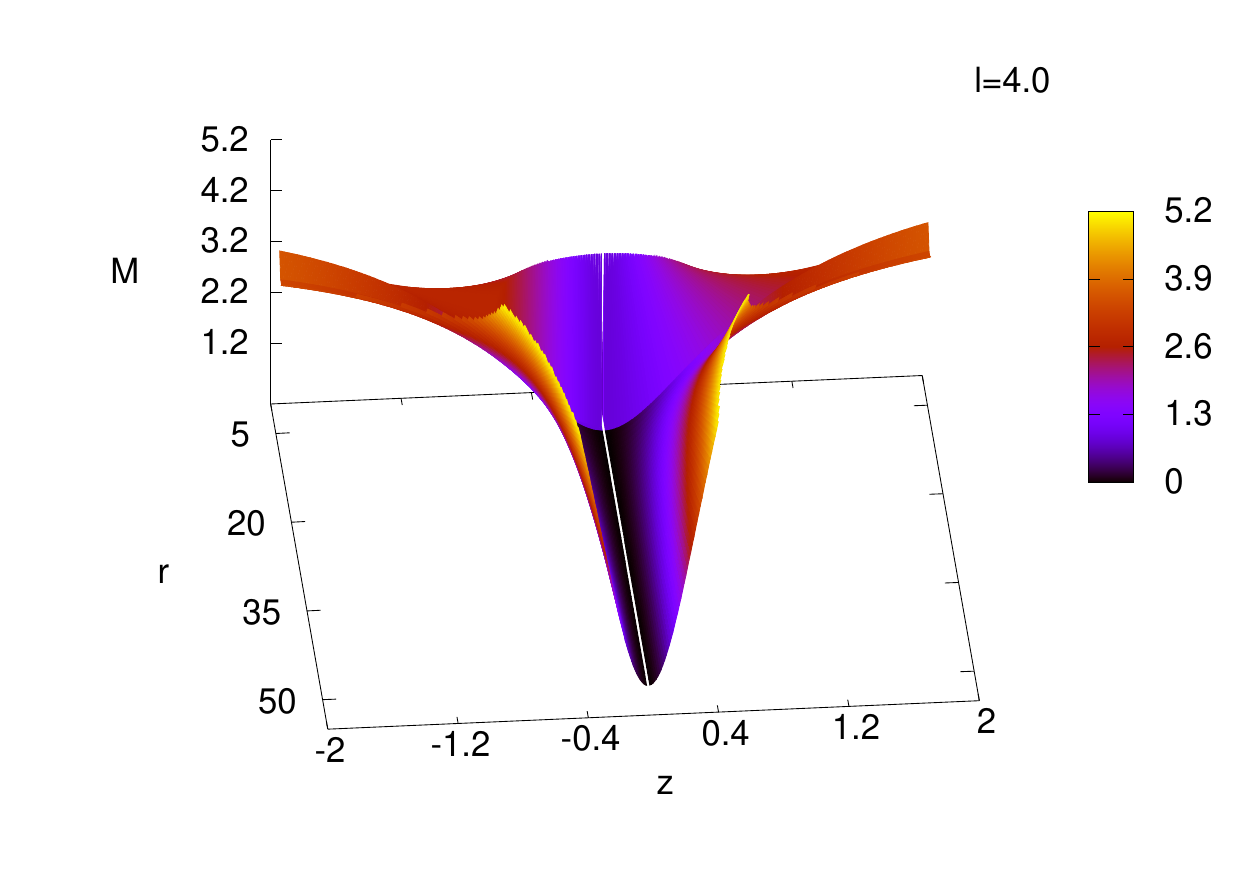} 
    \caption{The mass parameter $M$ for Hayward's BH $l/M < 0.7698$ (red$\dashrightarrow$gray color) is shown as a function of the frequency shift $z$ (redshift $z > 0$ and blueshift $z < 0$) and the radius $r$ of an eventual circular orbit of a photon emitter for $l=1$ and $l=2.5$. The green surface corresponds to the mass parameter for the globally regular spacetime sector $l/M> 0.7698$. The third and fourth plots corresponds to the mass parameter $M$ for solely globally regular (GR) spacetimes for $l=2.0,4.0$, one observes that $M$ increases with $l$; nonetheless, it is much smaller than the one for a BH.  Here, our numerical algorithm yields the bound that separates the BH from globally regular spacetime and agrees with the relationship given in \cite{Hayward:2005gi}.}
    \label{Masss_Hayward}
\end{figure}
In Fig. \ref{Cota_Hayward} one observes bounds of the frequency shifts for the two sectors: black-holes (upper plot) and globally regular spacetimes (middle plot). $\mathcal{D}_M^{BH}$ is a subset where all the conditions for existence of stable circular orbits of photons emitters outside the exterior event horizon are satisfied. It stretches out in the region where $z_{min}(r,l) < z < z_{max}(r,l)$ (see upper plot in figureFig. \ref{Cota_Hayward}) it happens that $z_{max}<0.75$ in all the domain $\mathcal{D}_M^{BH}$, this value is reached along the curve where the two bound surfaces merge. The corresponding subset for the global regular sector is shown in the middle plot in the same figure. The surface $z_{max}(r,l)$ for GR coincides with the $z_{min}(r,l)$ for the BH case, but in the case the surfaces spreads further and allows $z>0.75$ differing from the BH case. This is shown in the lower plot where we have superimposed both bounds for Hayward's BH and globally regular spacetimes.

As in the Bardeen working example, we scale $M$ and $r$ by an arbitrary multiple of the solar mass. Plots of $M=M(r,z)$ for different values of the Hayward parameter $l$ follows an analogous pattern as Bardeen's plots. This means that for $l=0.1$ the graph $M=M(r,z)$ is rather similar as the one for Schwarzschild BH. The gap in the $z$ axis that splits the regions for $z>0$ and $z<0$ also increases its width as $l$ increases. $z$ goes beyond the Schwarzschild bound $|z|<1/\sqrt{z}$, yet is smaller than unity. Figure \ref{Masss_Hayward} shows this scaled relationship $M=M(r,z)$ for $l=1.0$ and $2.5$. The mass parameter allowed for the globally regular spacetime sector (green color) is smaller than for BH sector (red-blue color). Since there is no event horizon, stable circular orbits of photon emitters may allow smaller radii than those for the BH sector. As $l$ increases, the surface $M(r,z)$ for the BH region, splits more notoriously into two parts for the red and blueshifts. We also observed that Hayward's parameter $l$ grows, the scaled mass parameter also increases. Working with the original Hayward's metric, the largest value of $M_o(r=50,l=0.1,z_{max}(50,0.1))  = 8.265$ for $l= 0.1$ and increases to 
$M_o(r=50,l=2,z_{max}(50,0.1)) = 8.3721$ for $l= 2$. This is a typical behavior of the mass parameter as $g$ grows. \\

{\bf B. Hayward spacetime as an exact solution of Einstein field equations with NED} \\

Considering Hayward spacetime as an exact solution of Einstein equationts with NED, the mass parameter expression (\ref{zfinalFBeffective}) must be employed. The effective metric is

\begin{equation}
     \widetilde{g}_{tt}= g_{tt}, \quad \widetilde{g}_{rr}= g_{rr}, \quad
    \widetilde{g}_{\theta \theta}= g_{\theta \theta}/\Theta, \quad  \widetilde{g}_{\phi \phi}= g_{\phi \phi}/\Theta, \nonumber
\end{equation}

where $\Theta(r,l)$ reads
\begin{equation}
    \Theta=1+\left [ \frac{1- \frac{7}{2} (\frac{2l^2 M}{r^2} )}{1+ (\frac{2 l^2 M}{r^2})}\right ], \nonumber
\end{equation}
(see the appendix). Since $\Theta \to 2$ as $l \to 0$, hereby the electromagnetic fields vanish, Hayward's effective metric does not become Schwarzschild metric in this limit, neither the expression for $\widetilde{M}(r,z,l)$ becomes the one for Schwarzschild. 

Substituting Hayward function $f(r)$ in (\ref{zfinalFBeffective}) one gets
\begin{equation}
    z^2 = \frac{Mr^2(4l^2M-r^3)}{S_1S_2(4l^2M(l^2M+r^3) + r^5(-3M+r))},
\end{equation}
where 
\begin{eqnarray}
    S_1 &=& 1 + \frac{r^3 - 7l^2M}{r^3 + 2l^2M},\nonumber \\
    S_2 &=& \frac{2r^2M}{2l^2M+r^2}-1, \nonumber
\end{eqnarray}
which relates the redshift $z$, the Hayward parameter $l$, the radius of the photon emitter $r$ and the mass parameter $M$,  From it, a four order polynomial for $M$ in terms of $z,r$ and $l$ is found
 \begin{equation}
     C_4M^4 +C_3 M^3 + C_2 M^2 + C_1 M + C_0=0,
 \end{equation}
 with 
 \begin{eqnarray}
     C_4 &=& 8l^6(5z^2(l^2-r^2)-2r^2), \nonumber \\
     C_3 &=& 2l^2r^3(z^2(22l^4+15r^4)-3l^2r^2(2+9z^2)), \nonumber \\
     C_2 &=& 3r^6z^2(2l^4 +l^2r^2-4r^4), \nonumber \\
     C_1 &=& r^9(r^2 + z^2(10r^2-7l^2)), \nonumber\\
     C_0 &=& -2r^{12}z^2. \nonumber
 \end{eqnarray}

\begin{figure}
   \includegraphics[scale=0.68]{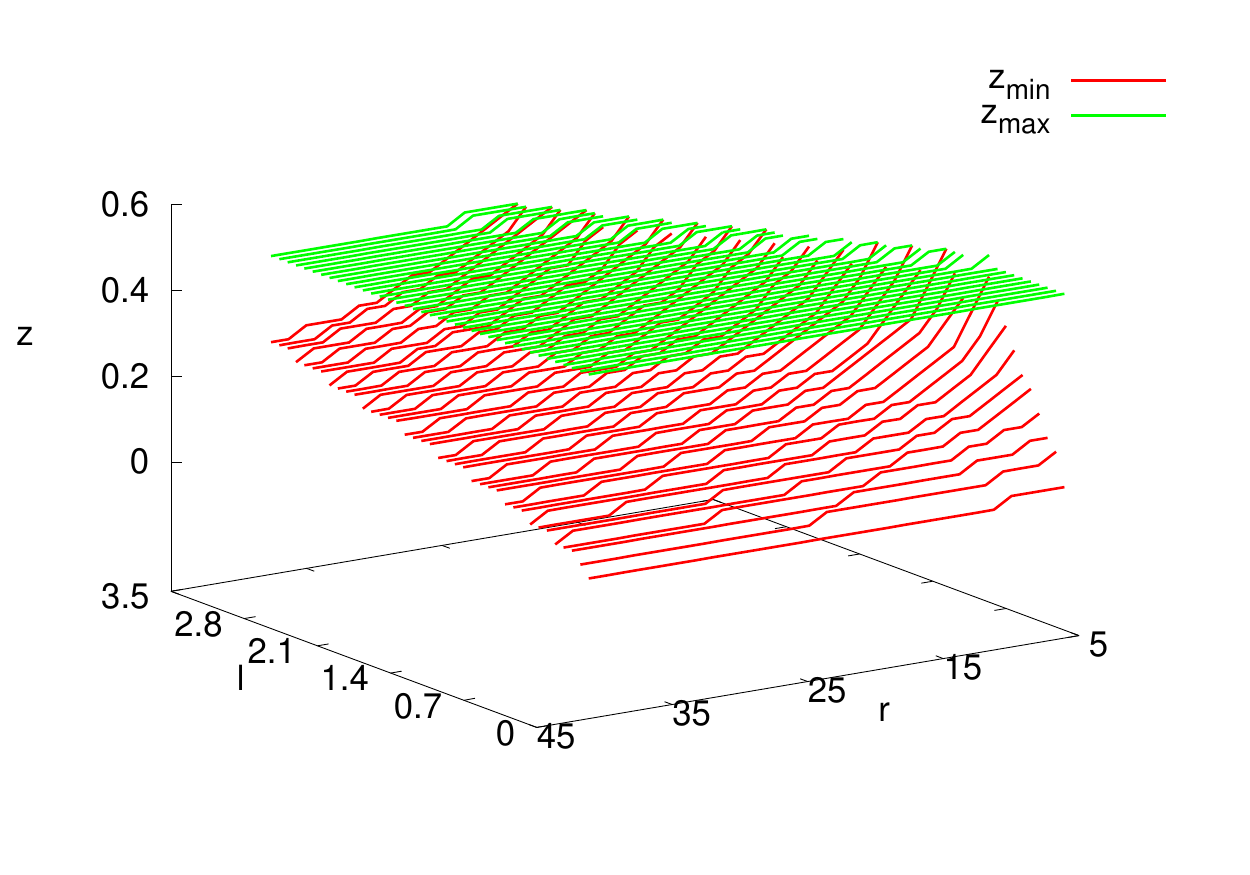}
    \includegraphics[scale=0.68]{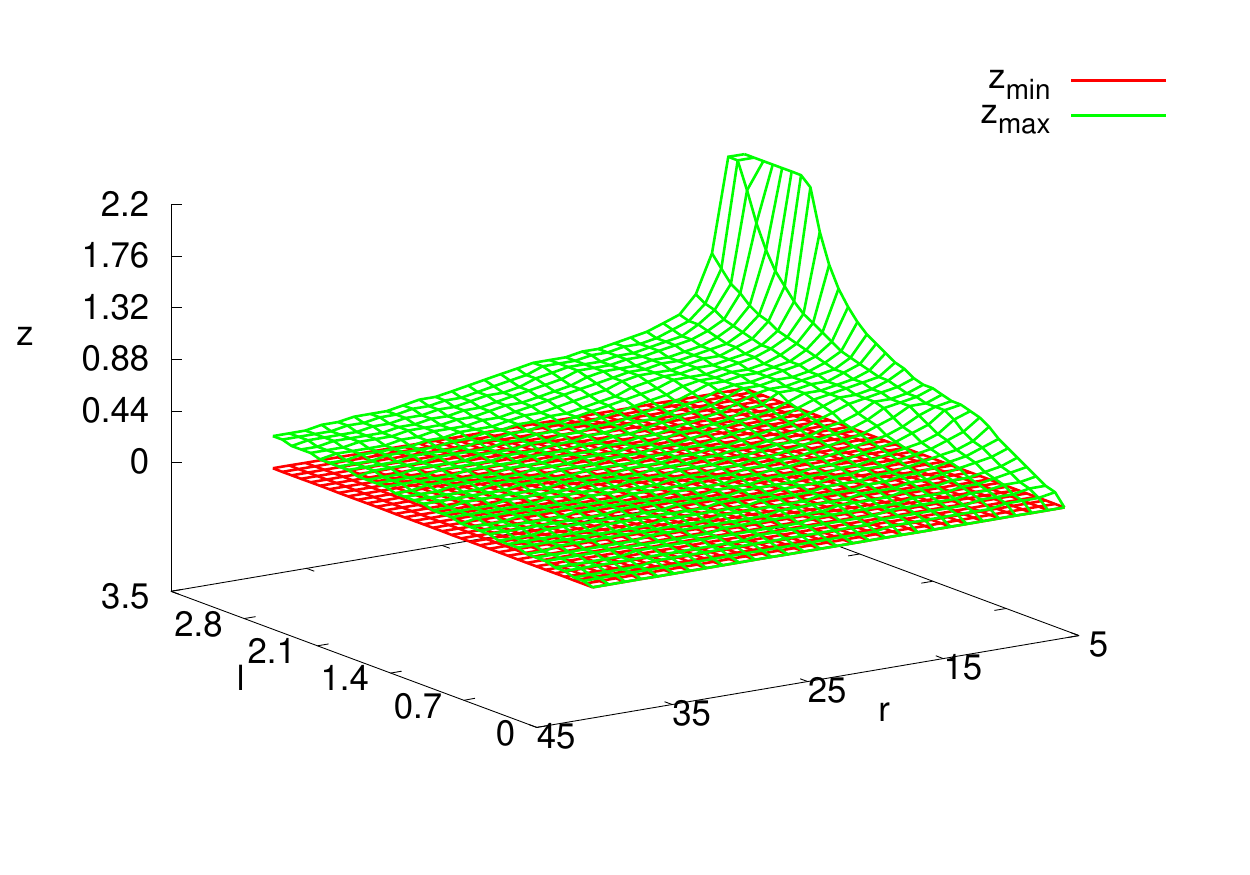}
     \includegraphics[scale=0.68]{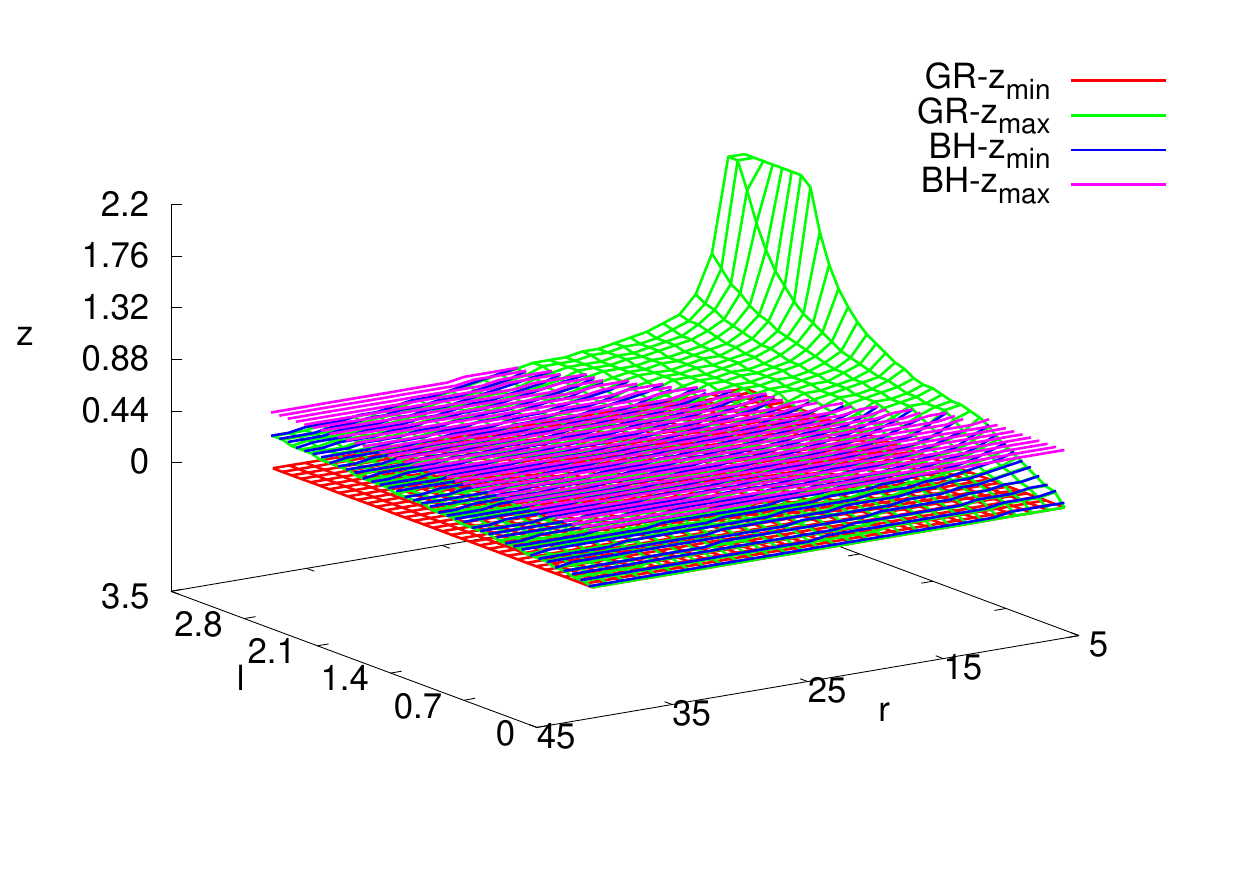}
    \caption{In the upper plot, we presents bounds of $z$ for Hayward's BH when using the {\it effective metric}. The only allowed frequency shifts that could be detected by a far away observer is located in the gap  $z_{min}(r,l) < z < z_{max}(r,l)$. It turns out that $z<0.53$. In the middle plot, we present bounds of $z$ for globally regular Hayward's spacetime. Only $z \in [z_{min},z_{max}]$ are allowed. The two surfaces $z_{max}$ and $z_{min}$ corresponding to Hayward's globally regular and BH spacetimes respectively, coincide when $z<0.53$. Whereas, when $z>0.53$ only globally regular spacetime is allowed.
    This is shown in the lower plot where we have superimposed both bounds for Hayward's BH and globally regular spacetimes}
    \label{Cota_Hayward_Effec}
\end{figure}

\begin{figure}[!h]
    \centering
    \includegraphics[scale=0.58]{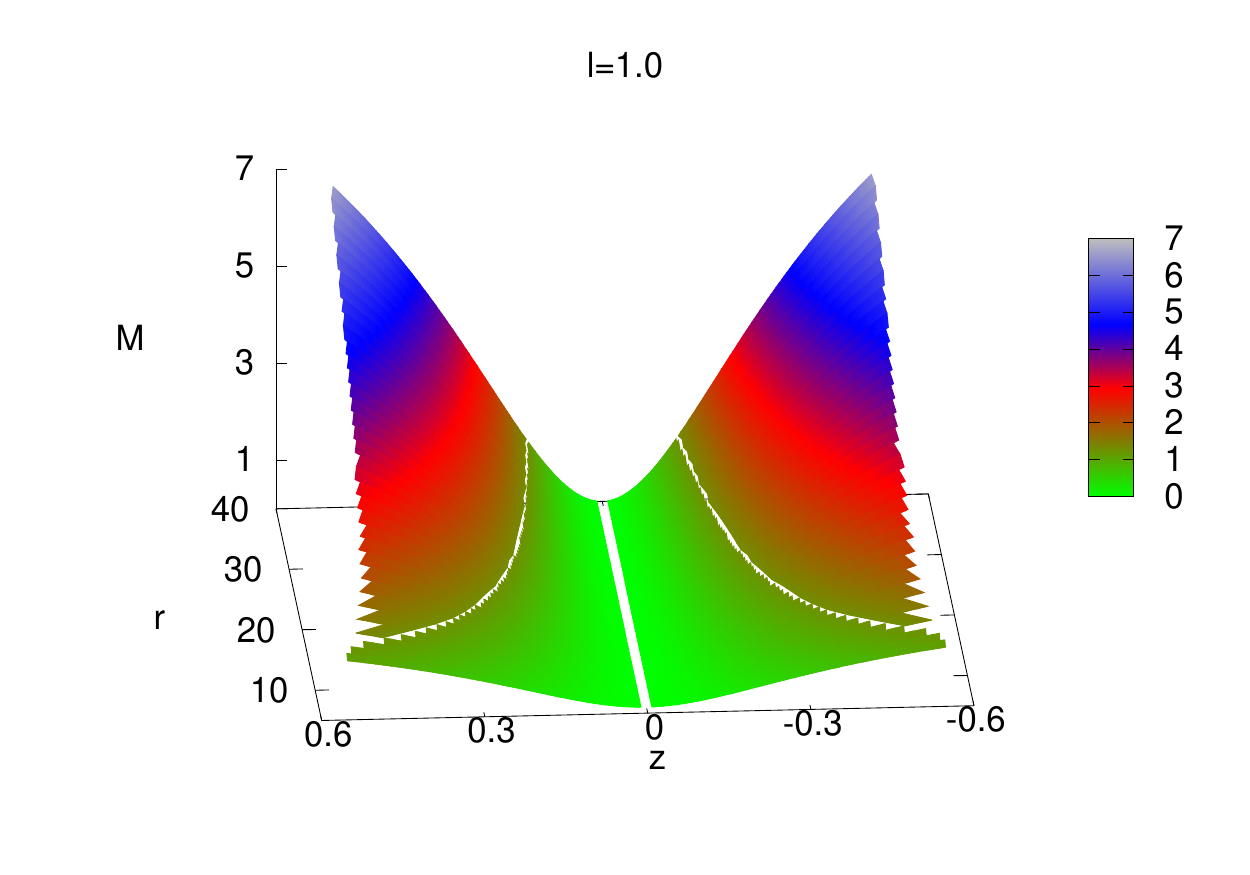} \\
    \vspace{-0.8cm}
    \includegraphics[scale=0.58]{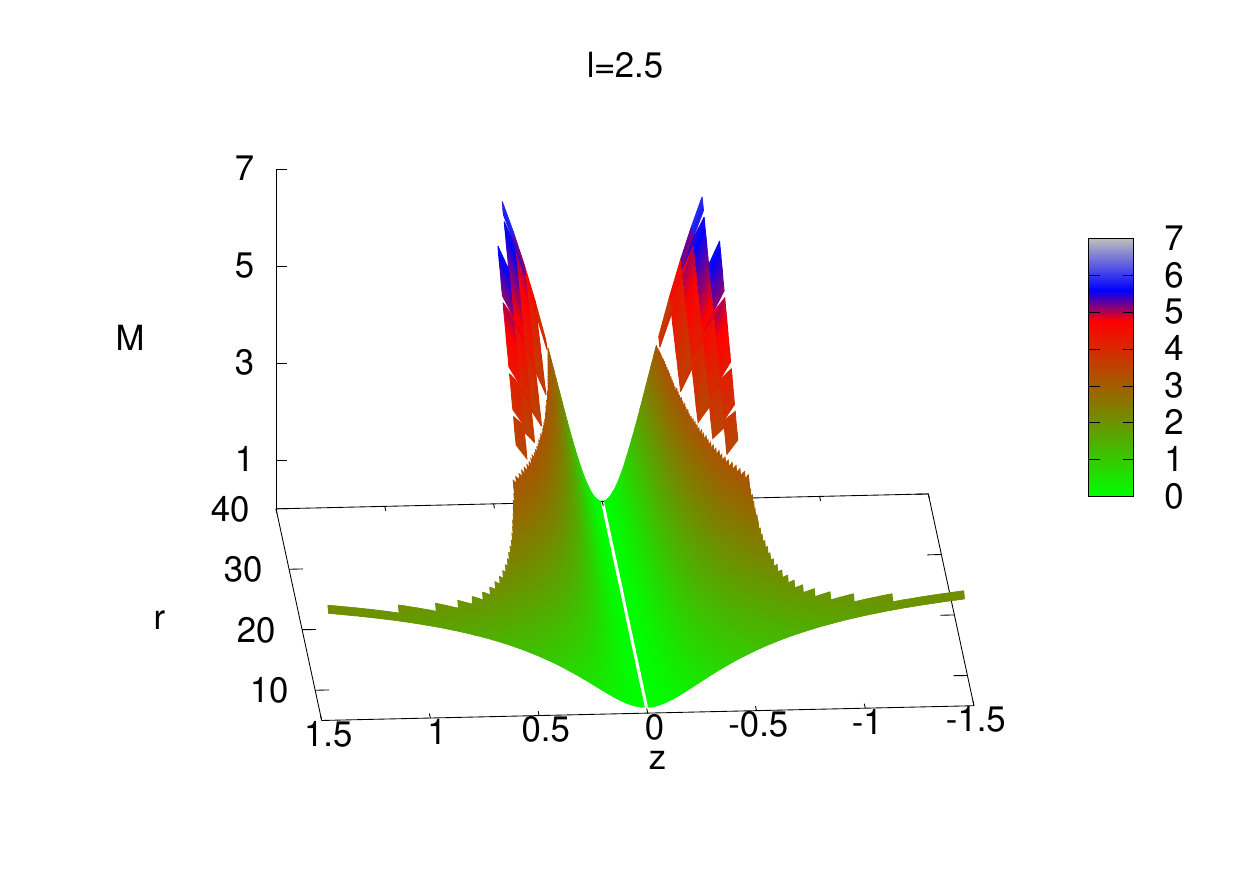} \\
    \vspace{-0.8cm}
    \includegraphics[scale=0.62]{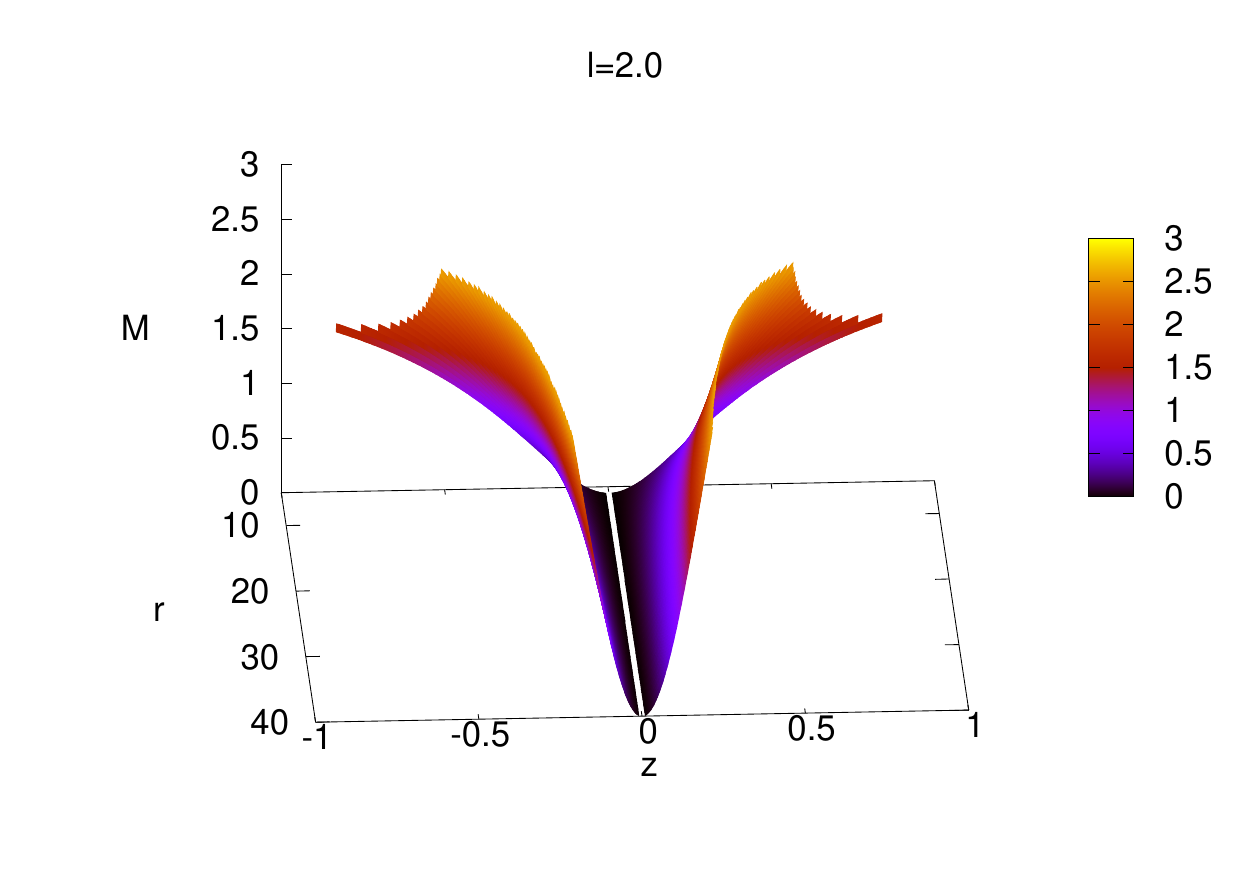} \\
    \vspace{-0.8cm}
    \includegraphics[scale=0.62]{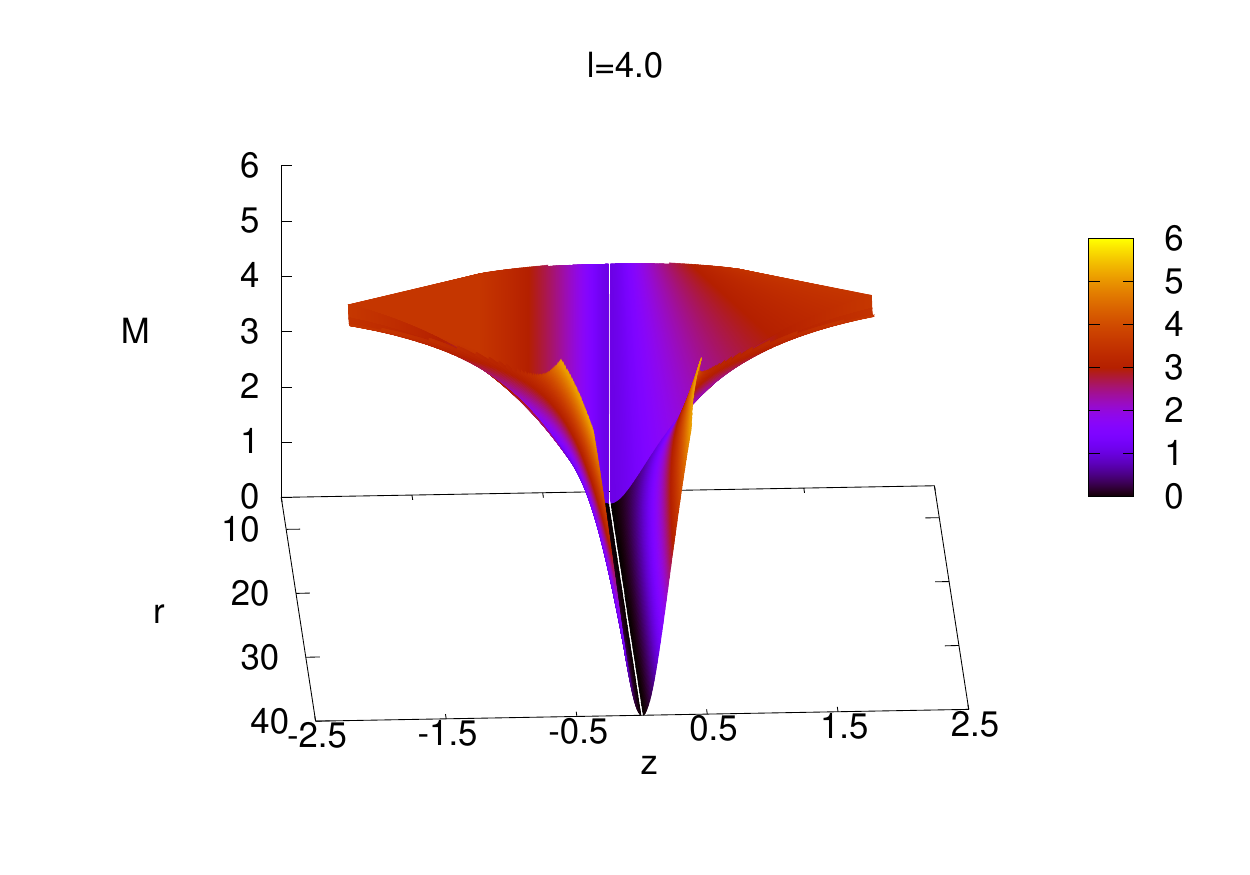} 
    \caption{The mass parameter $M=M(r,z,l)$ for Hayward's BH $l/M < 0.7698$ (red$\dashrightarrow$gray color) is shown for $l=1$ and $l=2.5$ when using the {\it effective metric}. The green surface corresponds to the mass parameter for the globally regular spacetime sector $l/M> 0.7698$. The third and fourth plots corresponds to the mass parameter $M$ for solely globally regular (GR) spacetimes for $l=2.0,4.0$, one observes that $M$ increases with $l$; nonetheless, it is much smaller than the one for a BH.}
    \label{Masss_Hayward_Effec}
\end{figure}

The conditions for existence of stable circular orbits remain the same (\ref{COHay}) and (\ref{COddV}).
A four degree polynomial may possess up to four real roots. We have verified numerically that only one real value of the mass parameter allows the conditions for stability of circular orbits to be fulfilled. Furthermore, these conditions are satisfied for $|z|<0.53$ considerably lower than the Schwarzschild fixed bound ($0.7071$). In Fig. \ref{Cota_Hayward_Effec} one observes bounds of the frequency shifts for the BH (upper plot) and GR (middle plot) sectors. The subset $\mathcal{D}_M^{BH}$ where the conditions for existence of stable  circular  orbits  of  photons emitters  outside  the  exterior  event  horizon  are  satisfied,   stretches out in the region where $z_{min}(r,l)< z < z_{max}(r,l)$ (see upper plot in Fig. \ref{Cota_Hayward_Effec}) it turns out that $|z| < 0.53$ in all the domain $\mathcal{D}_M^{BH}$. In Fig. \ref{Masss_Hayward_Effec} $M(r,z,l)$ is presented for $l=1,2.5$. The fact that photons travel in the corresponding effective metric was taken into account to numerically generate these plots. The green (lower section) surface corresponds to the mass parameter for the globally regular spacetime sector 
($g/M >0.7698$); the upper (blue section) surface corresponds to the BH sector ($g/M < 0.7698$). 
For the BH sector, $M_{eff}(r=40,z_{max}(40,0.1),l=0.1)=6.666$ and $M_{eff}(r=40,z_{max}(40,2),l=2)=6.676$, generally speaking, as $l$ grows, so does $M(r,z,l)$. When using the original metric $M_o(r=40,z_{max}(40,0.1),l=0.1)=6.6128$ and $M_o(r=40,z_{max}(40,2),l=2)=6.703$. In general, comparing both cases, as an exact solution or as a toy model, $M_{eff}(r,l,z_{max}(r,l)) > M_o(r,l,z_{max}(r,l))$, be aware that for both cases $z_{max}(r,l)$ most likely would be different.

To conclude this section, we calculate the angular velocity $\Omega$ of photon emitters orbiting along stable circular orbits. The relationship $\Omega=\sqrt{f^{\prime}/r^4}$ is employed, it yields for Hayward metric

\begin{equation}
    \Omega(r,z,l)=\frac{\sqrt{M(r^3-4l^2M)}}{2l^2M+r^3}.
\end{equation}

\noindent The angular velocity becomes a function of $l,r,z$ after inserting $M(l,r,z)$ solution of the cubic equation (\ref{MHay}). The effective metric does not play a role in computing $\Omega$ since it relates to geodesic particles, not photons. $\Omega$ is indeed a real quantity since $r^3 - 4 l^2 M>0$ in compliance with (\ref{COHay}). Once again, for $l=0$ the angular velocity for Schwarzschild metric is recovered.

\section{Ayon-Beato-Garc\'ia Regular Black Hole}
\label{ABG BH}

The previous two working examples ware originally regular {\it models} that avoid the BH singularity problem, not being truly solutions to the Einstein field equations at first. Later on, they were interpreted as a solution of Einstein field equations with NED. Unlike Bardeen and Hayward spacetimes, Ayon-Beato and Garc\'ia (ABG) constructed a singularity free {\it exact} solution of the Einstein field equations coupled to nonlinear electrodynamics satisfying the weak energy condition from the very beginning. The function $f(r)$ is given by

\begin{equation}
    f(r) = 1-\frac{2Mr^2}{R^{3/2}}+\frac{Q^2r^2}{R^2} \quad \text{with} \quad R=r^2+Q^2.
    \label{f_BG}
\end{equation}

This function asymptotically behaves as the one corresponding Reissner-Nordstr\"{o}m BH

\begin{equation}
    f(r) = 1- \frac{2M}{r} + \frac{Q^2}{r^2} + \mathcal{O}(1/r^3) ,
    \label{RN_limit}
\end{equation}

\noindent which possesses two event horizons given by $r_H= M \pm \sqrt{M^2-Q^2}$ that are real and distinct provided that $M^2 > Q^2$. ABG spacetime possesses an event horizon for certain values of $Q$ and $M$. In order to locate the event horizon one has to find the roots of $f(r)=0$. It is convenient to introduce the variables $\tilde{r}=r/Q$ and $\tilde{Q}=Q/(2M)$; hence, finding the roots of $f(r)=0$ is akin to finding the roots of

\begin{equation}
    (\tilde{r}^8+6\tilde{r}^6+11\tilde{r}^4+6\tilde{r}^2+1)\tilde{Q}^2-\tilde{r}^4(1+\tilde{r}^2)=0.
\end{equation}

\begin{figure}
    \includegraphics[scale=0.8]{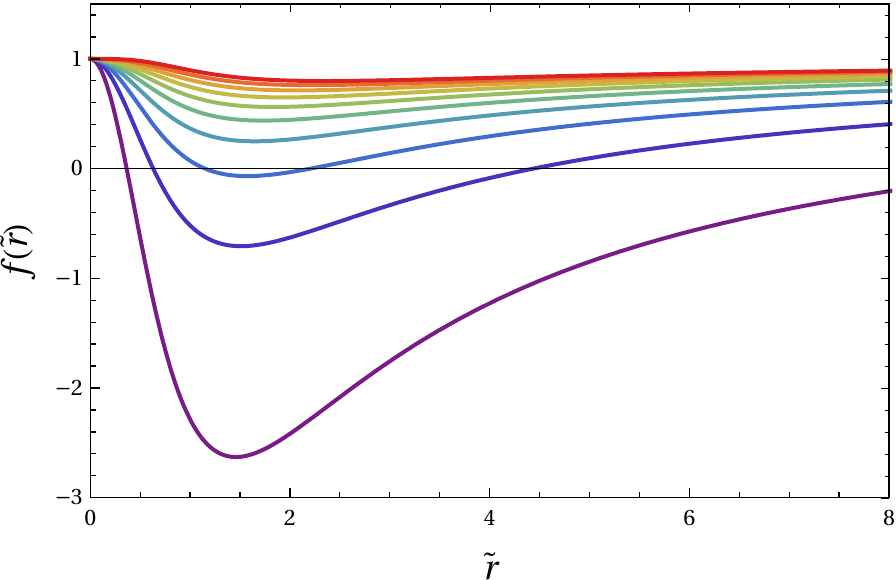}
    \caption{Plot of ABG's $f(\tilde{r})$ as a function of $\tilde{r}=r/Q$ for different values of parameter $\tilde{Q}$. Here, $\tilde{Q}= Q/2M$ changes from 0.1 (purple) to unity (red) in steps of 0.1.
    As $\tilde{Q}$ increases its value, the roots of $f(\tilde{r})$ get closer and then cease to exist. For $\tilde{Q} > 0.32$ approximately, $f(\tilde{r})$ is always positive.}
    \label{F_ABG}
\end{figure}

Figure \ref{fig:bounds1} shows the external $r_H^{ext}$ and internal $r_H^{int}$ event horizons. We work in the region outside the exterior horizon $r > r_H^{ext}$, whose value depends on $Q$ and $M$, that is $r_H^{ext}= r_H^{ext}(Q,M)$, this is a condition that we ought to keep in mind too.

We want to find an analytical formula for the mass parameter $M=M(z,r,Q)$, where $r$ is the radius of circular orbits followed by geodesics particles emitting photons whose frequency shift is $z$ as detected by a far away observer and $Q$ is the electric charge of the compact object. As in the previous two sections, we will distinguish two cases, first as if ABG spacetime were a toy model, that is to say, not considering the effective metric to determine the path followed by photons; secondly using the effective metric constructed. \\

{\bf A. ABG spacetime using the original metric} \\

For this first case, we insert the function (\ref{f_BG}) into (\ref{zfinalFB}), (\ref{energy}), (\ref{angularM}) and (\ref{segundita}), to find the analytic expressions for the redshift, the energy, the angular momentum and the second derivative of the effective potential 
\begin{widetext}
\begin{equation}
    z^2= \frac{r^2 R^2 \left [ Q^2 (Q^2-r^2)-M \sqrt{R} (2Q^2-r^2) \right ]}{\left [-2M^2 \sqrt{R}+(R^2+Q^2r^2)\right ] \left [-3Mr^4\sqrt{R}+(R^3+2Q^2r^4)\right ] } ,
    \label{zBG}
\end{equation}

\begin{equation}
    E^2= \frac{R^3 f^2(r)}{-3Mr^4 \sqrt{R} \left ( R^3+2Q^2r^4\right ) },
\end{equation}

\begin{equation}
    L^2=\frac{r^4 \left [ Q^2(Q^2-r^2)-M\sqrt{R} (2Q^2-r^2) \right ] }{-3Mr^4 \sqrt{R} \left ( R^3+2Q^2r^4 \right )},
\end{equation}

\begin{equation}
    V^{\prime \prime}_{eff}= \frac{2 \left [ A M^2 + B M + C \right ] }{R^2 \left [-2 M r^2 \sqrt{R}+(R^2+Q^2r^2) \right ] \left[ -3 M r^4 \sqrt{R} +(R^3+2Q^2r^4) \right ] },
    \label{VppABG}
\end{equation}

\noindent where $A= -6r^6R^2$ , \quad $B=R^{3/2}(r^8 19Q^2r^6+9Q^4r^4-8Q^6r^2-8Q^8)$ and 
 $C=4Q^4R(Q^6-3Q^2r^4-3r^6)$. 

\end{widetext}

Since the analytic expressions for the energy and angular momentum come from the condition for existence of circular orbits, then $E^2>0$, $L^2>0$ must hold simultaneously for circular orbits, their stability requires $V^{\prime \prime}_{eff}>0$. Of course $z^2>0$ must hold as well. Hence we have a set of three conditions to have stable circular orbits in addition to $z^2 >0$, these four conditions are explicitly given by

\begin{equation}
    -3Mr^4\sqrt{R} + (R^3 + 2 Q^2 r^4 ) > 0,
    \label{c1}
\end{equation}

\begin{equation}
    Q^2(Q^2-r^2) -M \sqrt{R} (2Q^2-r^2) > 0,
    \label{c2}
\end{equation}

\begin{equation}
    -2 M r^2 \sqrt{R} + (R^2 +  Q^2 r^2) > 0,
    \label{c3}
\end{equation}

\begin{equation}
    AM^2+BM+C > 0.
    \label{c4}
\end{equation}

With the coefficients $A,B$ and $C$ given in (\ref{VppABG}). It is by using (\ref{zBG}), that one finds a formula for the mass parameter of Ayon-Beato-Garc\'ia regular BH $M=M(z,r,Q)$. (\ref{zBG}) is equivalent to a quadratic equation for $M$, whose solution is given explicitly by

\begin{equation}
    M_{\pm}= \frac{ G(r,z,Q)  \pm \sqrt{H(r,z,Q)}}{12r^4z^2\sqrt{R}},
    \label{MBG}
\end{equation}

\noindent where the function $G(r,z,Q)$ is
\begin{eqnarray} 
G(r,z,Q)&=&g_0(r,Q)+g2(r,Q) z^2,  \nonumber \\
g_0(r,Q) &=& R^2 (r^2 - 2Q^2), \nonumber \\ 
g_2(r,Q) &=& 2 Q^6 + 9 Q^4 r^2 + 19 Q^2 r^4 + 5 r^6,
\label{MBG2}
\end{eqnarray}

and the function $H(r,z,Q)$ is
\begin{eqnarray}
H(r,z,Q) &=& (r^2-2Q^2)R^4-2R^2h_2(r,Q) z^2+h_4^2(r,Q) z^4, \nonumber \\
h_2(r,Q) &=& 4Q^8+16Q^6 r^2 +17 Q^4 r^4+ 3 Q^2 r^6 -5 r^8, \nonumber \\
h_4(r,Q) &=& 2Q^6+3 Q^4 r^2 +Q^2r^4 -r^6.
\label{MBG3}
\end{eqnarray}

As aforementioned, the ABG BH behaves asymptotically as the Reisnner-Nordstrom spacetime, in particular, in this limit, we recovered the expression found in \cite{Becerril:2016qxf}, namely

\begin{equation}
    M_{\pm}(r,z,Q) = r \mathcal{G}_{\pm}, 
    \label{RNMass}
\end{equation}

\noindent where

\begin{eqnarray}
    \mathcal{G}_\pm &=& \frac{1}{12z^2} \Bigg\{ (1+5z^2) + \frac{7Q^2z^2}{r^2} \nonumber \\
    & & \pm \left (1+10z^2+z^4+\frac{Q^2 z^2}{r^2}\left [ \frac{Q^2 z^2}{r^2}-2(z^2+5)\right ]\right)^{1/2} \Bigg\}. \nonumber \\
    \label{RNG}
\end{eqnarray}

\begin{figure}
    \includegraphics[scale=0.7]{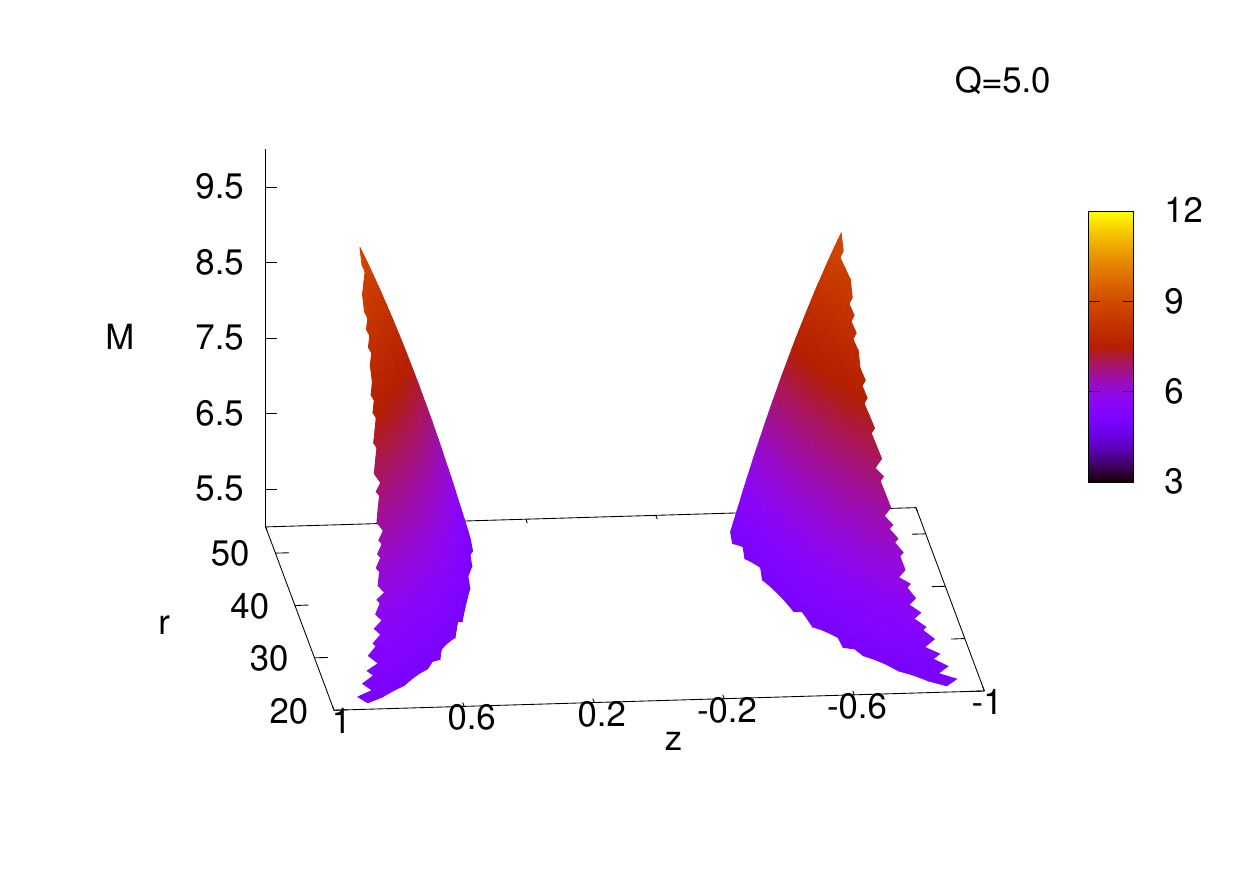}
    \caption{Scaled mass parameter of Reissner-Nordstr\"{o}m blak-hole with charge $Q=5$ as a function of $(r,z)$. Since the event horizon is given by $r_H=M \pm \sqrt{M^2-Q^2}$, the mass parameter must be larger than $Q$. Not all redshifts could be detected by a a faraway observer, none in the gap between the region for red and blueshifts.}
    \label{Cota_Ayon}
\end{figure}

If we take $Q=0$ (\ref{RNMass}) becomes the mass parameter expression for Schwarzschild. In  \cite{Becerril:2016qxf} it was proven that only $M_{-}$ in (\ref{RNMass}) was physically acceptable; hence, the uniqueness of the mass parameter was guaranteed. The external event horizon is given 
$r_H=M \pm \sqrt{M^2-Q^2}$, as a result, given a charge $Q$, the mass parameter must be greater than $|Q|$. This fact is illustrated in Fig. \ref{Cota_Ayon} for the case $Q=5$, as an example.

\begin{figure}[!h]
    \includegraphics[scale=0.7]{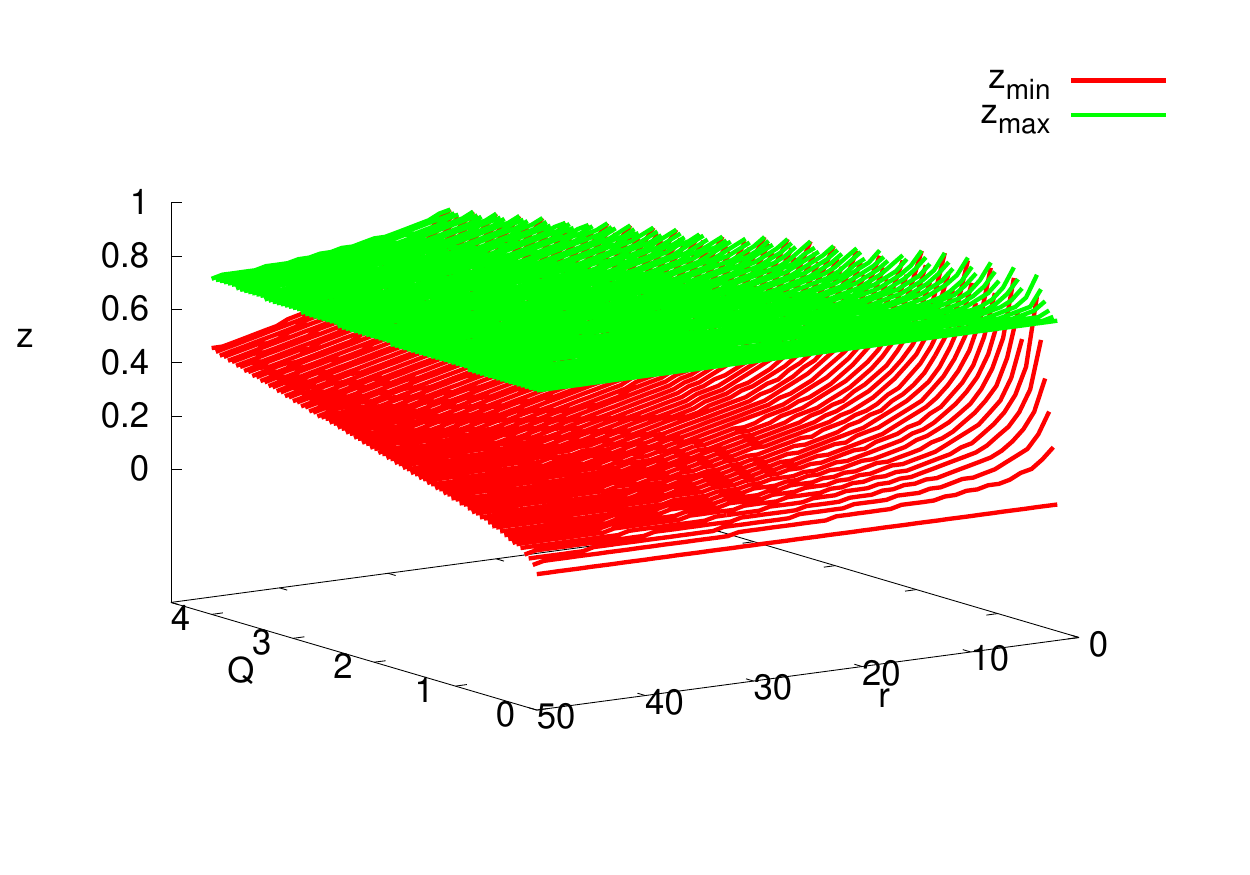}\\
    \vspace{-0.9cm}
    \includegraphics[scale=0.7]{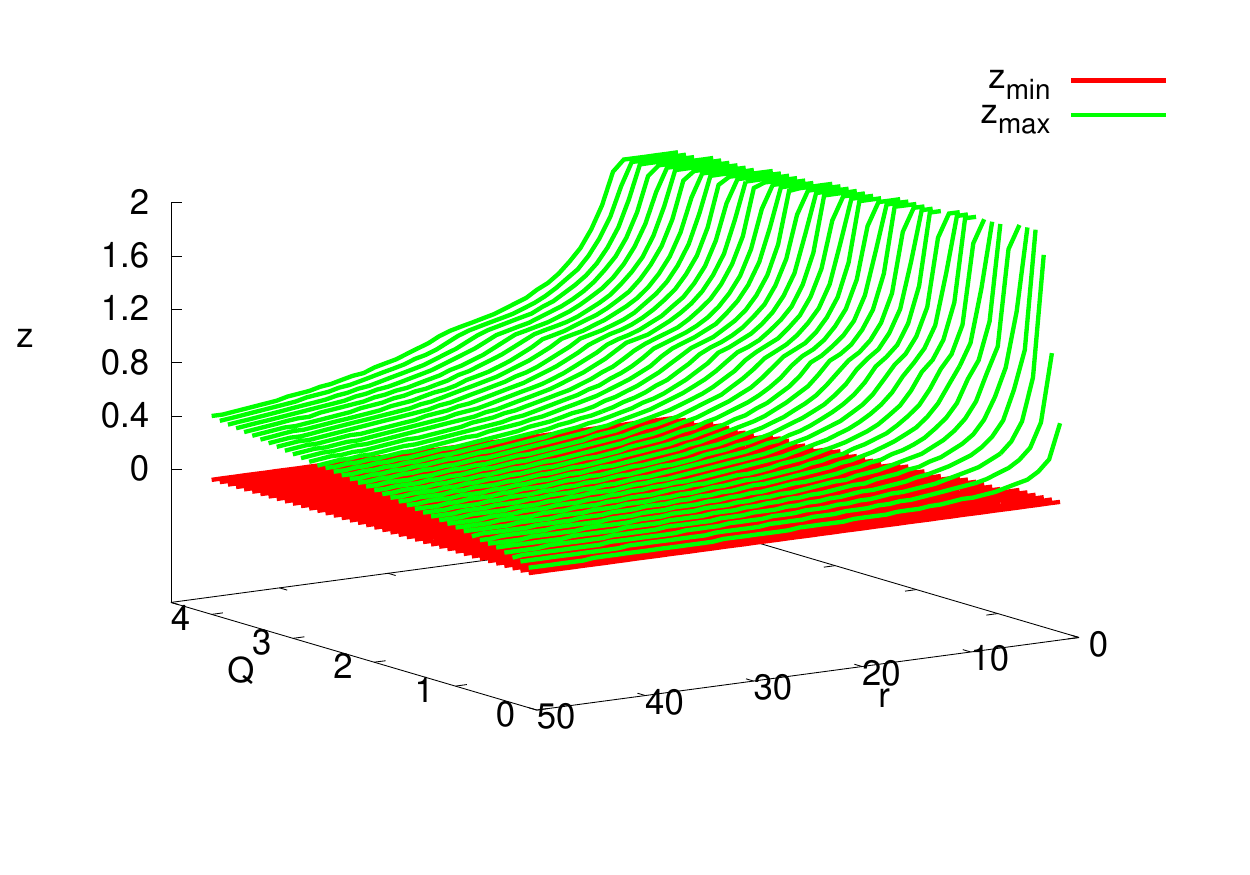}\\
    \vspace{-0.9cm}
     \includegraphics[scale=0.7]{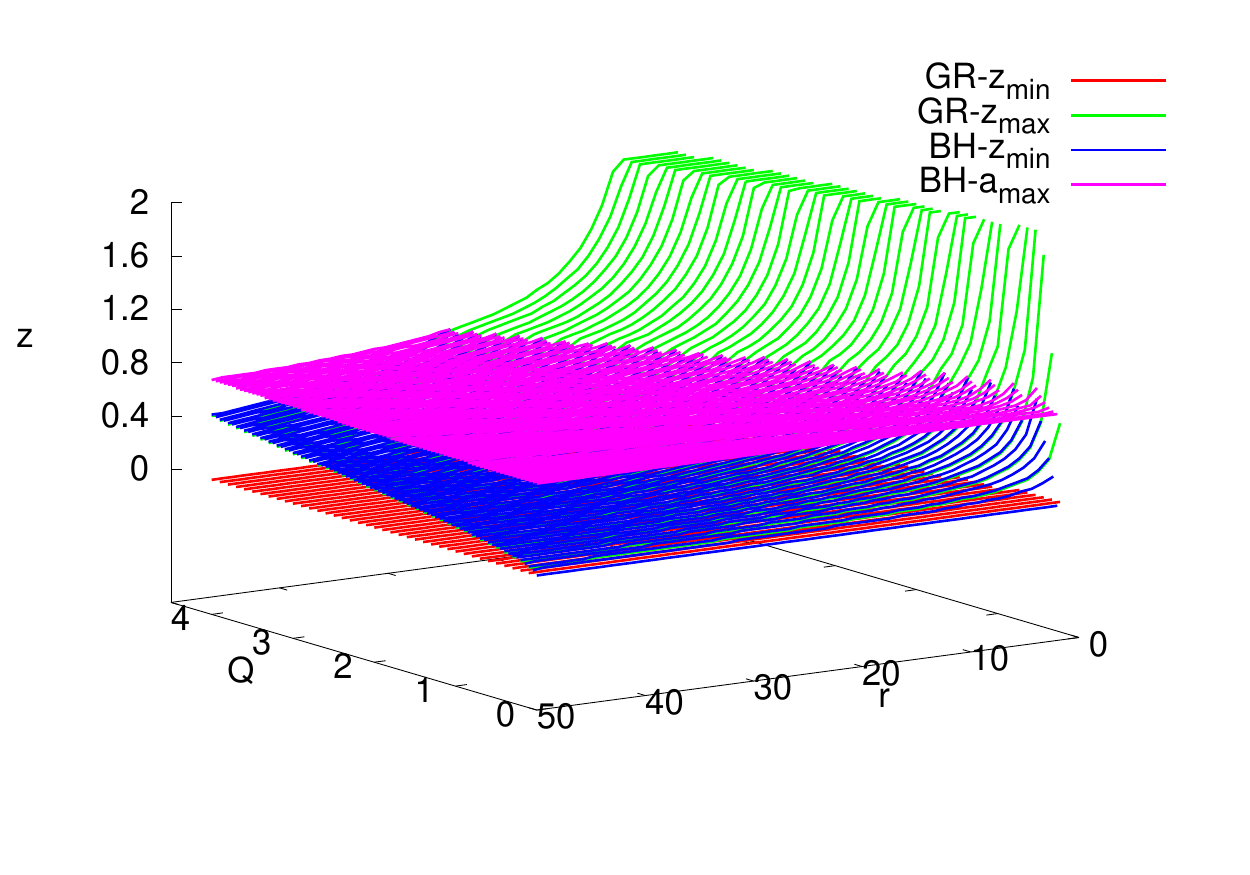}
    \caption{In the upper plot, we presents bounds of $z$ for ABG's BH. The only allowed frequency shifts that could be detected by a far away observer is located in the gap  $z_{min} < z < z_{max}$. In the middle plot, we present bounds of $z$ for globally regular ABG's spacetime. Only $z \in [z_{min},z_{max}]$ are allowed. The two surfaces $z_{max}$ and $z_{min}$ corresponding to ABG's globally regular and BH spacetimes respectively, coincide when $z<0.9$. Whereas, when $z>0.9$ only globally regular spacetime is allowed. This is shown in the lower plot where we have superimposed both bounds for ABG's BH and globally regular spacetimes}
    \label{Cota_Ayon1}
\end{figure}

In the ABG BH, the circular orbits stability condition $AM^2+BM+C > 0$ and the other aforementioned conditions (\ref{c1}), (\ref{c2}) and (\ref{c3}) should allow us to find bounds for $z$ and to determine whether for each point $P_{ijk}=(Q_i,r_j,z_k)$ there exists solely one mass $M$. We have performed the analysis following the numerical algorithm already described. 
It turns out that with $M_{+}$ there is not a single point in the domain $\mathcal{D}$ no matter how large we take it, for which all conditions are simultaneously fulfilled, on the other hand, working with $M_{-}$, there is a subset for the BH sector $\mathcal{D}_M^{BH} \subset \mathcal{D}$ where these conditions are simultaneously satisfied and those are considered physically acceptable. We partially show this subset in the upper plot of Fig.\ref{Cota_Ayon1}. $\mathcal{D}_M^{BH}$ spreads out between the surfaces  $z_{min}=z_{min}(Q,r)$ and $z_{max}=z_{max}(Q,r)$; at any rate, the shift is never greater than $0.9$. For $\tilde{Q}= Q/M > \tilde{Q}_c=0.634$, one encounters globally regular spacetimes, its corresponding $\mathcal{D}_M^{GR}$ is displayed in the middle plot of Fig. \ref{Cota_Ayon1}. The lower plot of Fig. \ref{Cota_Ayon1} presents the superimposed image of the above two.

\begin{figure*}
\begin{tabular}{cc}
\includegraphics[scale=0.7]{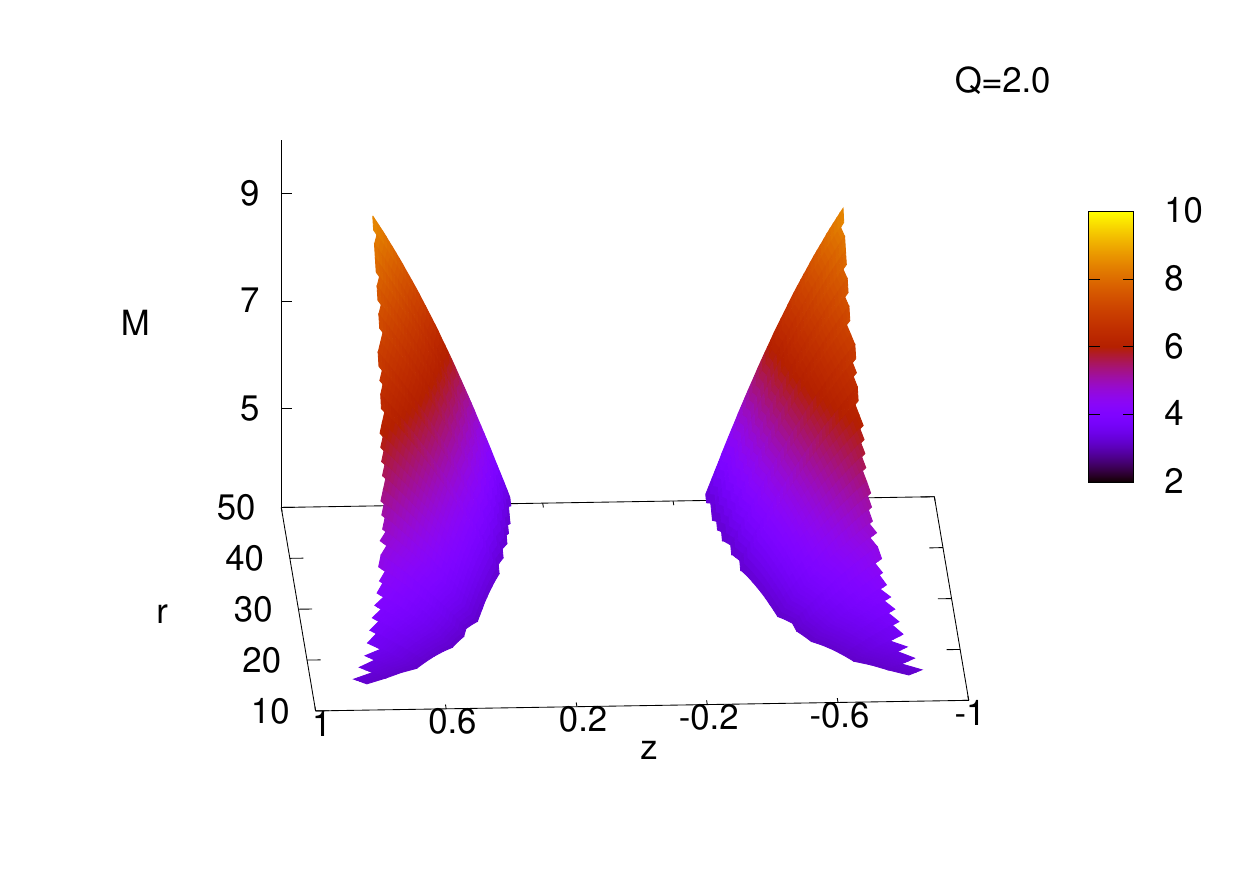} & \includegraphics[scale=0.7]{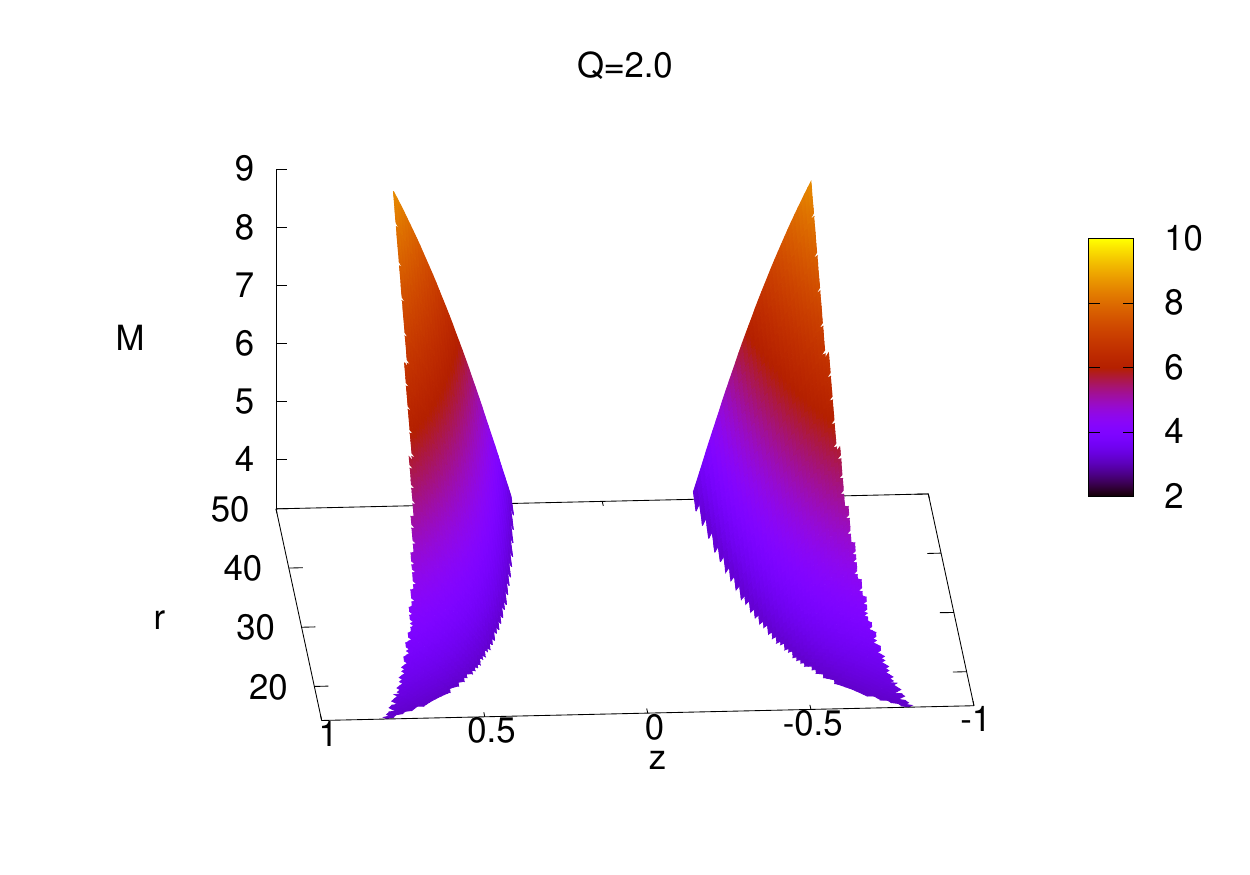}  \\
\includegraphics[scale=0.7]{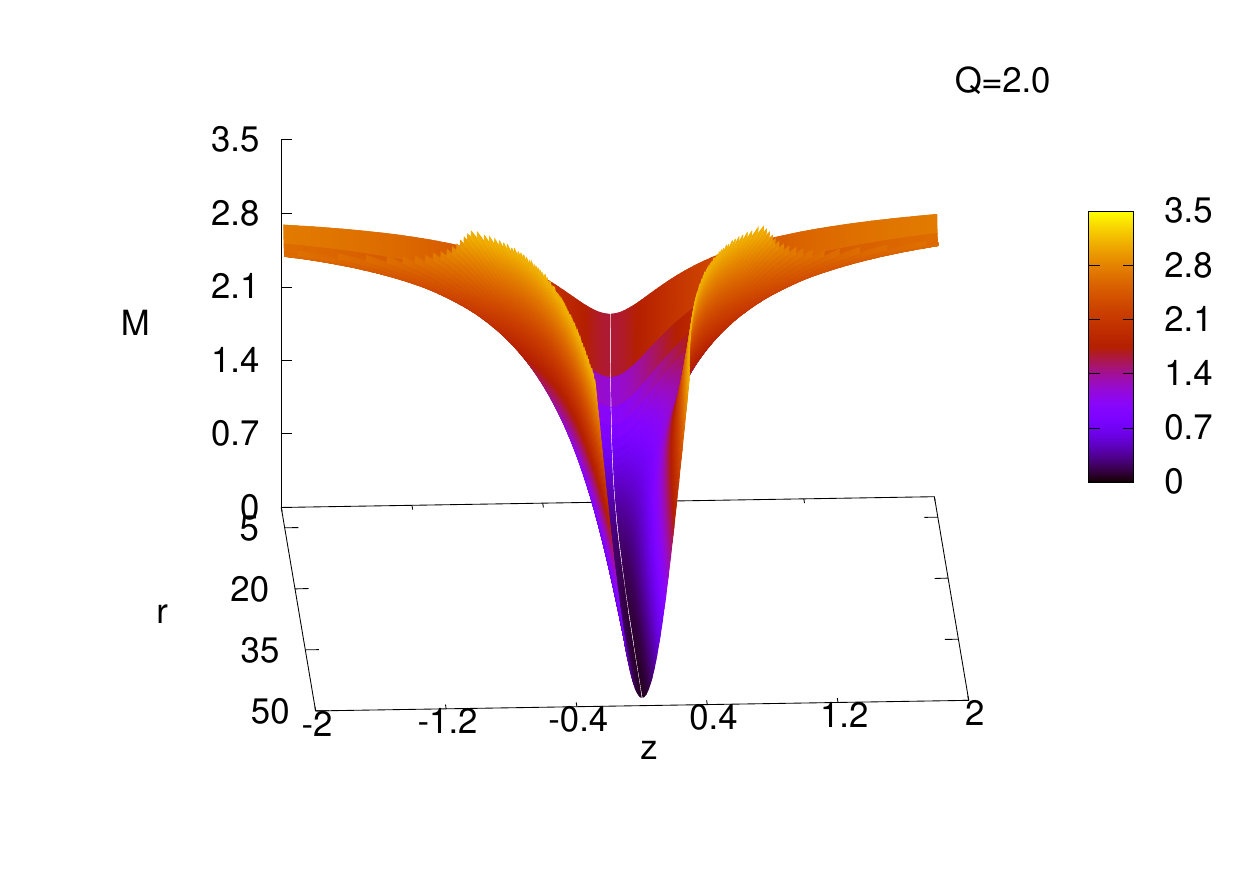} & \includegraphics[scale=0.7]{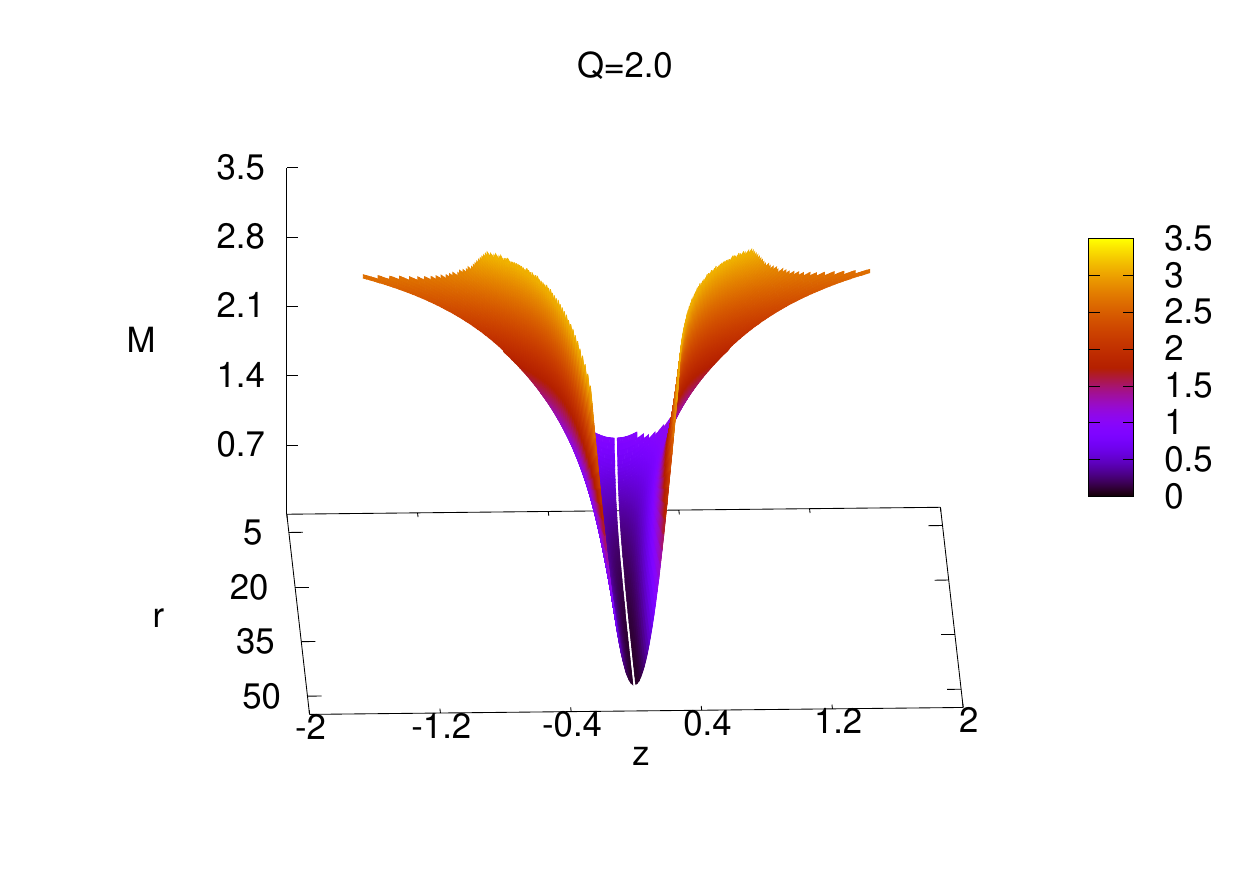} \\
\includegraphics[scale=0.7]{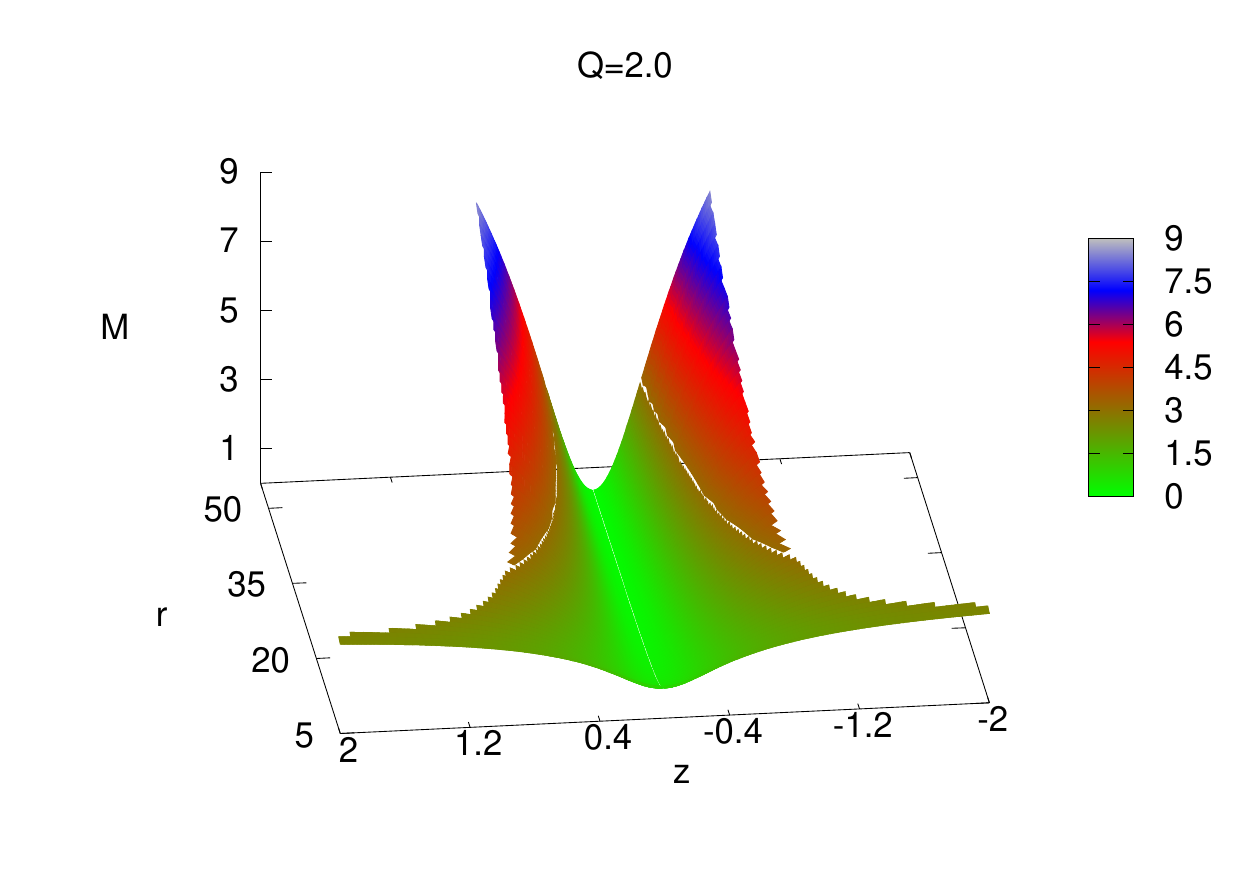} &  \includegraphics[scale=0.7]{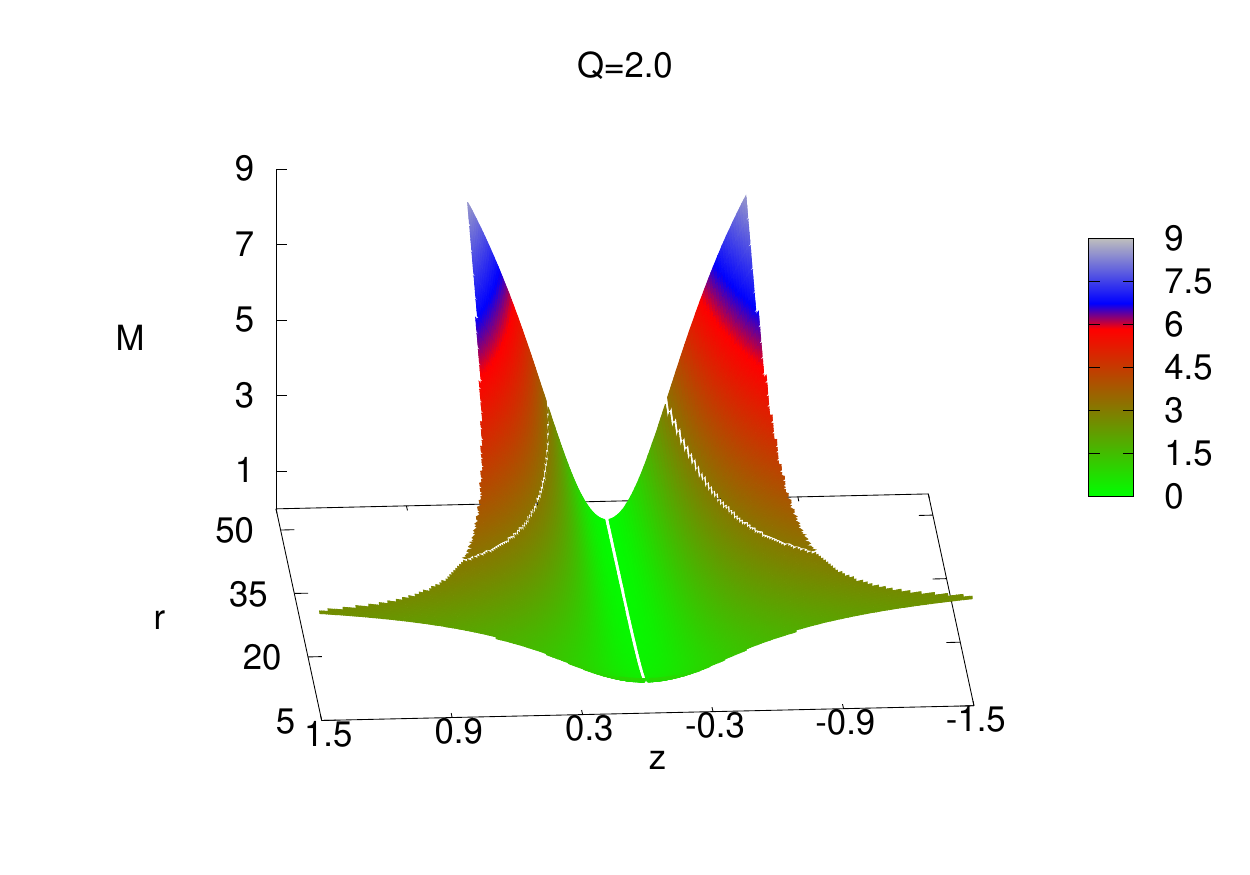}
\end{tabular}
\caption{The mass parameter $M=M(r,z,Q)$ computed with the {\it original} ABG-metric for $Q=2$ is presented on the left side of this figure, whereas $M=M(r,z,Q)$ constructed using the {\it effective} metrics is presented on the right side. $M(r,z,Q)$ for ABG's BH for $Q=2$ is shown in the first plot. The second plot displays $M(Q=2,r,z)$ for ABG's globally regular spacetime sector. The third plot superimposes both cases. Here, our numerical algorithm yields the bound that separates the BH from globally regular spacetime and agrees with the relationship $M= 1.57 Q$. On the left set of plot, we presented $M(r,z,Q=2)$ constructed with the effective metric, see text for details.}
\label{Masa_ABG}
\end{figure*}

The scaled mass parameter $M=M(Q,r,z)$ obtained when using the original metric for ABG spacetime is presented on the left set of plots in the Fig. \ref{Masa_ABG} for the value of the charge $Q=2$. The first plot corresponds to $M=M(Q=2,r,z)$ for ABG BH. As $Q$ increases, the gap between the zone for red and blueshifts broadens. The second plot (left side) in Fig. \ref{Masa_ABG} corresponds to the mass parameter $M(Q=2,r,z)$ for ABG's globally regular spacetime. The third plot we superimposed both cases. \\
 
{\bf B. ABG spacetime using the effective metric} \\

ABG regular spacetime was born as an exact solution of Einstein field equations in the context of nonlinear electrodynamics; hence, the corresponding effective metric should be employed, for this spacetime it reads

\begin{equation}
     \widetilde{g}_{tt}= g_{tt}/\Upsilon, \quad \widetilde{g}_{rr}= g_{rr}/\Upsilon, \quad \widetilde{g}_{\theta \theta}= g_{\theta \theta}, \quad \widetilde{g}_{\phi \phi}= g_{\phi \phi},
\end{equation}
with $\Upsilon$ given by

\begin{equation}
 \Upsilon =\frac{\eta^2\left[\left(r^2-5Q^2\right)+\frac{15M}{2}\eta\right]}{\left[\left(r^4-13Q^2r^2+10Q^4\right)+\frac{15M}{4}\left(3r^2-4Q^2\right)\eta\right]},\nonumber
\end{equation}

and
\begin{equation}
     \eta = \left(r^2+Q^2\right)^\frac{1}{2}.
\end{equation}

The relationship between $z$,$r$,$Q$ and $M$ reads for this case

\begin{figure}[!h]
    \includegraphics[scale=0.68]{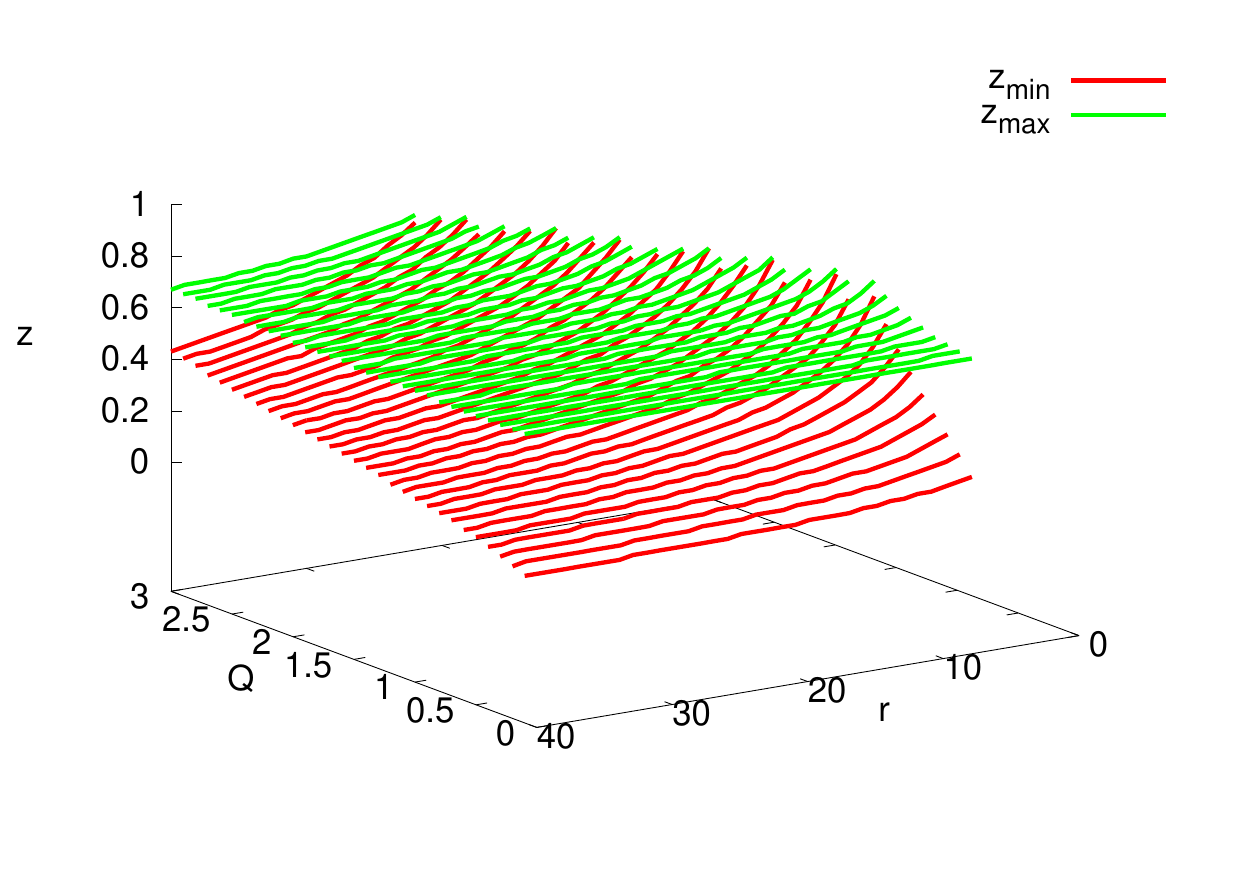}\\
    \vspace{-0.9cm}
    \includegraphics[scale=0.68]{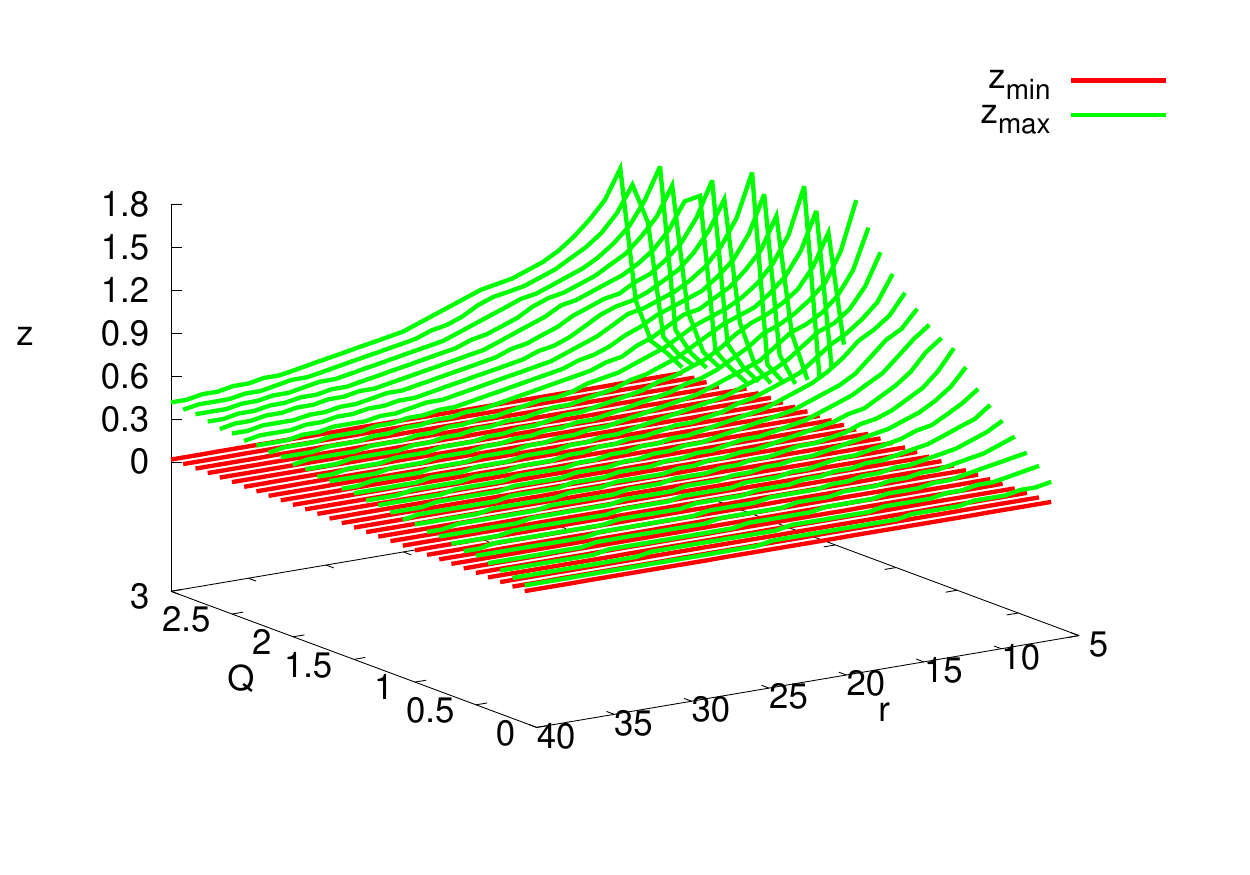}
    \caption{We presents bounds of $z$ for ABG's BH when using the {\it effective metric} in the upper plot. There exist a unique value of $M(r,z,Q)$ for those frequency shifts that are located in the gap  $z_{min} < z < z_{max}$. In the second plot, we present bounds of $z$ for globally regular ABG's spacetime, again, only $z \in [z_{min},z_{max}]$ are allowed. The two surfaces $z_{max}$ and $z_{min}$ corresponding to ABG's globally regular and BH spacetimes respectively, coincide when $z<0.81$. Whereas, when $z>0.81$ only a globally regular spacetime is allowed.}
    \label{Cota_Ayon_Effec}
\end{figure}

\begin{widetext}
\begin{equation}
    z^2 = \frac{2r^3\left(5Q^2-r^2-\frac{15}{2}M\sqrt{R}\right)(Q^2(r^2-Q^2)\sqrt{R}+M(2Q^2 +Q^2r^2-r^4)}{R^{5/2}\left(10Q^4+ r^2(r^2 -13Q^2)+\frac{15}{4}M\sqrt{R}(3r^2-4Q^2)\right)
    \left(\frac{4r^7}{R^3}+r^5\left(\frac{6M}{R^{5/2}}-\frac{4}{R^2}\right)-2r\right)
    \left(r^2\left(\frac{2M}{R^{3/2}}-\frac{Q^2}{R^2}\right) -1\right)},
\end{equation}

from it, $M=M(r,z,Q)$ can be attained by solving the cubic
\begin{equation}
    C_3M^3 + C_2M^2 +C_1 M + C_0=0,
\end{equation}

with 
\begin{eqnarray}
C_3 &=& 90r^6z^2(4Q^2-3r^2)R^{11/2},\nonumber\\
C_2 &=& 3r^2R^5(10(r^2-2Q^2)R^3 + z^2(67r^8+289Q^2r^6-325Q^4r^4-150Q^6r^2-40Q^8),\nonumber\\
C_1 &=& R^{9/2}(2r^2R^3(35Q^4-29Q^2r^2+2r^4)+z^2(60Q^{12}+395Q^{10}r^2+886Q^8r^4+765Q^6r^6-1127Q^4r^8-484Q^2r^{10}-25r^{12}),\nonumber\\
C_0 &=& -4(Q^2r^2R^7(5Q^4-6Q^2r^2+r^4)-R^4z^2(10Q^4-13Q^2r^2+r^4)(Q^6+3Q^4r^2+5Q^2r^4+r^6).\nonumber
\end{eqnarray}
\end{widetext}
We have also verified numerically, that only one real value of the mass parameter allows the conditions  for stability of circular orbits to hold. Moreover, these conditions are fulfilled only for $|z|<0.81$,
considerably higher than the Schwarzschild fixed bound of $1/\sqrt{2}$. Figure \ref{Cota_Ayon_Effec} shows bounds of the frequency shifts for the BH (upper plot) and GR (middle plot) sectors.

On the right side of Fig. \ref{Masa_ABG}, we show $M(r,z,Q)$ computed with the effective metric for $Q=2$. Choosing the point $(r=50,Q=0.1)$ and looking at $z_{max}(50,0.1)=0.62$ we observe that $M_{eff}(r=50,Q=0.1,z=0.62)=8.2717$, for another point with higher $Q$, $M_{eff}(r=50,Q=2,z_{max}(50,2))=8.5759$, generally speaking, as $Q$ climbs up, so does $M$. When employing the original metric, the largest values attained in the same points are $M_{o}(r=50,Q=0.1,z_{max}(50,0.1)=0.7) = 8.2662$ and $M_{o}(r=50,Q=2,z_{max}(50,2)=0.72) = 8.5566$. This is a typical behavior, we observed that $M_{eff}(r,Q,z_{max}(r,Q)) > M_0(r,Q,z_{max}(r,Q))$ generally holds, be aware that $z_{max}(r,Q)$ is usually different when is calculated with the original metric $g_{\mu \nu}$ or with the effective metric $\widetilde{g}_{\mu \nu}$.

 To close this section, the angular velocity $\Omega$ of photon emitters orbiting along stable circular orbits is computed. The relationship $\Omega=\sqrt{f^{\prime}/r^4}$ is employed, for ABG metric it reads

\begin{equation}
    \Omega(Q,r,z)=\frac{\sqrt{Q^2(Q^2-r^2)-MR^{1/2}(2Q^2-r^2)}}{R^{3/2}},
\end{equation}

\noindent where $R=Q^2+r^2$. The angular velocity becomes a function of $l,r,z$ after inserting $M_{-}(Q,r,z)$ solution of the quadratic equation (\ref{MBG}).  $\Omega$ is indeed a real quantity since
$Q^2(Q^2-r^2)-MR^{1/2}(2Q^2-r^2)>0$ in compliance with (\ref{c2}).

\section{Summary and Conclusions}
\label{Summary and Conclusion}

In this paper, we have determined the mass parameter of regular spacetimes (Bardeen, Hayward and Ayon-Beato-Garcia) in terms of the frequency shifts $z$ of light emitted by particles traveling along circular geodesics of radii $r$ orbiting in these spaces. We have sorted out our study in two categories, first when we use the original metric of these spacetimes, and second, when we use the effective metric that embodies the effects of the nonlinear electromagnetic fields, since light travels in null geodesics in effective geometries. In spite of this fact, one can find papers that study null geodesics in BH with NED without using the effective metric, we wanted to explore the difference that we could find in our study when one uses the original and the effective metrics, and then perform a comparison. 

We have found explicit formulas for the mass parameter using the original metric $M_o(r,z,p)$ and the effective metric $M_{eff}(r,z,p)$ for the three regular spacetimes considered in this paper, $p$ stands for the parameters $g,l,Q$ (Bardeen, Hayward and ABG parameters respectively). The three working examples have a BH sector which possess two event horizon, we work outside the exterior one, and a globally regular spacetime sector. 

Not all values of $z$ would be detected from a faraway observer, that is, there are bounds for the frequency shifts. For the BH sector, there is a subset $\mathcal{D}_M^{BH} \subset \;\mathcal{D}$ where all conditions for existence of photon emitters' stable circular orbits are simultaneously satisfied (together with the additional condition $r>r_H^{ext}$), in that subset, the mass parameter is unique, this was numerically proven. Either using the original $g_{\mu \nu}$ or the effective $\widetilde{g}_{\mu \nu}$ metric, in order to find $M=M(r,z,p)$ one has to solve a polynomial of order two, three or four. We found this domain for BH and GR spacetimes and plotted it for our three working examples. These domains (for BH and for GR spacetimes) are bounded by two surfaces $z_{min}(r,p)$ and $z_{max}(r,p)$, we observed that the gap between them narrows as $p$ increases. For the Bardeen BH using the original metric, we found that $z_{max}(r,g)<0.85$ and $z_{max}(r,g)<0.71$ using the effective metric. For the Hayward BH using the original metric, we found that $z_{max}(r,g)<0.75$ and $z_{max}(r,g)<0.53$ using the effective metric. For the ABG BH using the original metric, we found that $z_{max}(r,g)<0.9$ and $z_{max}(r,g)<0.81$ using the effective metric. Therefore, there is a considerable difference when using the original or the effective metric. In other words, the domain size $\mathcal{D}_M^{BH}=\{(r,z,p)\}$ where there exist a mass parameter satisfying all the conditions for existence of circular stable orbits of light emitters, vary significantly. However, the difference between $M_o(r,z,p)$ and $M_{eff}(r,z,p)$ is rather small. In the three working examples it happens that  $M_{eff}(r,p,z_{max}(r,p)) > M_0(r,p,z_{max}(r,p))$. 

The mass parameter for the BH sector is considerably larger than those for the GR sector. 
For the three cases, as the corresponding parameter increases, the gap between the region for red and blueshifts broadens. The angular velocity for photons emitters orbiting along to stable circular orbits were found for the three working examples. A study for rotating regular BHs is being carried out and we will report the findings somewhere else.

\begin{acknowledgments}
We would like to thank the anonymous referee for making critical comments which helped to improve the manuscript.
R.B. and U.N. acknowledges support of the present research from FORDECYT-PRONACES-CONACYT under grant No. CF-MG-2558591, \textit{Sistema Nacional de Investigadores} of the \textit{Consejo Nacional de Ciencia y Tecnolog\'{i}a} (SNI-CONACYT) and \textit{Coordinación de la Investigaci\'{o}n Cient\'{i}fica} of the \textit{Universidad Michoacana de San Nicol\'{a}s de Hidalgo} (CIC-UMSNH). U.N. also acknowledges support from PRODEP-SEP. P.S. would like to thank \textit{Programa de Desarrollo Profesional Docente} (PRODEP) of the \textit{Secretar\'{\i}a de Educaci\'on P\'{u}blica} (SEP) of the Mexican government, for providing the financial support. S.V-A and J.M.D acknowledges partial support from \textit{Secretaria de Investigaci\'on y Estudios Avanzados} of the \textit{Universidad Aut\'onoma del Estado de México} (SIEA-UAEM\'ex) under No. 5050/2020CIB project. S.V-A also acknowledges partial support from PRODEP-SEP, under No. 4025/2016RED project. 
\end{acknowledgments}

\appendix
\section{Effective geometry}
\label{effective_geo}

The action of gravitational field coupled to nonlinear electromagnetic field reads
\begin{equation}
    \mathcal{I} = \left(\frac{1}{16\pi}\right)\int{\sqrt{-g}\left[R-4\mathcal{L}(F)\right]dx^{4}},
\end{equation}
where scalar $F\equiv (1/4)F_{\alpha\beta}F^{\alpha\beta}$, Faraday tensor related to the vector potential via relation $F_{\alpha\beta}=\partial_{\alpha}A_{\beta}-\partial_{\beta}A_{\alpha}$, and $\mathcal{L}$ is the Lagrange function representing the NED.

The corresponding Einstein field equation derive from least action principle takes the form
\begin{equation}
    G_{\alpha\beta}=8\pi T_{\alpha\beta}=2\left[\mathcal{L}_{F}F_{\alpha\mu}F^{\mu}_{\beta}-g_{\alpha\beta}\mathcal{L}(F)\right],
\end{equation}
where $\mathcal{L}_{F}\equiv\partial_{F}\mathcal{L}(F)$ has been introduced.

Using symmetry of spacetime defined by Eq. (\ref{metric}) and a particular Lagrange function $\mathcal{L}(F)$ one can construct the corresponding spacetime metric in NED.

It has been argued in \cite{Novello:1999pg} that motion of photons around NED regular black holes is governed not by the null geodesics of the (original) spacetime geometry, but by the null geodesics of an effective geometry that incorporates
effects of the electrodynamic nonlinearities. 

These null geodesics that are defined in effective geometry follow from the Bianchi identities for the Faraday tensor $F_{\alpha\beta}$ which for the NED case read
\begin{eqnarray}
    &&\left(\mathcal{L}_{F}F^{\alpha\beta}\right)_{;\beta}=0,
    \label{Bian1}\\
    && F_{\alpha\beta;\mu}+F_{\mu\alpha;\beta}+F_{\beta\mu;\alpha}=0.
    \label{Bian2}
\end{eqnarray}
Using Bianchi identities (\ref{Bian1}) and (\ref{Bian2}), one can obtained the propagation equation
\begin{equation}
    k^{\mu}k_{\mu}-\left(\frac{\mathcal{L}_{FF}}{\mathcal{L}_{F}}\right)F^{\mu}_{\alpha}F^{\alpha\beta}k_{\mu}k_{\beta}=0,
    \label{k}
\end{equation}
where  $\mathcal{L}_{FF}\equiv\partial^{2}_{F}\mathcal{L}(F)$, wave vector $k_{\alpha}\equiv-\nabla_{\alpha}S$ and $S$ is the wave phase. Further, Eq. (\ref{k}) can be rearranged to give
\begin{equation}
    \left[g^{\mu\beta}-\left(\frac{\mathcal{L}_{FF}}{\mathcal{L}_{F}}\right)F^{\mu}_{\alpha}F^{\alpha\beta}\right]k_{\mu}k_{\beta}=0.
    \label{gen_k}
\end{equation}
This is the general equation describing motion of the photons in the spacetimes representing the solution of NED coupled with GRT. It is clearly seen from the above Eq. (\ref{gen_k}) that this is the modified normalization condition for photons related to the effective metric
\begin{equation}
g^{\alpha\beta}_{eff}\equiv\tilde{g}^{\alpha\beta}=g^{\alpha\beta}-\left(\frac{\mathcal{L}_{FF}}{\mathcal{L}_{F}}\right)F^{\alpha}_{\mu}F^{\mu\beta}.
\label{g_eff}
\end{equation}

The details of how to derive effective geometry can be found in
\cite{Novello:1999pg,2019ApJ,Schee_2019,2019EPJC...79...44S}

Now, we will discuss the Bardeen, Hayward and ABG  BHs effective spacetime cases respectively, depending upon the specific choice of $\mathcal{L}(F)$.

\textit{Case} I: For Bardeen BH, the Lagrange function $\mathcal{L}(F)$ is 

\begin{equation}
    \mathcal{L}(F)=\frac{3M}{q_m^3}\left ( \frac{\sqrt{2q_m^2 F}}{1+\sqrt{2q_m^2 F}} \right)^{\frac{5}{2}},
    \label{Lf_B}
\end{equation}

with the Faraday tensor of the form

\begin{equation}
    F_{\mu \nu}=2 \delta^{\theta}_{[\mu}\delta^{\phi}_{\nu}]q_m \sin{\theta}, 
    \label{Fab_B}
\end{equation}

that leads to the scalar of the electromagnetic field $F$
\begin{equation}
    F= \frac{q_m^{2}}{2r^4},
    \label{F_B}
\end{equation}
where the NED field is generated by a magnetic monopole charge $q_m$.
Now, using Eqs.~\eqref{g_eff}~--~\eqref{F_B}, the nonzero contravariant components of Bardeen metric in effective geometry are
\begin{align}
    \tilde{g}^{tt}_B &= g^{tt}_B,\label{eq:gttu}\\
    \tilde{g}^{rr}_B &= g^{rr}_B,\label{eq:grru}\\
    \tilde{g}^{\theta\theta}_B &= g^{\theta\theta}_B\Phi,\label{eq:gththu}\\
    \tilde{g}^{\phi\phi}_B &= g^{\phi\phi}_B\Phi,\label{eq:gppu}
\end{align}
and covariant components become
\begin{align}
    \tilde{g}_{tt}^B &= g_{tt}^B,\label{eqn:gttd}\\
    \tilde{g}_{rr}^B &= g_{rr}^B,\label{eqn:grrd}\\
    \tilde{g}_{\theta\theta}^B &= \frac{g_{\theta\theta}^B}{\Phi},\label{eqn:gththd}\\
    \tilde{g}_{\phi\phi}^B &= \frac{g_{\phi\phi}^B}{\Phi},\label{eqn:gppd}
\end{align}
where
\begin{equation}
    \Phi=1+2F\left(\frac{\mathcal{L}_{FF}}{\mathcal{L}_{F}}\right) = 1+\frac{1}{2}\left[\frac{r^2-6g^2}{r^2+g^2}\right],
\label{eqn:Phi}
\end{equation}
where $g\equiv q_{m}$.
In the limit, when $r \rightarrow \infty$, $\Phi$ becomes the value $3/2$.

\textit{Case} II: For Hayward BH, the Lagrange function $\mathcal{L}(F)$ is 
\begin{equation}
    \mathcal{L}(F)=\frac{\mathcal{A}}{\mathcal{B}}\frac{\left(\mathcal{B}F\right)^{\frac{3}{2}}}{\left[1+\left(\mathcal{B}F\right)^{\frac{3}{4}}\right]^2},
    \label{Lf_H}
\end{equation}
where the constants $\mathcal{A/B}\equiv 3/(2l^2)$ and $\mathcal{B}\equiv[2(2l^{2}M)^\frac{4}{3}]/q_{m}^{2}$. 

Now, using Eqs.~\eqref{g_eff}, \eqref{Fab_B}, \eqref{F_B} and \eqref{Lf_H}, the nonzero contravariant components of Hayward metric in effective geometry read as
\begin{align}
    \tilde{g}^{tt}_H &= g^{tt}_H,\label{eq:gttu_H}\\
    \tilde{g}^{rr}_H &= g^{rr}_H,\label{eq:grru_H}\\
    \tilde{g}^{\theta\theta}_H &= g^{\theta\theta}_H\Theta,\label{eq:gththu_H}\\
    \tilde{g}^{\phi\phi}_H &= g^{\phi\phi}_H\Theta,\label{eq:gppu_H}
\end{align}
and covariant components become
\begin{align}
    \tilde{g}_{tt}^H &= g_{tt}^H,\label{eqn:gttd_H}\\
    \tilde{g}_{rr}^H &= g_{rr}^H,\label{eqn:grrd_H}\\
    \tilde{g}_{\theta\theta}^H &= \frac{g_{\theta\theta}^H}{\Theta},\label{eqn:gththd_H}\\
    \tilde{g}_{\phi\phi}^H &= \frac{g_{\phi\phi}^H}{\Theta},\label{eqn:gppd_H}
\end{align}
where
\begin{equation}
    \Theta=1+\left[\frac{1-\frac{7}{2}\left(\frac{2l^{2}M}{r^{3}}\right)}{1+\left(\frac{2l^{2}M}{r^{3}}\right)}\right].
\end{equation}
Similar to Bardeen case, in the asymptotic limit (i.e., $r\rightarrow \infty$), $\Theta\rightarrow 2$.

\textit{Case} III: For ABG BH, the Lagrange function $\mathcal{L}$ is calculated from the Legendre transformation \cite{AyonBeato:1998ub}:
\begin{equation}
    \mathcal{H}\equiv 2F \mathcal{L}_{F}-\mathcal{L}.
    \label{eq:H}
\end{equation}
Now, using the definition
\begin{equation}
    P_{\mu\nu}\equiv \mathcal{L}_{F}F_{\mu\nu}=2 \delta^{t}_{[\mu}\delta^{r}_{\nu}]\frac{q_{e}}{r^2} ,
    \label{eq:P_ab}
\end{equation}
the scalar of $P_{\mu\nu}$ reads
\begin{equation}
P\equiv \frac{1}{4}P_{\mu\nu}P^{\mu\nu}=(\mathcal{L}_{F})^{2}F.
\label{eq:P}
\end{equation}
It is also shown in \cite{AyonBeato:1998ub} that $\mathcal{H}$ which is used to define the NED source for ABG BH as function of $P$, is given as
\begin{equation}
    \mathcal{H}(P)=P\frac{\left(1-3\sqrt{-2Q^{2}P}\right)}{\left(1+\sqrt{-2Q^{2}P}\right)^{3}}-\frac{3}{2Q^{2}s}\left(\frac{\sqrt{-2Q^{2}P}}{1+\sqrt{-2Q^{2}P}}\right)^{\frac{5}{2}},
    \label{eq:H_P}
\end{equation}
where $Q\equiv q_{e}$, $s\equiv\abs{Q}/2m$, $P=-Q^{2}/(2r^{4})$. Using Eqs. \eqref{eq:H} and \eqref{eq:P}, the functions $\mathcal{L}_{F}$ and  $\mathcal{L}_{FF}$  come out as
\begin{align}
    \mathcal{L}_{F} &=\frac{1}{\mathcal{H}_{P}},\label{eq:L_F}\\
     \mathcal{L}_{FF} &= -\frac{\mathcal{H_{PP}}}{\left(\mathcal{H}_{P}+2P\mathcal{H}_{PP}\right)\mathcal{H}_{P}^{3}},
     \label{eq:L_FF}
\end{align}
where $\mathcal{H}_{P}\equiv \partial_{P}\mathcal{H}(P)$ and $\mathcal{H}_{PP}\equiv \partial_{P}\mathcal{H_{P}}$. Finally,  the nonzero contravariant components of ABG metric in effective geometry, calculated using Eq. \eqref{g_eff} and \crefrange{eq:P_ab}{eq:L_FF}  are
\begin{align}
    \tilde{g}^{tt}_{ABG} &= g^{tt}_{ABG}\Upsilon,\label{eq:gttu_ABG}\\
    \tilde{g}^{rr}_{ABG} &= g^{rr}_{ABG}\Upsilon,\label{eq:grru_ABG}\\
    \tilde{g}^{\theta\theta}_{ABG} &= g^{\theta\theta}_{ABG},\label{eq:gththu_ABG}\\
    \tilde{g}^{\phi\phi}_{ABG} &= g^{\phi\phi}_{ABG},\label{eq:gppu_ABG}
\end{align}
and covariant components become
\begin{align}
    \tilde{g}_{tt}^{ABG} &= \frac{g_{tt}^{ABG}}{\Upsilon},\label{eqn:gttd_ABG}\\
    \tilde{g}_{rr}^{ABG} &= \frac{g_{rr}^{ABG}}{\Upsilon},\label{eqn:grrd_ABG}\\
    \tilde{g}_{\theta\theta}^{ABG} &= g_{\theta\theta}^{ABG},\label{eqn:gththd_ABG}\\
    \tilde{g}_{\phi\phi}^{ABG} &= g_{\phi\phi}^{ABG},\label{eqn:gppd_ABG}
\end{align}
where
\begin{align}
        \Upsilon &=\frac{\eta^2\left[\left(r^2-5Q^2\right)+\frac{15M}{2}\eta\right]}{\left[\left(r^4-13Q^2r^2+10Q^4\right)+\frac{15M}{4}\left(3r^2-4Q^2\right)\eta\right]},\\
    \text{and}~
    \eta &= \left(r^2+Q^2\right)^\frac{1}{2}.
\end{align}
Further, for the limiting cases:
\begin{itemize}
    \item when $r\rightarrow\infty$, we have $\Upsilon=1$.
    \item And also, when parameter $Q\rightarrow 0$, $\Upsilon=\left[r+(15/2)M\right]/\left[r+(45/4)M\right]$ which again in the asymptotic limit $r\rightarrow\infty$ becomes unity.
\end{itemize}

It is important to note here that the effective metric components found here for Bardeen, Hayward and ABG BHs   
are different from what have been already published in the literature \cite{2019ApJ,Schee_2019,2019EPJC...79...44S} by the multiplication of a conformal factor proportional to  $\mathcal{L}_{F}^{-1}$. This fact, of course, does not affect the light trajectories because the geodesic equations on the effective spacetime is invariant under a conformal transformation. Interestingly, the conformal effective metrics for Bardeen and Hayward BHs found by us are asymptotically flat except for a deficit solid angle (the study of this class of asymptotic behavior was done in (\cite{Nucamendi_1997})) and for the corresponding ABG BHs are asymptotically flat.

\bibliography{my}

\end{document}